\documentclass{aa}
\usepackage{graphicx}
\usepackage{txfonts,textcomp}
\usepackage{multirow}
\usepackage{subcaption}
\usepackage{adjustbox}
\usepackage{amssymb}
\usepackage{xcolor}
\usepackage{colortbl}
\usepackage[colorlinks=true,allcolors=blue]{hyperref}

\graphicspath{{./}{figures/}}

\begin{document}

	\title{HYPERION. Coevolution of supermassive black holes and galaxies at $z>6$ and the build-up of massive galaxies}
 \titlerunning{HYPERION. Coevolution of SMBHs and galaxies at $z>6$}

	\author{R. Tripodi\fnmsep\thanks{\email{roberta.tripodi@inaf.it}}
		\inst{1,2,3,4}
		\and C. Feruglio
		\inst{2,3}
		\and F. Fiore
		\inst{1,2,3}
		\and L. Zappacosta
		\inst{5}
		\and E. Piconcelli
		\inst{5}
		\and M. Bischetti
		\inst{1}
		\and A. Bongiorno
		\inst{5}
		\and S. Carniani
		\inst{6}
		\and F. Civano
		\inst{7}
		\and C.-C. Chen
		\inst{8}
		\and S. Cristiani
		\inst{2,3}
		\and G. Cupani
		\inst{2,3}
		\and F. Di Mascia
		\inst{6}
		\and V. D'Odorico
		\inst{1,3,6}
		\and X. Fan
		\inst{9}
		\and A. Ferrara
		\inst{6}
		\and S. Gallerani
		\inst{6}
		\and M. Ginolfi
		\inst{10,11}
		\and R. Maiolino
		\inst{12,13,14}
		\and V. Mainieri
		\inst{15}
		\and A. Marconi
		\inst{10}
		\and I. Saccheo
		\inst{5}
		\and F. Salvestrini
		\inst{2}
		\and A. Tortosa
		\inst{5}
		\and R. Valiante
		\inst{5}
	}
	
	\institute{Dipartimento di Fisica, Università di Trieste, Sezione di Astronomia, Via G.B. Tiepolo 11, I-34131 Trieste, Italy
		\and
		INAF - Osservatorio Astronomico di Trieste, Via G. Tiepolo 11, I-34143 Trieste, Italy
		\and
		IFPU - Institute for Fundamental Physics of the Universe, via Beirut 2, I-34151 Trieste, Italy
		\and
		University of Ljubljana, Department of Mathematics and Physics, Jadranska ulica 19, SI-1000 Ljubljana, Slovenia
		\and 
		INAF - Osservatorio Astronomico di Roma, Via Frascati 33, I-00040 Monte Porzio Catone, Italy
		\and 
		Scuola Normale Superiore, Piazza dei Cavalieri 7, 56126, Pisa, Italy
		\and 
		NASA Goddard Space Flight Center, Greenbelt, MD 20771, USA
		\and
		Academia Sinica Institute of Astronomy and Astrophysics (ASIAA), No. 1, Sec. 4, Roosevelt Road, Taipei 10617, Taiwan
		\and 
		Steward Observatory, University of Arizona, Tucson, Arizona, USA
		\and
		Università di Firenze, Dipartimento di Fisica e Astronomia, via G. Sansone 1, 50019 Sesto F.no, Firenze, Italy
		\and 
		INAF - Osservatorio Astrofisico di Arcetri, Via Largo E. Fermi 5, 50125 Firenze, Italy
		\and
		Institute of Astronomy, University of Cambridge, Madingley Road, Cambridge CB3 0HA, UK
		\and
		Kavli Institute for Cosmology, University of Cambridge, Madingley Road, Cambridge CB3 0HA, UK
		\and 
		Department of Physics and Astronomy, University College London, Gower Street, London WC1E 6BT, UK
		\and
		European Southern Observatory, Karl–Schwarzschild–Straße 2, D-85748 Garching bei München, German
		\\
	}
	
	%  \date{Received September 15, 1996; accepted March 16, 1997}

	\abstract{We used low- to high-frequency ALMA observations to investigate the cold gas and dust in ten quasistellar objects (QSOs) at $z\gtrsim 6$. Our analysis of the CO(6-5) and CO(7-6) emission lines in the selected QSOs provided insights into their molecular gas masses, which average around $10^{10}\ \rm M_\odot$. This is consistent with typical values for high-redshift QSOs. Proprietary and archival ALMA observations in bands 8 and 9 enabled precise constraints on the dust properties and star formation rate (SFR) of four QSOs in our sample for the first time. The examination of the redshift distribution of dust temperatures revealed a general trend of increasing $T_{\rm dust}$ with redshift, which agrees with theoretical expectations. In contrast, our investigation of the dust emissivity index indicated a generally constant value with redshift, suggesting shared dust properties among sources. We computed a mean cold dust spectral energy distribution considering all ten QSOs that offers a comprehensive view of the dust properties of high-$z$ QSOs. The QSOs marked by a more intense growth of the supermassive black hole (HYPERION QSOs) showed lower dust masses and higher gas-to-dust ratios on average, but their $\rm H_2$ gas reservoirs are consistent with those of other QSOs at the same redshift. The observed high SFR in our sample yields high star formation efficiencies and thus very short gas depletion timescales ($\tau_{\rm dep}\sim 10^{-2}$ Gyr). Beyond supporting the paradigm that high-$z$ QSOs reside in highly star-forming galaxies, our findings portrayed an interesting evolutionary path at $z>6$. Our study suggests that QSOs at $z\gtrsim 6$ are undergoing rapid galaxy growth that might be regulated by strong outflows. In the $M_{\rm BH}-M_{\rm dyn}$ plane, our high-$z$ QSOs lie above the relation measured locally. Their inferred evolutionary path shows a convergence toward the massive end of the local relation, which supports the idea that they are candidate progenitors of local massive galaxies. The observed pathway involves intense black hole growth followed by substantial galaxy growth, in contrast with a symbiotic growth scenario. The evidence of a stellar bulge in one of the QSOs of the sample is further aligned with that typical of local massive galaxies.}
	\maketitle
	
	\nolinenumbers
	
	%--------------------------------------------
	%--------------------------------------------
	\section{Introduction}
	\label{sec:intro}
	
	In the past few decades, many advances have been made in unveiling the properties of the host galaxies of the first quasi-stellar-objects (QSOs) from theory and observations. The Atacama Large Millimeter/sub-millimeter Array (ALMA), along with the Northern Extended Millimeter Array (NOEMA), the Very Large Array (VLA), and \textit{Herschel}, have probed the cold gas and dust of the QSO host galaxies. The dust continuum was detected in many $z \sim 6$ QSOs with estimated far-infrared (FIR) luminosities of $L_{\rm FIR}=10^{11-13}\ \rm L_{\odot}$ and dust masses of about $M_{\rm dust}= 10^{7-9}\ \rm M_{\odot}$ \citep{decarli2018, carniani2019, shao2019}. 
	
	The rest-frame FIR continuum emission originates from dust heated by the ultraviolet (UV) radiation from young and massive stars \citep{decarli2018, venemans2020, neeleman2021} and the active galactic nucleus (AGN) radiation \citep{schneider2015, dimascia2021, walter2022}. The latter contribution is usually neglected when modeling the FIR spectral energy distribution (SED) of $z\sim 6$ QSOs, although the AGN heating can contribute $30-70$\% of the FIR luminosity \citep{schneider2015, duras2017}. Moreover, dust masses are often determined with huge uncertainties and only rely on single-frequency continuum detections. However, if multifrequency ALMA and/or NOEMA observations are available in the rest-frame FIR, which probe both the peak and the Rayleigh Jeans tail of the dust SED, the dust temperature and mass can be constrained with statistical uncertainties $<10-20\%$ (e.g., \citealt{carniani2019, Tripodi2022,tripodi2023a,witstok2023}). This results in a high accuracy in the determination of the star formation rate (SFR).  
	
	Accurate estimates of the dust masses would also allow us to derive the molecular gas mass of the host galaxy through the gas-to-dust ratio (GDR). Although it is possible to directly probe the molecular reservoirs of the QSO host galaxies using the rotational transitions of the CO (e.g., \citealt{vallini2018, madden2020}), very few high-$z$ QSOs are observed in CO because this emission line is typically faint at high $z$. The GDR has indeed often been assumed in order to compute the gas mass, resulting in a high degree of uncertainty in its estimate. Studies of $z\sim 2.4-4.7$ hyperluminous QSOs showed that the GDR spans a broad range of values, 100-300, with an average GDR $\sim 180$ \citep{bischetti2021}. This is consistent with the results found for submillimeter (submm) galaxies to $z\sim3-5$ with a GDR$\sim150-250$ (e.g., \citealt{saintonge2013,miettinen2017}). Recent studies suggested that the GDR for SMGs may be lower than 100 \citep{birkin2021,liao2023}. In low-z galaxies, a GDR$\sim 100$ is typically observed \citep{draine2007, leroy2011}, implying that the GDR increases with redshift. However, the combination of gas-mass estimates from CO with the dust mass from finely sampled FIR SEDs results in an accurate determination of the GDR, and this allows us to study its evolution with redshift.
	
	Another central question is to understand the efficiency of large halos in forming stars during the first cosmic billion year. In this context, luminous QSOs are efficient signposts of large halos at high redshift \citep[e.g.,][]{wang2024}, which allows us to investigate the complex physics at play in the building of massive galaxies. The efficiency of a galaxy in forming stars is an intricate mechanism that depends on the efficiency with which cold gas is formed from baryons and on the efficiency with which the cold gas is converted into stars. The former is influenced by the capability of gas cooling and the capability of gas removal or gas heating by feedback. The gas star formation efficiency (SFE) has been investigated for decades in both the local and high-redshift universe. It is parameterized by the Kennicutt-Schmidt (KS) relation \citep[see e.g.,][]{kennicutt1998,genzel2010,speagle2014,calabro2024}. Because spatially resolved observations and of robust estimates of the SFR are still lacking, the KS relation is still poorly studied in high-z QSOs. Many questions about the SFE in these objects are therefore still unanswered.
	
	Targeting the QSO host galaxies at these redshifts provides a unique opportunity to characterize the formation and concurrent build-up of SMBHs and their host galaxies, and also the physical properties of the ISM in the first billion year of the Universe \citep[e.g.,][]{decarli2018,venemans2020,neeleman2021}. The local correlation between massive BHs and bulges is indeed tight, suggesting that the same process that assembled galaxy bulges may cause most of the growth of massive BHs. QSOs at $z\gtrsim 6$ appear to lie above the local $M_{\rm BH}-M_*$ correlation, and thus, the BH growth seems to precede that of its host galaxy \citep{ding2023,stone2023,yue2023}. However, before the advent of the James Webb Space Telescope (JWST), it was difficult to obtain reliable and accurate estimates of the stellar mass in QSO hosts at high $z$, even with deep HST observations \citep{metchtley2012,marshall2020}. Therefore, instead of the stellar mass, the dynamical mass is used at high $z$, exploring the $M_{\rm BH}-M_{\rm dyn}$ relation. For instance, \citet{feruglio2018} and \citet{pensabene2020} studied the $M_{\rm BH}-M_{\rm dyn}$ relation in luminous high-$z$ QSO hosts and reported that this relation evolves with redshift and that high-$z$ QSOs lie above the local relations. This implies that the SMBHs formed significantly faster than their hosts. In this context, \citet{izumi2018,izumi2019} studied the $M_{\rm BH}-M_{\rm dyn}$  using a sample of seven $z> 6$ low-luminosity quasars. They found that while the luminous quasars typically lie above the local relation \citep{kormendy2013} with BHs overmassive compared to local AGNs, the discrepancy becomes less evident for low-luminosity quasars \citep{maiolino2023c}. In order to avoid being biased by the properties of higher luminosity sources, it is essential to study the whole population of QSOs and galaxies at high $z$, including the low-luminosity sources. This is now possible with data from JWST \citep[see e.g.][]{santini2023, harikane2023}, which enable us to determine the stellar mass especially in low-luminosity sources based on a sharper PSF (especially in the short-wavelength bands) and longer-wavelength coverage out to 5 micron compared to HST, which is less affected by dust attenuation. Recently, \citet{harikane2023} analyzed a first statistical sample of faint type 1 AGNs at $z>4$ identified by JWST/NIRSpec deep spectroscopy. They compared their AGNs to the AGNs at $z\sim 0$ \citep{reines2015} and to high-$z$ QSOs. Their AGNs have similar BH masses, but systematically lower stellar masses than the local AGNs. Moreover, \citet{maiolino2023c,maiolino2023} showed even more extreme $M_{\rm BH}-M_*$ ratios in their sample of 12 AGNs at $z=4-7$ that belong to the JADES survey. Similar results were also obtained in previous studies with a smaller number of AGNs \citep{ubler2023}. This indicates that the BH grows faster than its host galaxy at high redshift. The fast BH growth was also suggested by previous studies at $z\sim 2$ \citep[e.g.,][]{zhang2023c}. These overmassive (compared to their host stellar masses) BHs are indeed predicted in some theoretical models simulations \citep[e.g., ][]{toyouchi2021,trinca2022,inayoshi2022,hu2022,zhang2023b,pacucci2023,pacucci2024}. In summary, recent observations showed that high-$z$ QSOs may lie above the local $M_{\rm BH}$ - $M_{*}$ correlation, and they are thus likely to follow the BH-dominance growth or BH-dominance evolutionary path \citep[i.e., the green line in Fig. 3 of][]{volonteri2012}.  
	
	The significance of our work becomes evident in the context of these prior studies, as we aim to explore the interconnected growth of supermassive black holes and their host galaxies. Additionally, we seek to understand whether the high-redshift hosts of quasars serve as the progenitors of the massive galaxies observed in the local Universe. These two queries are pivotal to the study of galaxy evolution and are tightly interconnected. They can only be addressed through a reliable and precise understanding of the properties of QSOs and their massive hosts. 
	
	In this work, we present the analysis of the cold-dust SEDs of four QSOs at $z>6$, PSO J036.5078+03.0498 \citep[hereafter J036+03, ][]{venemans2015} at $z=6.5405$, VDESJ022426.54-471129.4 \citep[hereafter J0224-4711, ][]{reed2017} at $z=6.5222$, PSO J231.6576-20.8335 \citep[hereafter J231-20, ][]{mazzucchelli2017} at $z=6.587$, and SDSS J205406.49-000514.8 \citep[hereafter J2054-0005, ][]{reed2017} at $z=6.0391$, based on new ALMA band 8 (B8) and/or band 9 (B9) observations, archival ALMA observations from band 3 (B3) to band 6 (B6), and a NOEMA observation at 3 mm. These observations allowed us to retrieve reliable and accurate estimates of the dust parameters and the SFR of the QSO host galaxies in our sample. Furthermore, from the analysis of archival observations, we were also able to detect the CO(7-6) and [CI] emission lines for J0224-4711 and the CO(6-5) emission line for J2054-0005, from which we were able to estimate the gas mass for these objects. 
	
	In order to investigate the dust properties, SFR, GDR, and gas SFE in a statistically sound sample of QSOs at $z>6$, we considered another six high-$z$ QSOs whose dust properties and SFR were already derived with high accuracy. The sample is presented in the following section. 
	
	The paper is organized as follows. In Sect. \ref{sec:sample}, we describe our sample of QSOs at $z>6$. In Sect. \ref{sec:obs}, we describe the observations used in this work. In Sect. \ref{sec:res}, we present the results for the continuum and line emissions. In Sect. \ref{sec:analysis}, we analyze the cold-dust SED of the QSOs in our sample, and we derive the gas mass for QSOs J0224-4711, J1319+0950, and J2054-0005. In Sect. \ref{sec:disc}, we contextualize our findings regarding the dust and gas in high-$z$ QSO, and we discuss the observed scenario for the evolutionary paths of these objects. In Sect. \ref{sec:sum}, we summarize the content of this work.
	
	Throughout the paper, we adopt the $\Lambda$CDM cosmology from \citet{planck2018}: $H_0=67.4\ \rm km\ s^{-1}\ Mpc^{-1}$, $\Omega_m = 0.315$, and $\Omega_{\Lambda} = 0.685$. Thus, the angular scale is $5.66$ kpc/arcsec at $z=6.3$.
	
	\section{Sample}
	\label{sec:sample}
	
	We selected from the currently known $\sim$300 QSOs at $z>6$ \citep[see, e.g.,][]{fan2022,jiang2016,mortlock2011,wang2019b,inayoshi2020} all the QSOs at $z>6$ for which we were able to investigate and compare the evolutionary path of the SMBHs and their host galaxies, and we linked it to feedback processes. This implied that we should be able derive or retrieve accurate estimates of the SMBH properties, dust properties, SFR and gas mass, and to detect outflowing emission. In other words, all these QSOs have high-quality rest-frame optical-UV spectra and ALMA and/or NOEMA observations targeting the continuum emission from low to high frequencies (in the range of observed frequency $\sim$ 100-600 GHz), that is, probing both the Rayleigh-Jeans and peak region of the cold-dust SED, and targeting the CO and [CII] line emissions \citep[see, e.g.,][and other refs throughout this work]{zappacosta2023,dodorico2023,wang2013,neeleman2019,witstok2023,decarli2022}. Moreover, we selected QSOs with $L_{\rm bol}>10^{47}\rm erg \ s^{-1}$. Consequently, our findings are not directly applicable to the entire population of AGNs at $z>6$; instead, they specifically pertain to high-luminosity sources. Nevertheless, this luminosity bias aligns with our objectives, as we are particularly interested in investigating the role of AGN feedback in influencing the SMBH-galaxy evolution, and notably, high-luminosity QSOs exhibit compelling evidence of powerful outflows \citep[see, e.g., ][]{bischetti2022, shao2022, tripodi2023c, salak2023}.
	
	Our final sample consisted of ten QSOs at $z>6$. Together with the four QSOs presented in the previous section (J036+03, J0224-4711, J231-20, and J2054-0005), we considered six other QSOs: SDSS J010013.02+280225.8 (hereafter J0100+2802), SDSS J231038.88+185519.7 (hereafter J2310+1855), J1319+0950, ULAS J134208.10+092838.35 (hereafter J1342+0928), SDSS J114816.64+525150.3 (hereafter J1148+5251), and PSO J183.1124+05.0926 (hereafter J183+05).

	\begin{table*}
		\caption{General properties of the QSOs in our sample}
		\centering
		\begin{tabular}{lccccc}
			\hline
			QSO & z & $\log(M_{\rm BH}/M_\odot)$ & $\log(L_{\rm bol}/{\rm erg\ s^{-1}})$ & Refs & Results\\
			\hline
			J0100+2802 & 6.327 & 10.04 $\pm$ 0.27 & 48.24 $\pm$ 0.04 & [1],[2] & SED$^{(*)}$\\
			J036+03 & 6.540 & 9.49 $\pm$ 0.12 & 47.33 $\pm$ 0.05 & [1],[2] & SED, $M_{\rm H_2}$ \\
			J0224-4711 & 6.522 & 9.36 $\pm$ 0.08 & 47.53 $\pm$ 0.01 & [1] & SED, $M_{\rm H_2}$ \\
			J231-20 & 6.587 & 9.50 $\pm$ 0.09 & 47.31 $\pm$ 0.01 & [1],[2] & SED, $M_{\rm H_2}$ \\
			J1342+0928 & 7.540 & 8.90 $\pm$ 0.14 & 47.19 $\pm$ 0.01 & [1],[2] & SED \\
			J1148+5251 & 6.419 & 9.74 $\pm$ 0.03 & 47.57 $\pm$ 0.01 & [1] & --\\
			\hline
			\hline
			J2310+1855 & 6.003 & 9.67$^{+0.06}_{-0.08}$ & 47.49$^{+0.10}_{-0.13}$ & [3],[4] & SED$^{(**)}$\\[0.1cm]
			J2054-0005$^{(***)}$ & 6.390 & 9.17 & 47.03 & [2],[5] & SED, $M_{\rm H_2}$ \\[0.1cm]
			J1319+0950 & 6.133 & 9.53$^{+0.05}_{-0.11}$ & 47.30$^{+0.07}_{-0.08}$ & [3],[2] & $M_{\rm H_2}$\\[0.1cm]
			J183+05 & 6.439 & 9.41$^{+0.21}_{-0.41}$ & 47.20$^{+0.16}_{-0.25}$ & [3],[2] & SED\\[0.1cm]            			
			\hline
		\end{tabular}
		\label{tab:sample}
		\flushleft 
		\footnotesize {{\bf Notes.}  Columns: QSO name; redshift from the [CII] emission line; BH mass; bolometric luminosity; references for z, $M_{\rm BH}$, and $L_{\rm bol}$; results presented in this work: cold-dust SED along with the dust properties and SFR (SED), gas mass from the CO(6-5) or CO(7-6) emission lines ($M_{\rm H_2}$). $^{(*)}$, $^{(**)}$: results already presented in \citet{tripodi2023b} and in \citet{Tripodi2022}, respectively, which are reported in this work. $^{(***)}$ The error on the BH mass for J2054-0005 was not provided in the reference works, and we therefore considered an average systematic error on the BH mass of 0.3 dex. The QSOs above the double black line belong to the HYPERION sample. The errors for the BH masses and bolometric luminosities of the HYPERION QSOs are taken from Tortosa et al. in prep. References:  [1] \citet{zappacosta2023}; [2] \citet{neeleman2021}; [3] \citet{mazzucchelli2023}; [4] \citet{Tripodi2022}; [5] \citet{wang2013}.}
	\end{table*}
	
	We divided the sample into two subsamples of six and four QSOs each, hereafter called HYPERION QSOs and $z>6$ QSOs, respectively (see Tab. \ref{tab:sample}).  
	
	The first subsample is composed of the six QSOs belonging to the sample called HYPerluminous QSOs at the Epoch of ReionizatION (HYPERION). 
	HYPERION comprises the titans of $z>6$ QSOs \citet{zappacosta2023}, that is, those whose SMBH had the fastest mass growth history. In particular, these QSOs were selected so that the SMBHs powering them required to form at least a 1000 $M_{\odot}$ BH seed (at $z=20$) under the hypothesis of a continuous exponential accretion at the Eddington rate. These SMBHs likely assembled from the largest BH seeds, or alternatively, they experienced peculiar, possibly super-Eddington, mass accretion histories. Of the $\sim 300$ QSOs known at the epoch of reionization (EoR), the HYPERION QSOs comprise 18 QSOs with a mean redshift $z \sim 6.7$, and average $\rm \log(L_{\rm bol}/\rm erg/s) \sim 47.3$, and a BH mass in the range $10^9-10^{10}~\rm M_\odot$. 
	HYPERION is based on a 2.4 Ms XMM-Newton Multi-Year Heritage Programme (PI: Zappacosta) to provide for the first time for such a large sample of $z>6$
	QSOs a uniform high quality X-ray spectral characterization for a detailed investigation of the nuclear properties and the accretion and ejection processes that are  tied to the fast build-up experienced by their SMBH. The first results suggest a genuine redshift evolution of their X-ray spectral slopes, which appear to be steeper than reported in $z<6$ QSOs with a similar luminosity and accretion rate. This supports a different regime for the X-ray nuclear properties of the first quasars that might be linked to the presence of fast disk-driven winds \citep{zappacosta2023}. While the nuclear properties of HYPERION QSOs are well constrained, as are the dynamical masses of most QSOs based on archival [CII] observations, their dust properties and SFR are still mostly unconstrained. 
	
	The $z>6$ QSO subsample comprises the QSOs that did not satisfy the criteria for belonging to the HYPERION sample \citep[see Sect. 2 of][]{zappacosta2023}, that is, they did not experience a rapid SMBH mass growth. 
	
	We divided our final sample based on the HYPERION survey selection criteria in order to assess whether the BH accretion history has strong implications for the properties of the host galaxy. 
	
	Seven QSOs in our sample\footnote{These QSOs are J0100+2802, J036+03, J0224-4711, J231-20, J2310+1855, J1319+0950, and J183+05.} are also part of the XQR-30 ESO Large Program \citep{dodorico2023}. Therefore, they have high S/N X-Shooter spectra, from which \citet{mazzucchelli2023} derived BH masses using the C IV and Mg II emission lines. The Mg II BH masses derived for J0100+2802, J036+03, J0224-4711, and J231-20 agree well with those in \citet{zappacosta2023}. 
	
	The properties of the ten QSOs in our final sample are summarized in Tab. \ref{tab:sample}, and a more detailed presentation of each object can be found in Appendix \ref{app:a}.
	
	\section{Observations}
	\label{sec:obs}
	
	All ALMA observations targeting QSOs J036+03, J0224-4711, J231-20, J1319+0950, J2054-0005 were calibrated and imaged as outlined in the following.
	
	The visibility calibration of the observations was executed by the ALMA science archive. The imaging was performed through the Common Astronomy Software Applications (CASA; \citealt{mcmullin2007}), version 5.1.1-5. We applied \texttt{tclean} using natural weighting and a $3\sigma$ cleaning threshold. We imaged the continuum using the multifrequency synthesis (MFS) mode in all line-free channels. To image the line emissions, we used the CASA task \texttt{uvcontsub} to fit the continuum visibilities in the line-free channels with a first-order polynomial for QSO J0224-4711, since the continuum showed a non-negligible slope, and zeroth-order polynomial for the other QSOs. We then obtained continuum-subtracted cubes to be used in our analyses. We produced line maps using the MFS mode in the channels enclosing the emission lines (see Tab. \ref{tab:c2-table-obs}).
	
	We analyzed all archival observations available for J036+03 and J0224-4711. We also used the results on the continuum emission at $\sim 107$ GHz of J036+03, obtained from a NOEMA observation (project ID: S17CD).
	
	Because many ALMA observations are available for J2054-0005, especially in B6, we considered for each band those whose continuum sensitivity and resolution were suited for our analysis of the SED. That is, when multiple observations per band were available, we considered the observation with the highest continuum sensitivity and with a resolution that allowed us to spatially resolve the source and/or that was as close as possible to the angular resolution of the B9 observation, in order to ensure a reliable and consistent analysis of the cold-dust SED (for more details, see Sect. \ref{subsec:cont}). 
	
	For J231-20, we used the results presented in \citet{Pensabene2021}, who analyzed all the ALMA observations from B3 to B6 available for this QSO, and we independently analyzed a new B8 ALMA observation targeting this QSO (ID: 2021.2.00064.S). Since J231-20 was found to have a close companion (at a distance $<10$ kpc), we discuss the analysis of the B8 observation in more detail in Sect. \ref{subsec:cont}.
	
	Information about the project ID, synthesized beam, lines detected, line channels, and root mean square (r.m.s) noise of the continuum map and of the continuum-subtracted cube for each observation analyzed in this work is reported in Tabs. \ref{tab:c3-table-obs} and \ref{tab:c2-table-obs}.
	
	\section{Data analysis}
	\label{sec:res}
	
	In the following, we report the results for our proprietary observations targeting the continuum (Sect. \ref{subsec:cont}) and/or line emissions (Sect. \ref{subsec:emlines}) in ALMA bands 3, 8, and/or 9 of QSOs J036+03, J0224-4711, J1319+0950, and J2054-0005. As an exception, we discuss the archival observation targeting the continuum emission in B8 of J231-20 because it was found to have a close companion in lower-frequency observations. Details of the analysis of archival observations of the continuum emissions of QSOs J036+03, J0224-4711, and J2054-0005 can be found in Appendix \ref{app:b}. The results obtained from the analysis of archival observations are reported in Tab. \ref{tab:c3-table-obs}.
	
	\subsection{Continuum emission}
	\label{subsec:cont}
	
	\begin{figure*}
		\centering
		\includegraphics[width=0.45\linewidth]{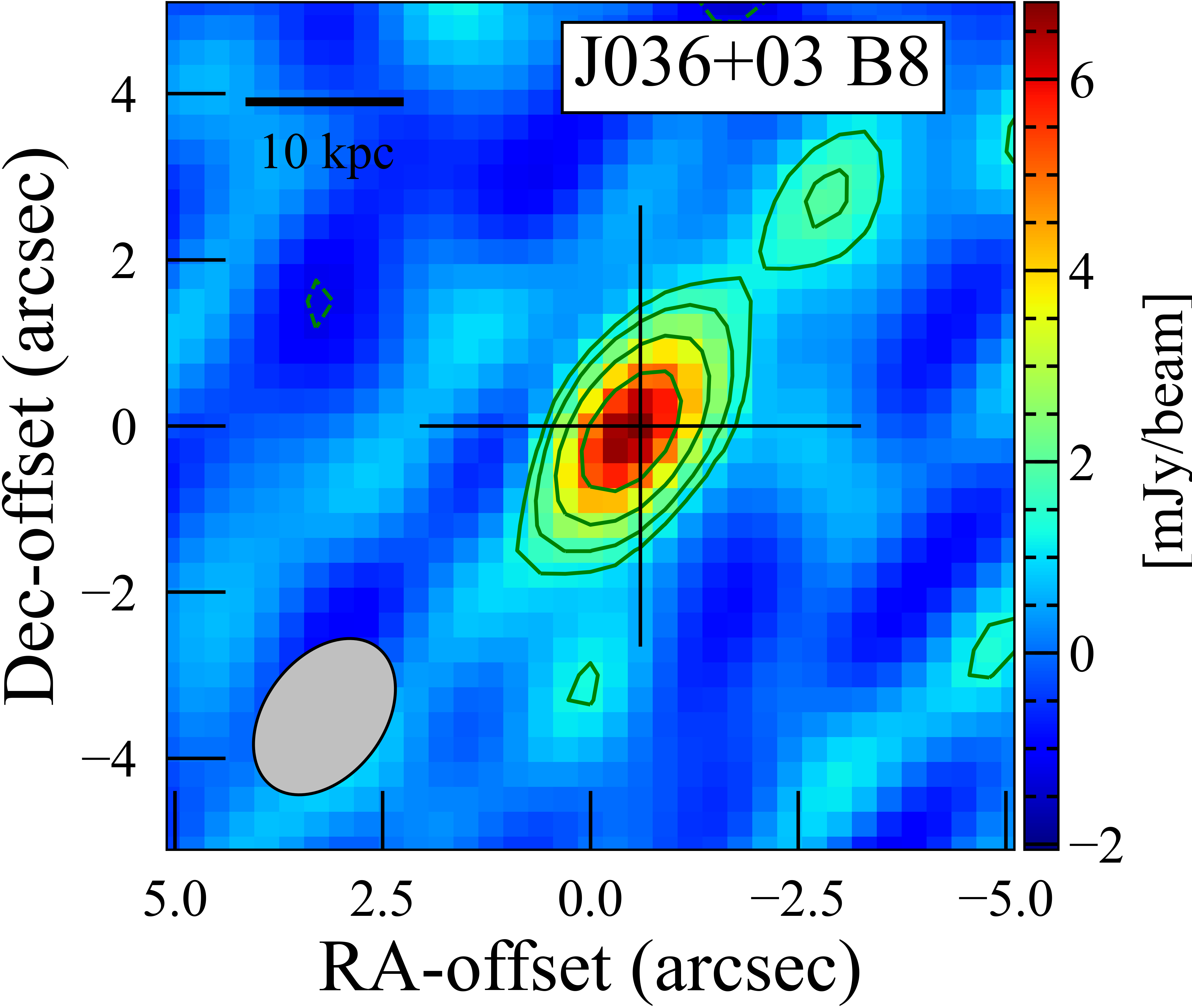}
		\includegraphics[width=0.45\linewidth]{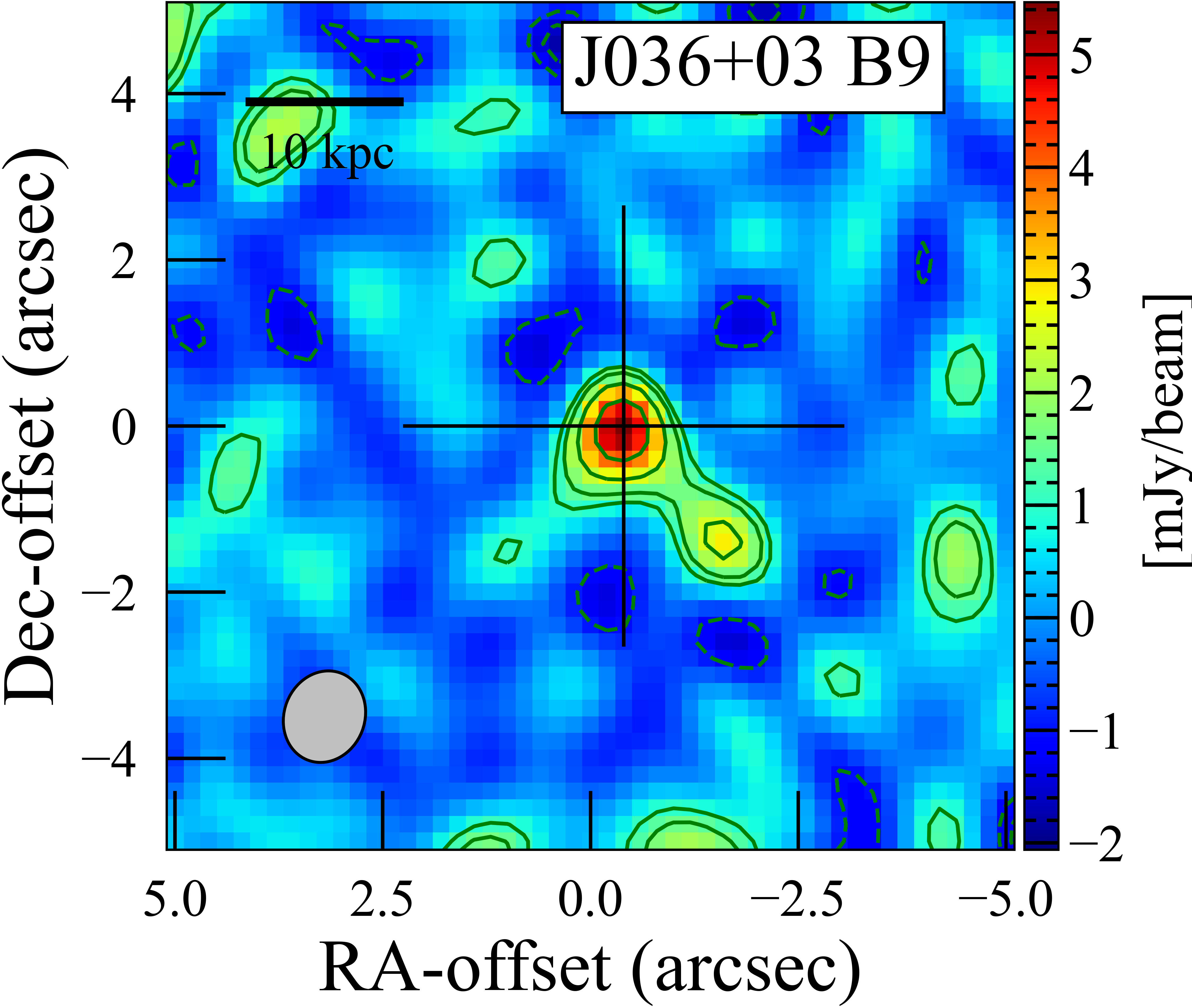}
		\includegraphics[width=0.45\linewidth]{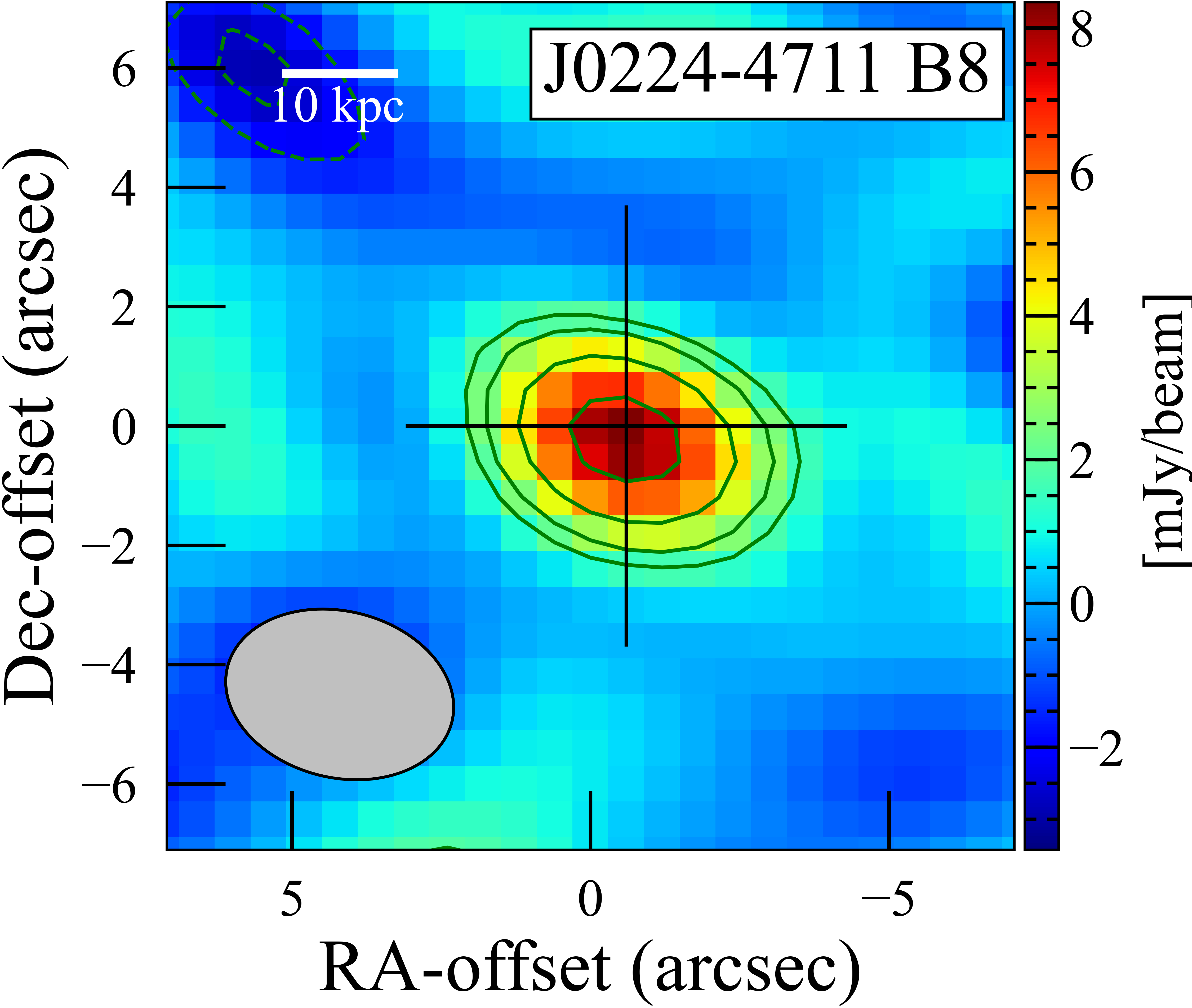}
		\includegraphics[width=0.45\linewidth]{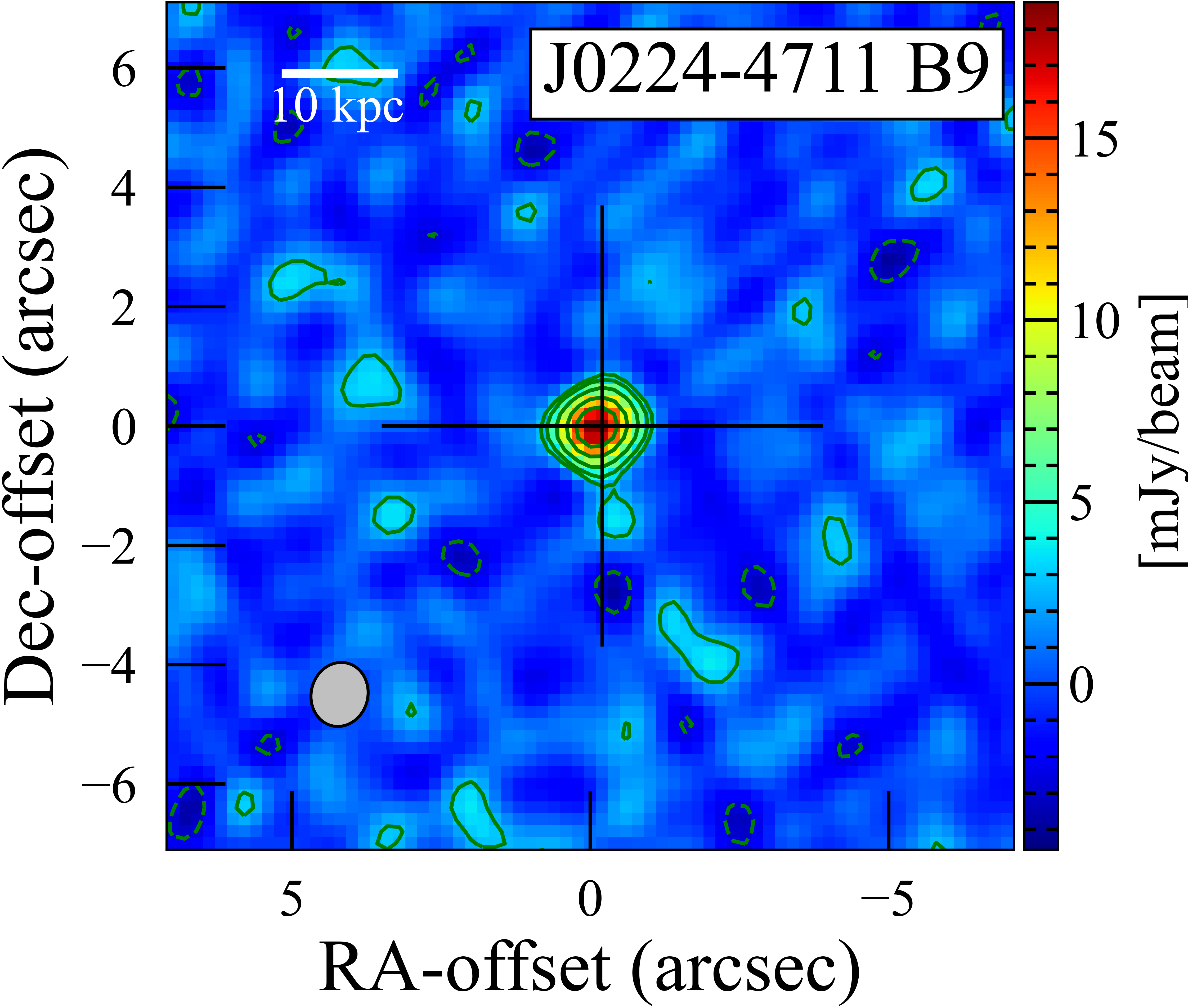}
		\caption{Dust continuum maps. Top left panel: 404.9 GHz dust continuum map of QSO J036+03 (levels $-3,-2,2,3,5,\text{and }8\sigma$, $\sigma = 0.6$ mJy/beam). The clean beam ($2.10\times 1.43\rm \ arcsec^2$, PA=37.40$^\circ$) is indicated in the lower left corner of the diagram. The cross indicates the position of the continuum peak. Top right panel: 670.9 GHz dust continuum map of QSO J036+03 (levels $-3,-2,2,3,5,\text{and }8\sigma$, $\sigma = 0.5$ mJy/beam). The clean beam ($1.12\times 0.97\rm \ arcsec^2$, PA=21.11$^\circ$) is indicated in the lower left corner of the diagram. The cross indicates the position of the continuum peak. Bottom left panel: 405.2 GHz dust continuum map of QSO J0224-4711 (levels $-3,-2,2,3,5,\text{and }8\sigma$, $\sigma = 0.88$ mJy/beam). The clean beam ($3.87\times 2.79\rm \ arcsec^2$, PA=-76.45$^\circ$) is indicated in the lower left corner of the diagram. The cross indicates the position of the continuum peak. Bottom right panel: 670.9 GHz dust continuum map of QSO J0224-4711 (levels $-3,-2,2,3,5,8,\text{and }11\sigma$, $\sigma = 1.3$ mJy/beam). The clean beam ($1.08\times 0.95\rm \ arcsec^2$, PA=14.05$^\circ$) is indicated in the lower left corner of the diagram. The cross indicates the position of the continuum peak.}
		\label{fig:c3-contj036}
	\end{figure*}

	\subsubsection{QSO J036+03} 
	
	\noindent The top left and right panels of Fig. \ref{fig:c3-contj036} show the continuum emission at 404.9 GHz and 670.9 GHz of J036+03, respectively. Neither emission is spatially resolved, and therefore, we considered the peak flux as the total flux of the source, that is, 6.63 $\pm$ 0.39 mJy/beam at 404.9 GHz and 5.60 $\pm$ 0.69 mJy/beam at 670.9 GHz. The emission at 670.9 GHz presents a secondary peak at RA,DEC=[-1.5,-1.5], which likely is an artifact due to the low resolution of the B9 observation. This interpretation is supported by the absence of line and continuum emission at the same spatial position in all lower-frequency observations and by the presence of another dimmer peak that is located symmetrically with respect to the QSO position (RA,DEC=[1.0,-1.5]).

	\subsubsection{QSO J0224-4711}

	\noindent  The continuum emission of J0224-4711 at 405.2 GHz and 670.9 GHz is shown in the bottom left and right panels of Fig. \ref{fig:c3-contj036}, respectively. Neither emission is spatially resolved, and therefore, we consider the peak flux as total flux of the source, which is 8.73 $\pm$ 0.38 mJy/beam at 405.2 GHz and 19.9 $\pm$ 0.96 mJy/beam at 670.9 GHz.

	\subsubsection{QSO J2054-0005}
	
	\noindent We show the  continuum map at 92.3 GHz underlying the CO(6-5) emission line in the bottom row of Fig. \ref{fig:c2-co65J2054}. Performing a 2D Gaussian fit, we found that the peak flux is 0.066$\pm$0.004 mJy/beam, the integrated flux is $0.082 \pm 0.009$ mJy, and the emission is spatially resolved with a size of $(0.27\pm 0.07)\times (0.07\pm 0.09)$ arcsec$^2$.

	\subsubsection{QSO J231-20}
	\label{subsec:j231}
	
	\noindent  J231-20 was found to have a close companion that is detected in multiple emission lines and in the continuum emission from $\sim 100$ GHz up to $\sim 250$ GHz at a distance of $\sim 2$ arcsec from the QSO \citep{Pensabene2021, decarli2017, neeleman2019}. \citet{Pensabene2021} were able to disentangle the continuum emission of the QSO from that of the companion using $\sim 1$ arcsec resolution ALMA observations. Unfortunately, the low resolution of the new band 8 observation (beam of $4.3\times 2.9$ arcsec$^2$) did not allow us to distinguish the two emissions (see Fig. \ref{fig:app-j231}). Performing a 2D fit with CASA on the continuum map, we found a peak flux of $8.43\pm 0.39$ mJy/beam, and the source was unresolved. The B8 flux is indeed contaminated by the continuum emission of the companion, and this can therefore bias the SED fitting and the determination of the dust properties. We corrected for the companion contribution using the results found by \citet{Pensabene2021}. In their SED modeling, the flux of the companion at $\sim 400$ GHz is more than 0.6 dex lower than that of the QSO. Conservatively considering the highest-temperature model for the companion, that is, that the flux of the companion is 0.6 dex lower than that of the QSO, we can correct our flux in band 8. We obtain 6.74$\pm$0.31. We considered this flux as a detection in the SED fitting, but it might also be seen as a lower limit to the flux of the QSO. 
	
	\subsection{Emission lines}
	\label{subsec:emlines}
	
	\subsubsection{CO(7-6) and [CI] emission lines in J0224-4711}
	\label{subsec:co76}
	
\begin{figure*}
		\centering
		\includegraphics[width=0.9\linewidth]{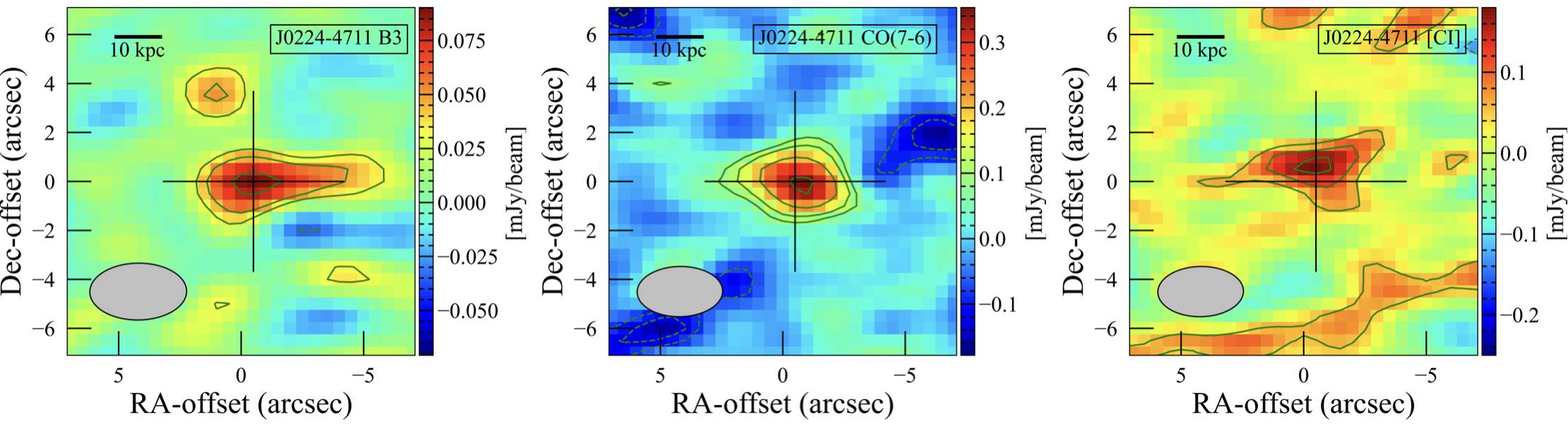}
		\caption{Dust continuum and emission line maps of QSO J0224-4711. Left panel: 95.33 GHz dust continuum map of QSO J0224-4711 (levels $-3,-2,2,3,\text{and }5\sigma$, $\sigma = 0.016$ mJy/beam). The clean beam ($3.95\times 2.32\rm \ arcsec^2$, PA=-89.43$^\circ$) is indicated in the lower left corner of the diagram. The cross indicates the position of the continuum peak. Central panel: CO(7-6) emission line map of QSO J0224-4711 (levels $-3,-2,2,3,5,\text{and }8\sigma$, $\sigma = 0.042$ mJy/beam). The clean beam ($3.49\times 2.05\rm \ arcsec^2$, PA=-89.06$^\circ$) is indicated in the lower left corner of the diagram. The cross indicates the position of the B3 continuum peak. Right panel: [CI] emission line map of QSO J0224-4711 (levels $-3,-2,1,2,\text{and }3\sigma$, $\sigma = 0.054$ mJy/beam). The clean beam ($3.49\times 2.05\rm \ arcsec^2$, PA=-89.06$^\circ$) is indicated in the lower left corner of the diagram. The cross indicates the position of the B3 continuum peak.}
		\label{fig:c2-co76}
	\end{figure*}

	We used the continuum-subtracted data cube in B3 to study the CO(7-6) and [CI] emission lines of J0224-4711. The central and right panels of Fig. \ref{fig:c2-co76} present the maps of CO(7-6) and [CI], respectively, imaged using the MFS mode in the channels specified in Tab. \ref{tab:c2-table-obs}. The CO(7-6) and [CI] emission are detected with a statistical significance of $\sim 8\sigma$ and $\sim 3\sigma$, respectively, and neither is spatially resolved. Performing a 2D Gaussian fit, we obtained a peak flux of 0.36 $\pm$ 0.02 mJy/beam for CO(7-6) and of $0.19 \pm 0.03$ mJy/beam for [CI]. 
	
\begin{figure*}
		\centering
		\includegraphics[width=1\linewidth]{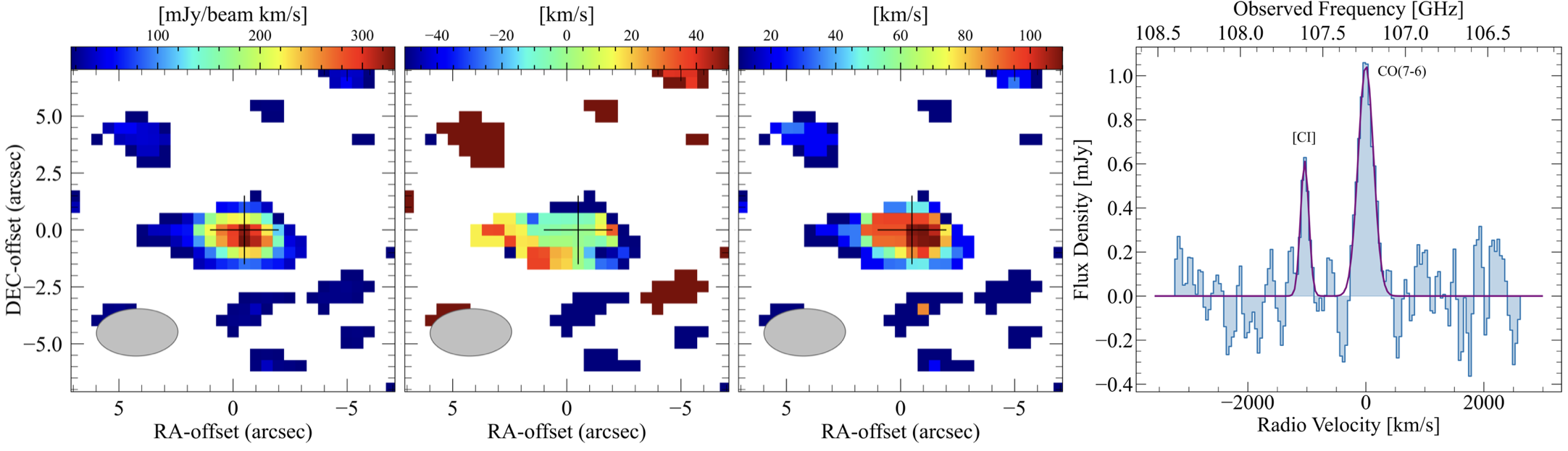}
		\caption{Moment maps of the CO(7-6) emission line and spectrum of CO(7-6) and [CI] emission lines of J0224-4711. From left to right: Integrated flux, mean velocity map, velocity dispersion map, and continuum-subtracted spectra of CO(7-6) and [CI]. The clean beam is plotted in the lower left corner of the moment maps. The cross indicates the peak position of the Band 3 continuum emission.
			The spectrum was extracted from the region included within $\geq 2\sigma$ in the CO(7-6) map.}
		\label{fig:c2-momj0224}
	\end{figure*}

	Fig. \ref{fig:c2-momj0224} shows the moment -0, -1, and -2 maps of the CO(7-6) emission, obtained by applying a 2.5$\sigma$ threshold to the continuum-subtracted cube in the line channels, and the spectrum extracted from the region with S/N$>2$ in the CO(7-6) map. The moment-0 map shows a velocity gradient oriented east to west  with $\Delta v=100$ $\rm km\ s^{-1}$, and the moment-2 map shows a range of the velocity dispersion\footnote{The maximum value of the velocity dispersion toward the nucleus is usually affected by beam smearing \citep{davies2011}.} between 20 and 120 $\rm km\ s^{-1}$. The CO(7-6) line profile peaks at a frequency of 107.239 GHz, corresponding to $z=6.5220\pm 0.0002$, consistent with previous determinations \citep{reed2017} and with the redshift derived from MgII emission line \citep{dodorico2023}. From the fit with a single Gaussian, the FWHM of the line is $307 \pm 25$ $\rm km\ s^{-1}$ and the integrated flux is 0.34 $\pm$ 0.05 Jy km s$^{-1}$, which corresponds to a luminosity of $L_{\rm CO(7-6)}= (1.6 \pm 0.2) \times 10^8 \rm \ L_\odot$ and $L'_{\rm CO(7-6)}= (9.6\pm 1.4)\times 10^9\rm K ~km ~s^{-1} ~pc^2$ \citep[following Eq.1 and Eq.3 of ][]{solomon2005}. The [CI] line is slightly blueshifted, peaking at 107.610 GHz, which corresponds to $z=6.5211 \pm 0.0004$. Performing a single-Gaussian fit, we obtained that the FWHM of the line is $165 \pm 29$ $\rm km\ s^{-1}$ and the integrated flux is 0.108 $\pm$ 0.036 Jy km s$^{-1}$, which corresponds to a luminosity\footnote{All luminosities were computed following Eq.1 and Eq.3 of \citet{solomon2005}} of $L_{\rm [CI]}= (5.1 \pm 1.7) \times 10^7 \rm \ L_\odot$ and $L'_{\rm [CI]}= (3.0\pm 1.0)\times 10^9\rm K ~km ~s^{-1} ~pc^2$. All line properties are reported in Tab. \ref{tab:c2-colines}.
	
	\subsubsection{CO(6-5) emission line in J1319+0950}
	
	In order to analyze the CO(6-5) emission line of J1319+0950, we rebinned the continuum-subtracted data cube in B3 to 50 km s$^{-1}$ .
	
	The top row of Fig. \ref{fig:c2-co65J2054} presents the map of the CO(6-5) emission line of J1319+0950 and its underlying continuum, imaged using the MFS mode in the channels specified in Tab. \ref{tab:c2-table-obs}. The CO(6-5) is detected with a statistical significance of $\sim 15 \sigma$, and performing a 2D Gaussian fit, we obtained a peak flux of 0.303 $\pm$ 0.014 mJy/beam and an integrated flux of 0.570 $\pm$ 0.038 mJy. The emission is spatially resolved with a size of $(0.35 \pm 0.03)\times (0.23\pm 0.03)$ arcsec$^2$. 
	
	The top row of Fig. \ref{fig:c2-momj2054} shows the moment -0, -1, and -2 maps of the CO(6-5) emission of J1319+0950, obtained by applying 3$\sigma$ threshold to the continuum-subtracted cube in the line channels, and the spectrum extracted from the region with S/N$>2$ in the CO(6-5) map. The moment-0 shows a velocity gradient oriented northeast to southwest  with $\Delta v=400$ $\rm km\ s^{-1}$, and the moment-2 map shows a range of the velocity dispersion between 20 and 160 $\rm km\ s^{-1}$. The CO(6-5) line profile peaks at a frequency of 96.939 GHz, corresponding to $z=6.1331\pm 0.0004$. 
	Performing a single-Gaussian fit to the line spectral profile, we obtained an FWHM of the line of $529 \pm 41$ $\rm km\ s^{-1}$ and an integrated flux of 0.662 $\pm$ 0.094 Jy km s$^{-1}$, which corresponds to a luminosity of $L_{\rm CO(6-5)}= (2.4 \pm 0.3) \times 10^8 \rm \ L_\odot$ and $L'_{\rm CO(6-5)}= (2.3 \pm 0.3)\times 10^{10}\rm K ~km ~s^{-1} ~pc^2$.

	\subsubsection{CO(6-5) emission line in J2054-0005}
	\label{subsec:co65}
	
	In order to analyze the CO(6-5) emission line of J2054-0005, we rebinned to 50 km s$^{-1}$ the continuum-subtracted data cube in B3. 
	
\begin{figure*}
		\centering
		\includegraphics[width=0.48\linewidth]{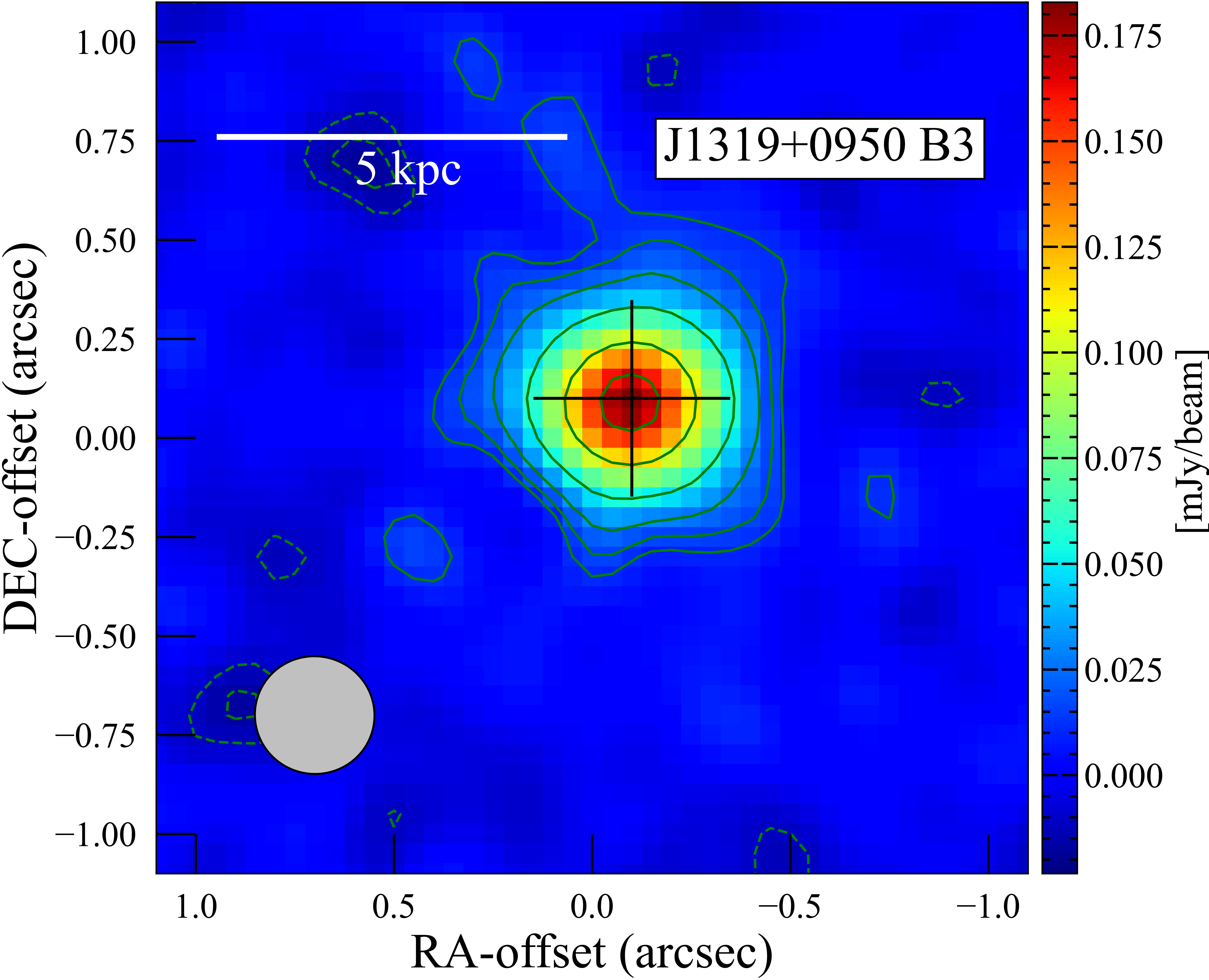}
		\includegraphics[width=0.48\linewidth]{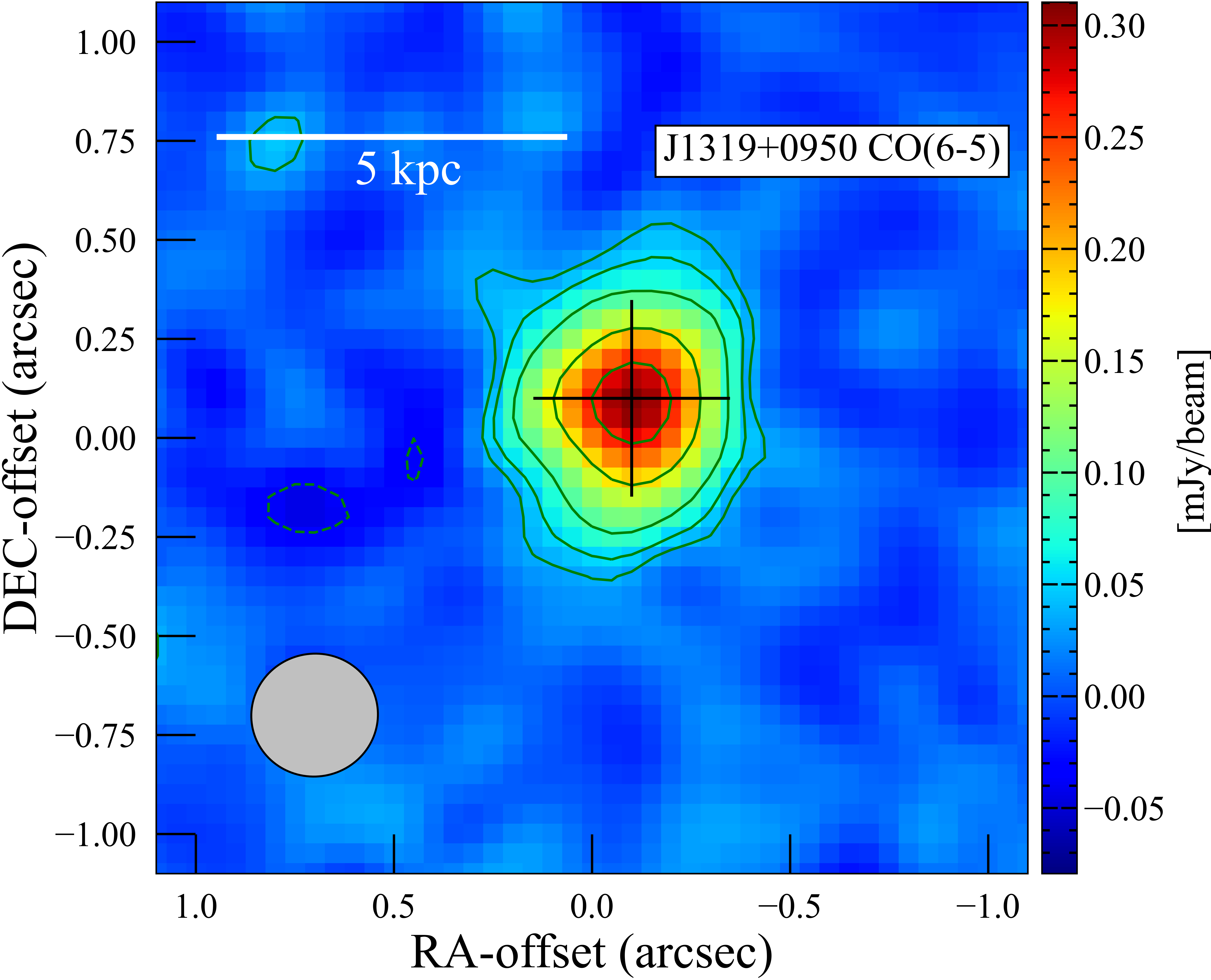} 
		
		\includegraphics[width=0.48\linewidth]{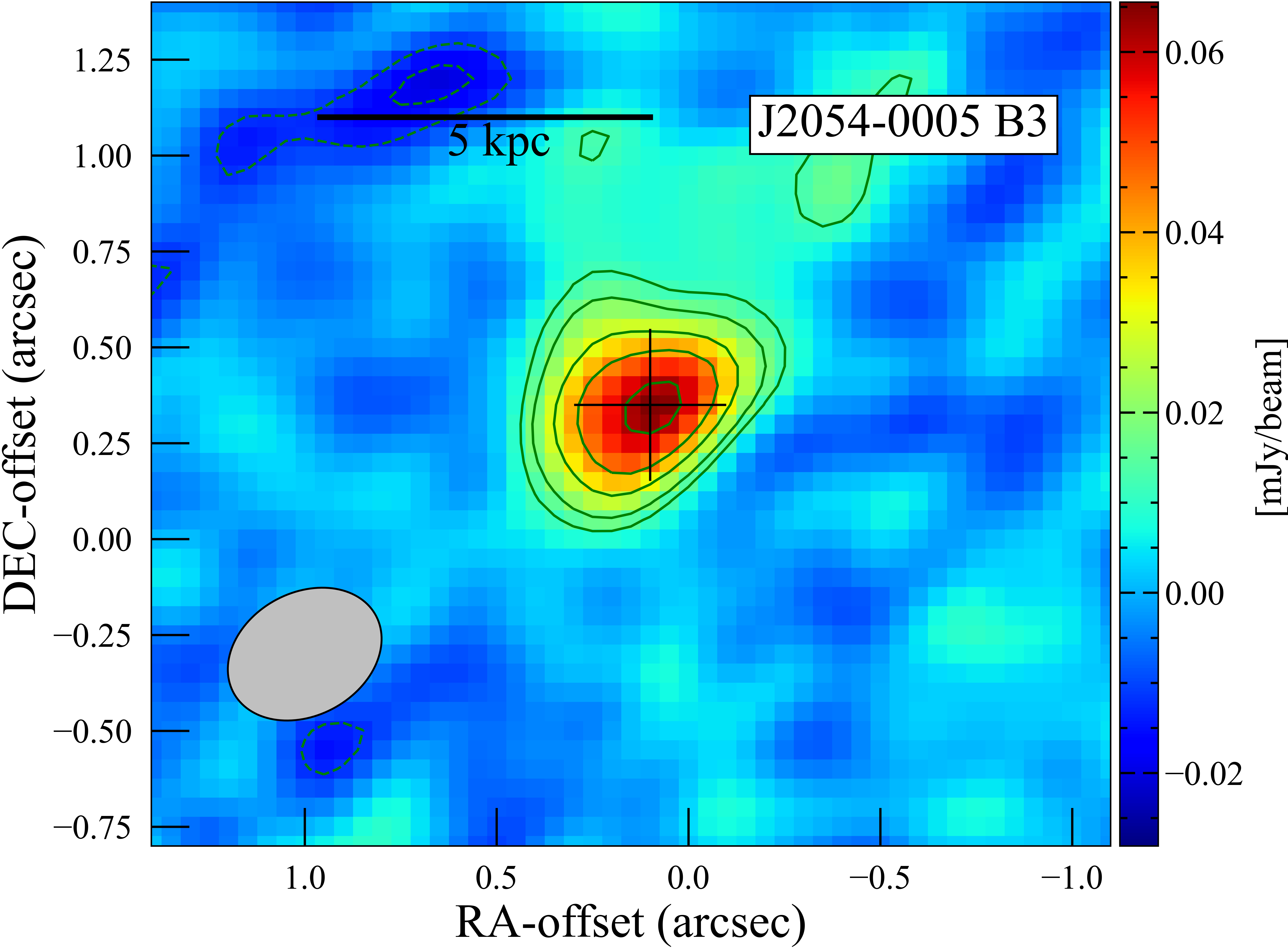}
		\includegraphics[width=0.48\linewidth]{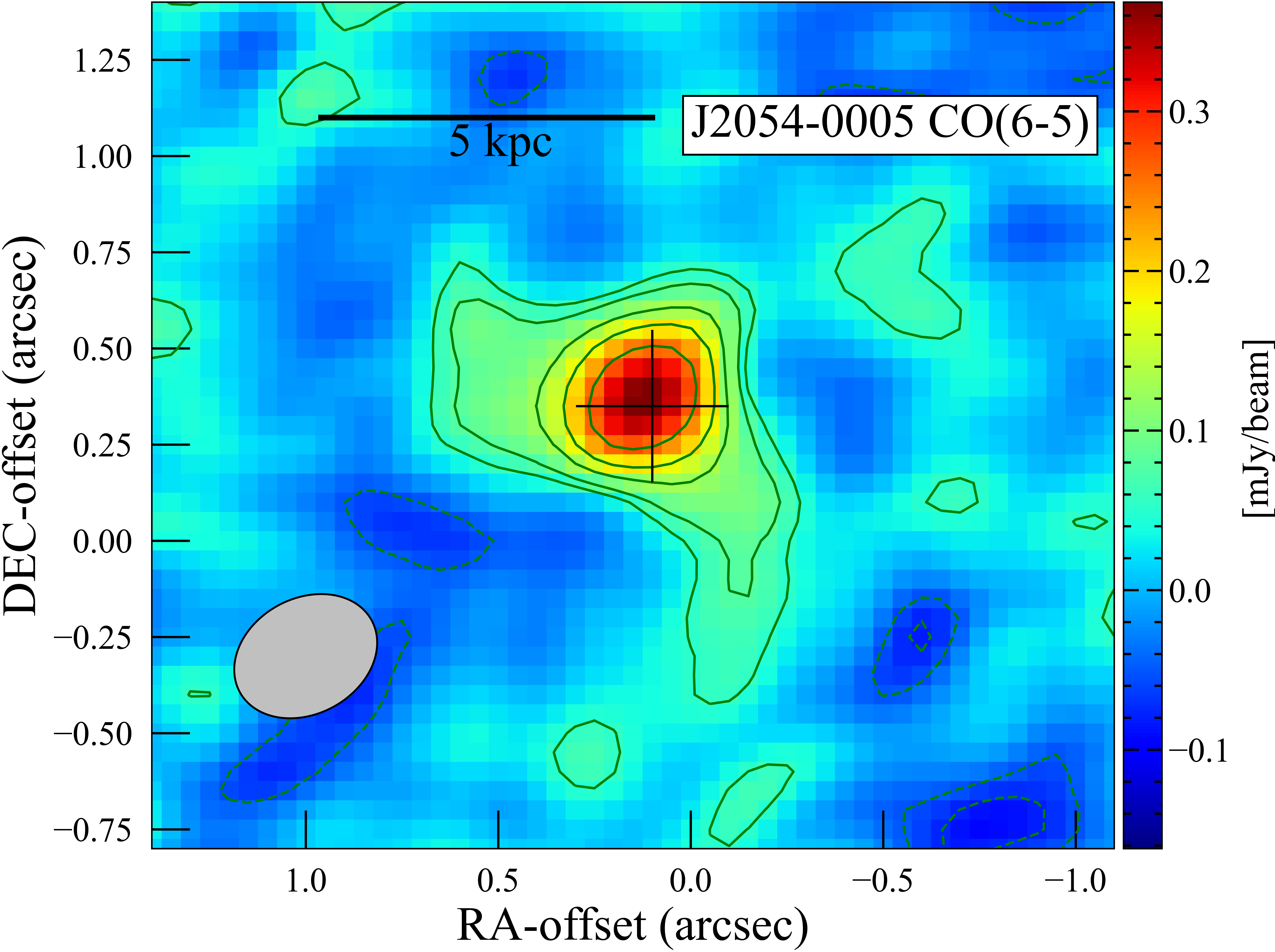} 
		\caption{Dust continuum and CO(6-5) emission line maps of QSOs J1319+0950 and J2054-0005. Top left panel: 103 GHz dust continuum map of QSO J1319+0950 (levels $-3,-2,2,3,5,10,\text{and }20\sigma$, $\sigma = 0.005$ mJy/beam). The clean beam ($0.30\times 0.30\rm \ arcsec^2$, PA=-78.56$^\circ$) is indicated in the lower left corner of the diagram. Top right panel: CO(6-5) map of QSO J1319+0950 (levels $-3,-2,2,3,5,10,\text{and }15\sigma$, $\sigma = 0.017$ mJy/beam). The clean beam ($0.32\times 0.31\rm \ arcsec^2$, PA=-78.56$^\circ$) is indicated in the lower left corner of the diagram. Bottom left panel: 92 GHz dust continuum map of QSO J2054-0005 (levels $-3,-2,2,3,5,7\text{and }10\sigma$, $\sigma = 0.006$ mJy/beam). The clean beam ($0.42\times 0.32\rm \ arcsec^2$, PA=-61.3$^\circ$) is indicated in the lower left corner of the diagram. Bottom right panel: CO(6-5) map of QSO J2054-0005 (levels $-3,-2,2,3,5,7\text{and }10\sigma$, $\sigma = 0.025$ mJy/beam). The clean beam ($0.39\times 0.30\rm \ arcsec^2$, PA=-61.4$^\circ$) is indicated in the lower left corner of the diagram. The cross indicates the position of the continuum peak for each source. }
		\label{fig:c2-co65J2054}
	\end{figure*}

	The bottom row of Fig. \ref{fig:c2-co65J2054} presents the map of the CO(6-5) emission line of J2054-0005 and its underlying continuum, imaged using the MFS mode in the channels specified in Tab. \ref{tab:c2-table-obs}. The CO(6-5) is detected with a statistical significance of $\sim 10 \sigma$. Performing a 2D Gaussian fit, we obtained a peak flux of 0.349 $\pm$ 0.036 mJy/beam, an integrated flux of 0.496 $\pm$ 0.080 mJy, and the emission is spatially resolved with a size of $(0.30 \pm 0.09)\times (0.08\pm 0.16)$ arcsec$^2$.

\begin{figure*}[t]
		\centering
		\includegraphics[width=1\linewidth]{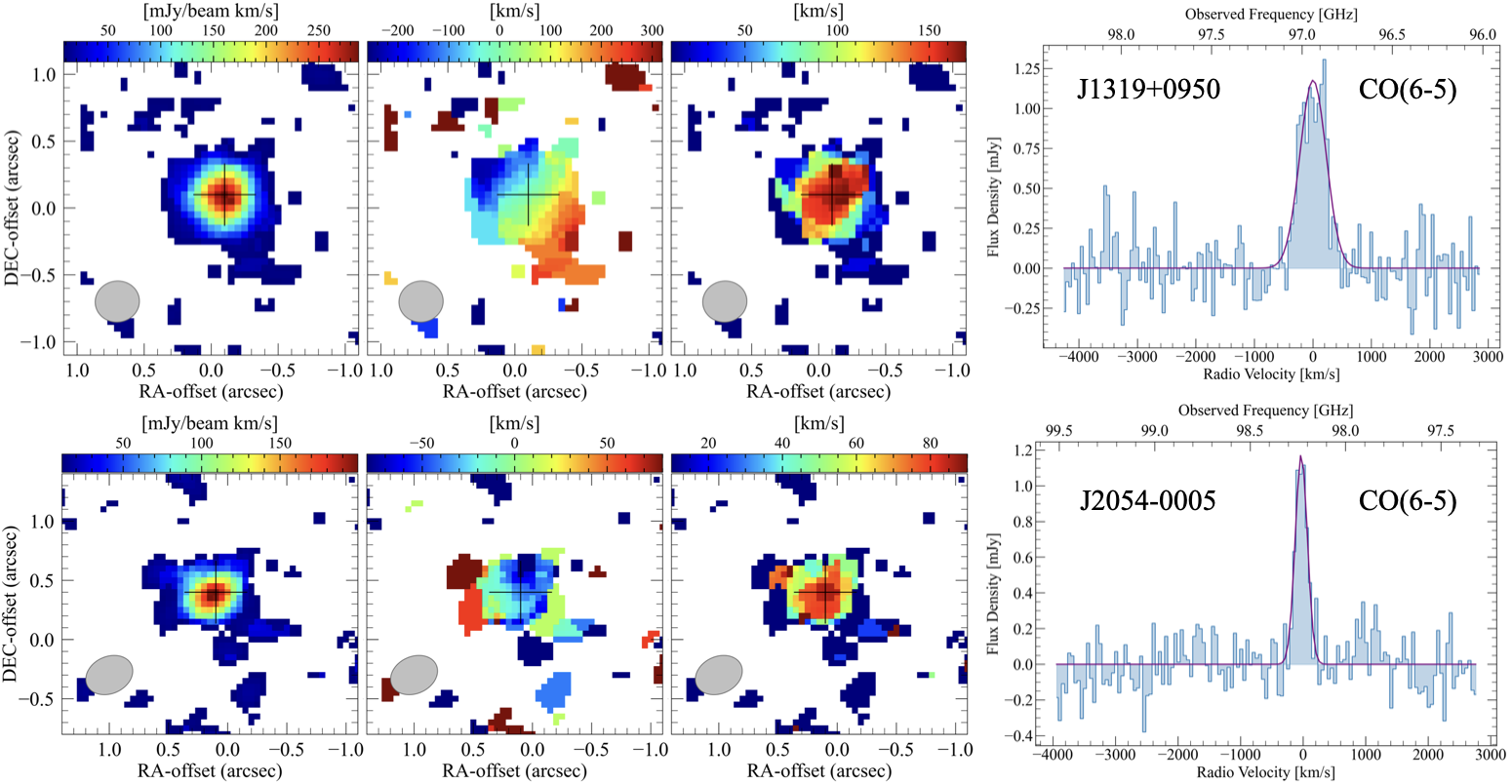}
		\caption{Moment maps of the CO(6-5) emission lines and spectra of J1319+0950 (top row), J2054-0005 (bottom row). From left to right: integrated flux, mean velocity map, and velocity dispersion map, continuum-subtracted spectrum of CO(6-5). The clean beam is plotted in the lower left corner of the moment maps. The cross indicates the peak position of CO(6-5).
			The spectra were extracted from the region included within $\geq 2\sigma$ in the CO(6-5) map of each source.}
		\label{fig:c2-momj2054}
	\end{figure*}

	The bottom row of Fig. \ref{fig:c2-momj2054} shows the moment-0, -1, and -2 maps of the CO(6-5) emission of J2054-0005, obtained by applying 3$\sigma$ threshold to the continuum-subtracted cube in the line channels, and the spectrum extracted from the region with S/N$>2$ in the CO(6-5) map. The moment-0 shows a velocity gradient oriented southeast to northwest  with $\Delta v=120$ $\rm km\ s^{-1}$, and the moment-2 map shows a range of the velocity dispersion between 20 and 80 $\rm km\ s^{-1}$. The CO(6-5) line profile peaks at a frequency of 98.233 GHz, corresponding to $z=6.0391\pm 0.0002$. 
	Performing a single-Gaussian fit to the line spectral profile, we obtained a FWHM of the line of $229 \pm 20$ $\rm km\ s^{-1}$ and an integrated flux of 0.288 $\pm$ 0.047 Jy km s$^{-1}$, which corresponds to a luminosity of $L_{\rm CO(6-5)}= (1.0 \pm 0.2) \times 10^8 \rm \ L_\odot$ and $L'_{\rm CO(6-5)}= (9.8\pm 1.6)\times 10^9\rm K ~km ~s^{-1} ~pc^2$.
	
	\begin{table*}
		\vspace{0.2cm}
		\caption{Line properties of QSOs J0224-4711, J1319+0950, and J2054-0005}
		\centering
		\begin{tabular}{lccc}
			\hline
			\hline
			& J0224-4711 & J1319+0950 & J2054-0005 \\
			\hline
			CO transition & CO(7-6) & CO(6-5) & CO(6-5) \\
			$z_{\rm CO}$ &  $6.5220\pm0.0002$ & 6.1331 $\pm$ 0.0004 & 6.0391 $\pm$ 0.002 \\
			FWHM$_{\rm CO}$ [$\rm km\ s^{-1}$] &  $307\pm25$ & 529 $\pm$ 41 & 229 $\pm$ 20 \\ 
			F$_{\rm CO}$ [Jy $\rm km\ s^{-1}$] &  $0.342\pm 0.052$ & 0.662 $\pm$ 0.094 & 0.288 $\pm$ 0.047 \\
			L$^\prime_{\rm CO}$ [$\rm K ~km ~s^{-1} ~pc^2$]&  $(9.6\pm 1.4)\times 10^{9}$ & $(2.3 \pm 0.3)\times 10^{10}$ & $(9.8\pm 1.6)\times 10^{9}$\\
			L$_{\rm CO}$ [L$_\odot$] & $(1.6\pm 0.2)\times 10^{8}$ & $(2.4 \pm 0.3)\times 10^{8}$ &  $(1.0\pm 0.2)\times 10^{8}$ \\
			M$_{\rm H2, CO}^{(\rm a)}$ [M$_\odot$] & (1.0 $\pm$ 0.1) $\times 10^{10}$ & $(1.5 \pm 0.2)\times 10^{10}$ & (6.4 $\pm$ 1.0) $\times 10^9$\\[0.1cm]
			$z_{\rm [CI]}$ & $6.5211\pm0.0004$ & -- & -- \\
			FWHM$_{\rm [CI]}$ [$\rm km\ s^{-1}$] & $165\pm29$ & -- & --  \\ 
			F$_{\rm [CI]}$ [Jy $\rm km\ s^{-1}$]& $0.108 \pm 0.036$ & -- & --  \\
			L$^\prime$[CI] [$\rm K ~km ~s^{-1} ~pc^2$]& $(3.0\pm 1.0)\times 10^{9}$ & -- & --  \\  
			L[CI] [L$_\odot$]& $(5.1\pm 1.7)\times 10^{8}$ & -- & -- \\
			\hline
		\end{tabular}
		\label{tab:c2-colines}
		\flushleft
		\footnotesize {{\bf Notes.} $^{(\rm a)}$: assuming a conversion factor $\alpha_{\rm CO}$ = 0.8 $\rm M_{\odot} (K\,km \, s^{-1} \,pc^{2})^{-1} $ and assuming a brightness temperature ratio $\rm r_{76}=CO(7-6)/CO(1-0)=0.76$  for J0224-4711, $\rm r_{65}=CO(6-5)/CO(1-0)=1.23$  for J1319+0950, and J2054-0005.} 
	\end{table*}

	\section{Analysis}
	\label{sec:analysis}
	
	\subsection{Dust properties and star formation rate}
	\label{sec:sed}

	In order to ensure a self-consistent analysis of the cold-dust SEDs of J036+03, J0224-4711, J231-20 and J2054-0005, we performed an SED fitting of the flux densities reported in Tab. \ref{tab:c3-table-obs} considering the tapered fluxes for the higher-resolution observations if needed (see Appendix \ref{app:b} for details of the tapering procedure). The observations of J231-20 in \citet{Pensabene2021} did not need additional tapering because their resolution was well matched and low enough to account for the fainter and more extended emission. Moreover, for J231-20, we considered the flux corrected for the contribution of the companion to the QSO emission as explained in Sect. \ref{sec:res}.
	
	\noindent We modeled the dust continuum in an optically thick regime with a modified black body (MBB) function given by
	\begin{equation}\label{eq:MBB}
		S_{\nu_{\rm obs}}^{\rm obs} = S_{\nu/(1+z)}^{\rm obs} = \frac{\Omega}{(1+z)^3}[B_{\nu}(T_{\rm dust}(z))-B_{\nu}(T_{\rm CMB}(z))](1-e^{-\tau_{\nu}}),
	\end{equation}
	
	\noindent where $\Omega = (1+z)^4A_{\rm gal}D_{\rm L}^{-2}$ is the solid angle as a function of the surface area of the galaxy, $A_{\rm gal}$, and of the luminosity distance of the galaxy, $D_{\rm L}$. The dust optical depth is
	\begin{equation}
		\tau_{\nu}= \frac{M_{\rm dust}}{A_{\rm gal}}k_\nu= \frac{M_{\rm dust}}{A_{\rm gal}}k_0\biggl(\frac{\nu}{\nu_0}\biggr)^{\beta},
	\end{equation}
	\noindent where $\beta$ is the emissivity index, $k_{\nu}$ is the opacity, $k_0 = 0.45\  \rm cm^{2}\ g^{-1}$ is the mass absorption coefficient, and $\nu_0= 250$ GHz. The latter two terms define the opacity model adopted here \citep{beelen+2006, carniani2019}. In this model, $k_\nu$ carries huge systematics because the actual composition of the dust is currently unknown. Variations in the opacity model can, in principle, lower dust masses by  a factor of $\sim$ 3 or increase them by a factor of $\sim 1.5$. The solid angle for each source was estimated using the continuum mean size from resolved observations (see Tab. \ref{tab:c3-table-obs}). The effect of the CMB on the dust temperature is given by
	\begin{equation}
		T_{\rm dust}(z)=((T_{\rm dust})^{4+\beta}+T_0^{4+\beta}[(1+z)^{4+\beta}-1])^{\frac{1}{4+\beta}},
	\end{equation}
	\noindent with $T_0 = 2.73$ K.
	We also considered the contribution of the CMB emission given by $B_{\nu}(T_{\rm CMB}(z)=T_0(1+z))$ \citep{dacunha2013}.

\begin{figure*}[t]
		\centering
		\includegraphics[width=0.35\linewidth]{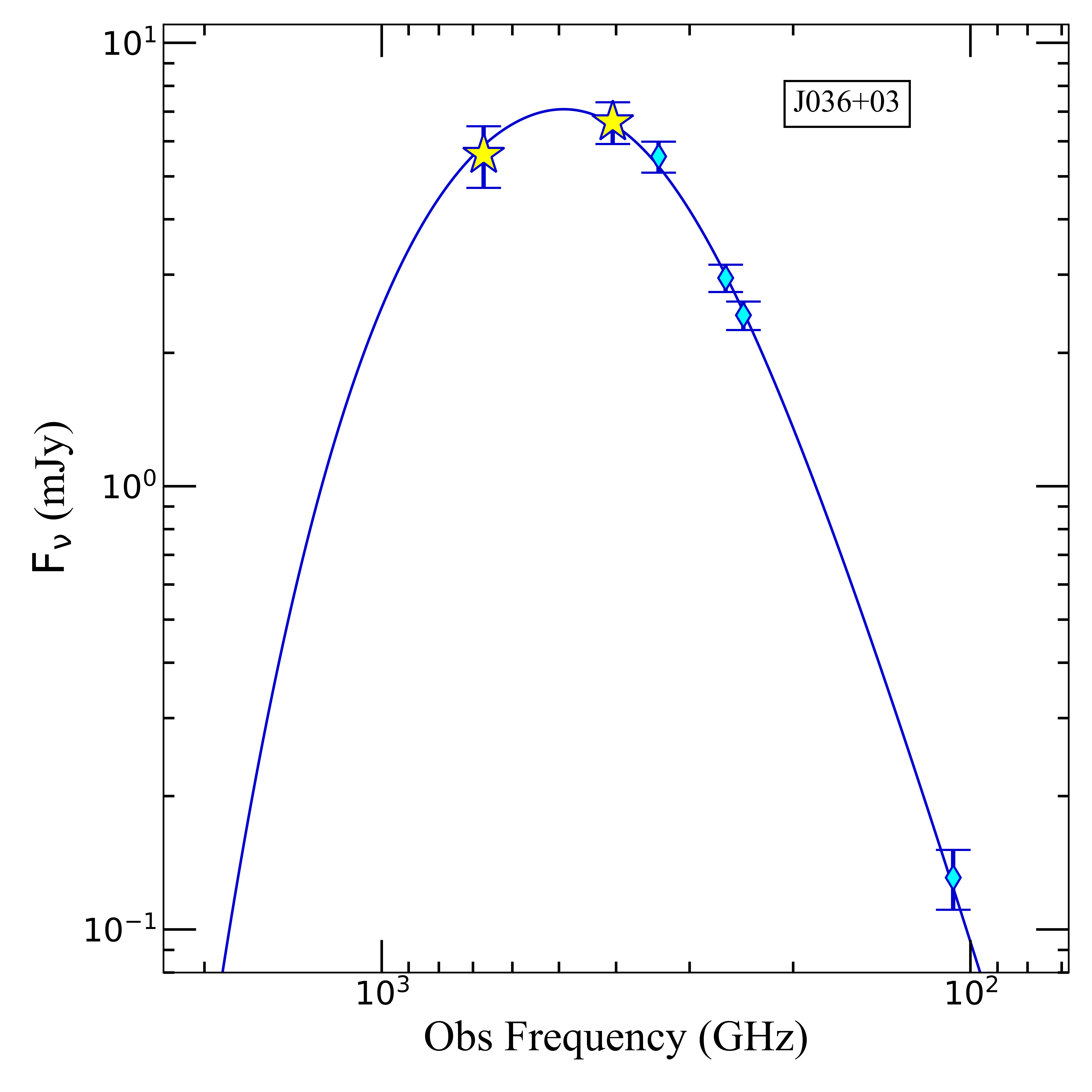}
		\includegraphics[width=0.35\linewidth]{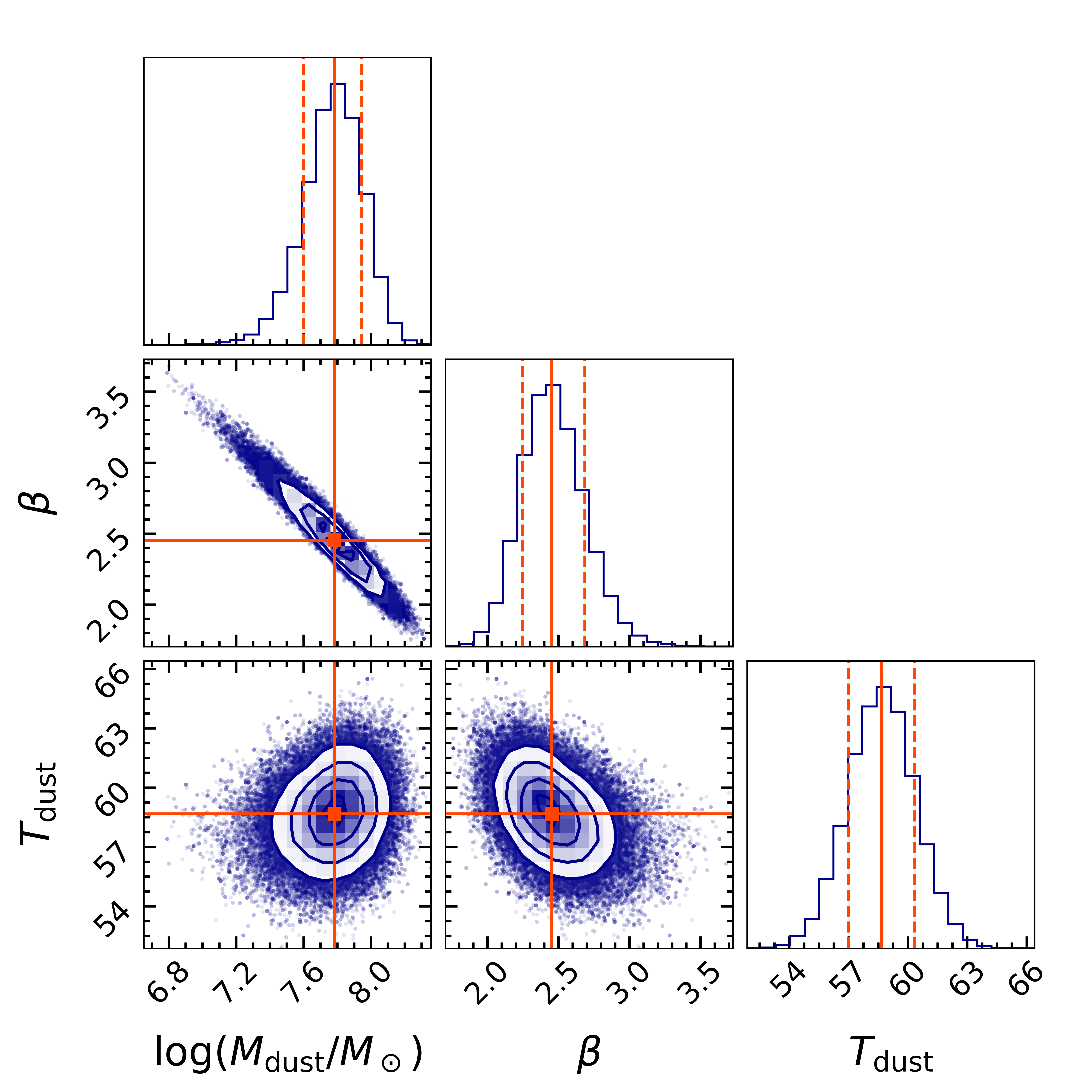} 
		\includegraphics[width=0.35\linewidth]{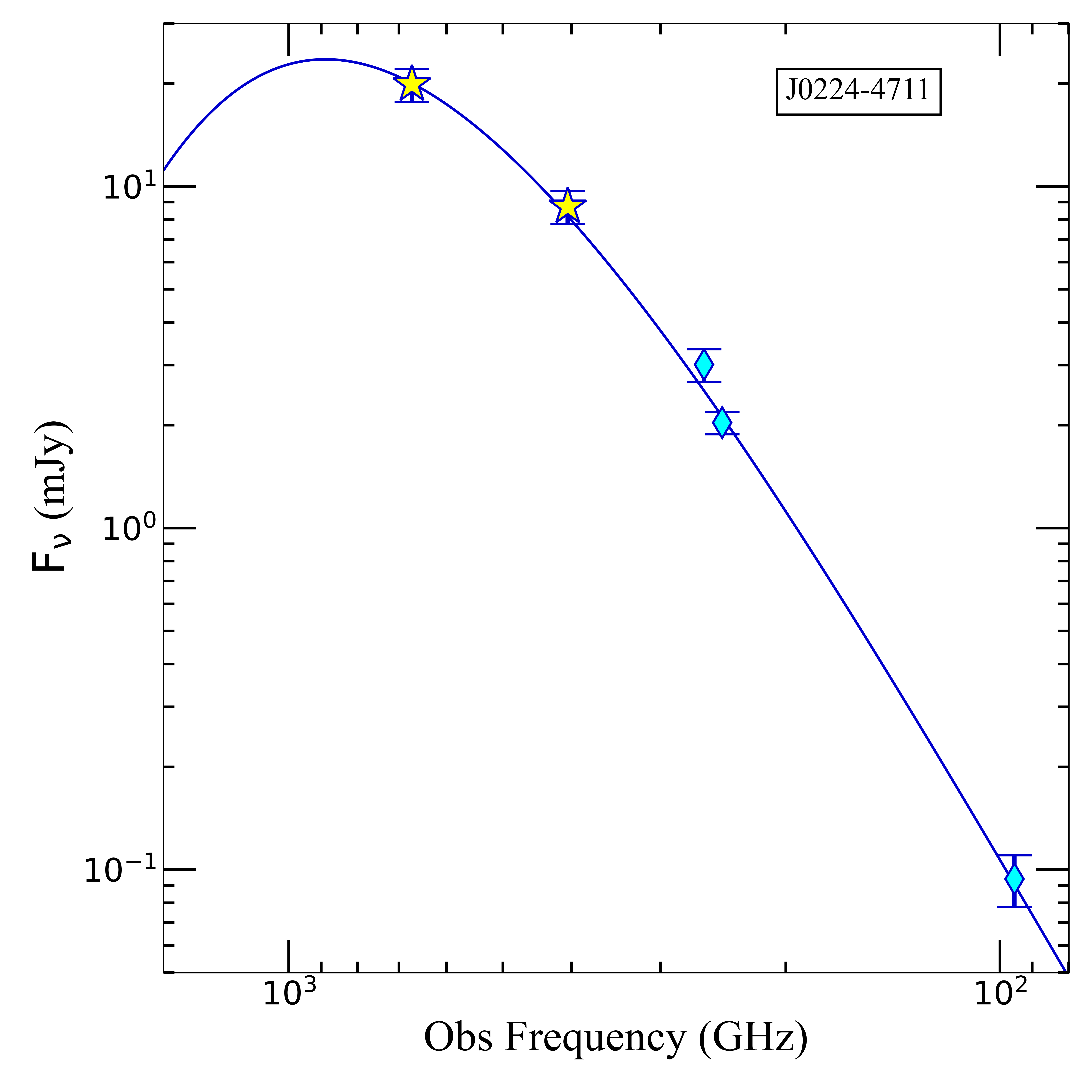}
		\includegraphics[width=0.35\linewidth]{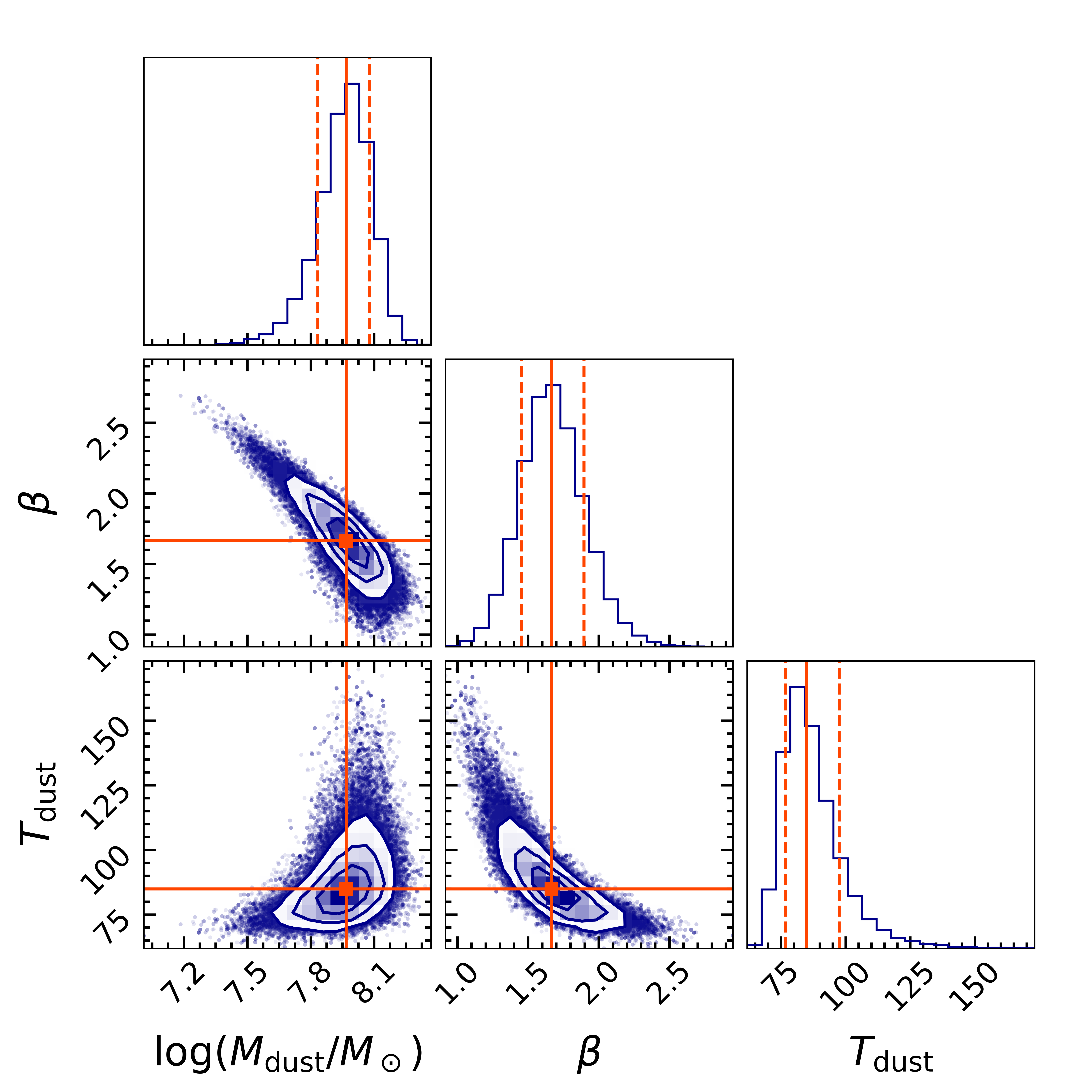}
		\includegraphics[width=0.35\linewidth]{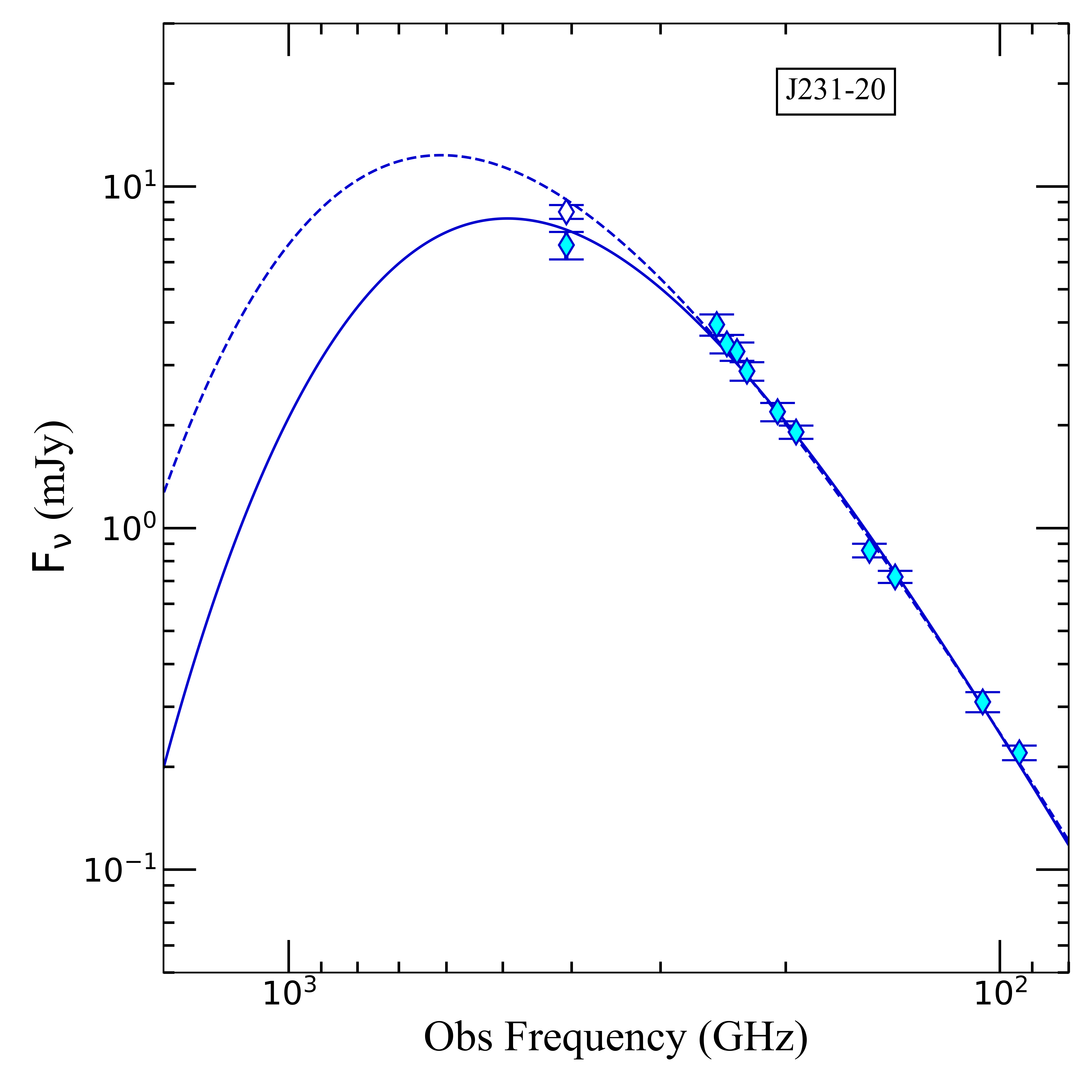}
		\includegraphics[width=0.35\linewidth]{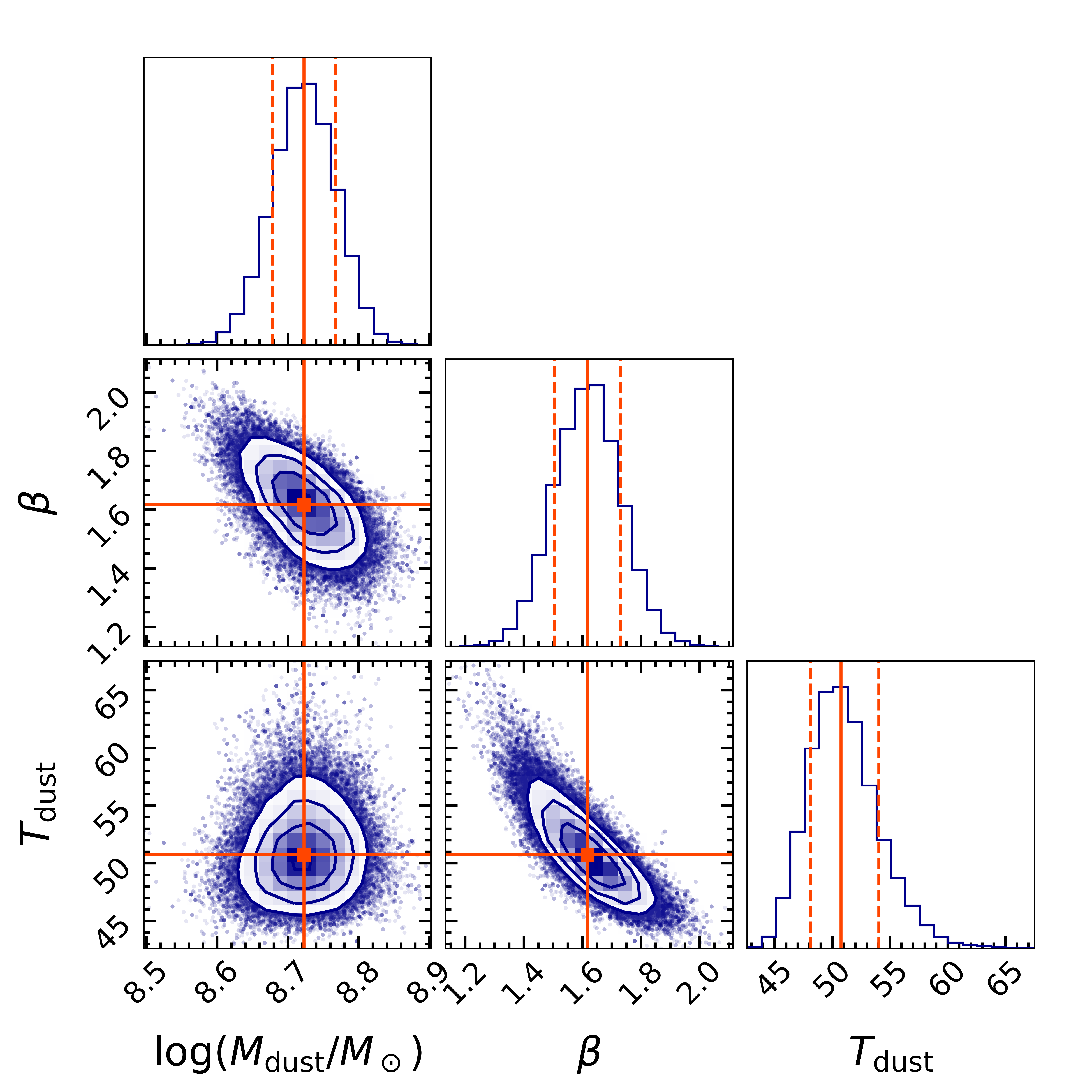}
		\caption{Observed SEDs and best-fitting results. Left panels: Observed SEDs of QSOs J036+03, J0224-4711, and J231-20 (top, central, and bottom row). Our new ALMA B8 and B9 data are shown as yellow stars, and other archival observations are plotted as cyan diamonds. The best-fitting curve is shown as the solid blue line. Right panels: Corner plot showing the posterior probability distributions of $T_{\rm dust}, M_{\rm dust}, {\rm ~and} ~\beta$. The solid orange lines indicate the best-fitting value for each parameter, and the dashed lines mark the 16th and 84th percentiles for each parameter. For J231-20, the empty diamond is the flux in B8, not corrected for the presence of the companion, and the dashed line is the best-fitting SED considering this flux (for details, see Sect. \ref{sec:sed}).}
		\label{fig:c3-sedj036}
	\end{figure*}

\begin{figure*}[t]
		\centering
		\includegraphics[width=0.45\linewidth]{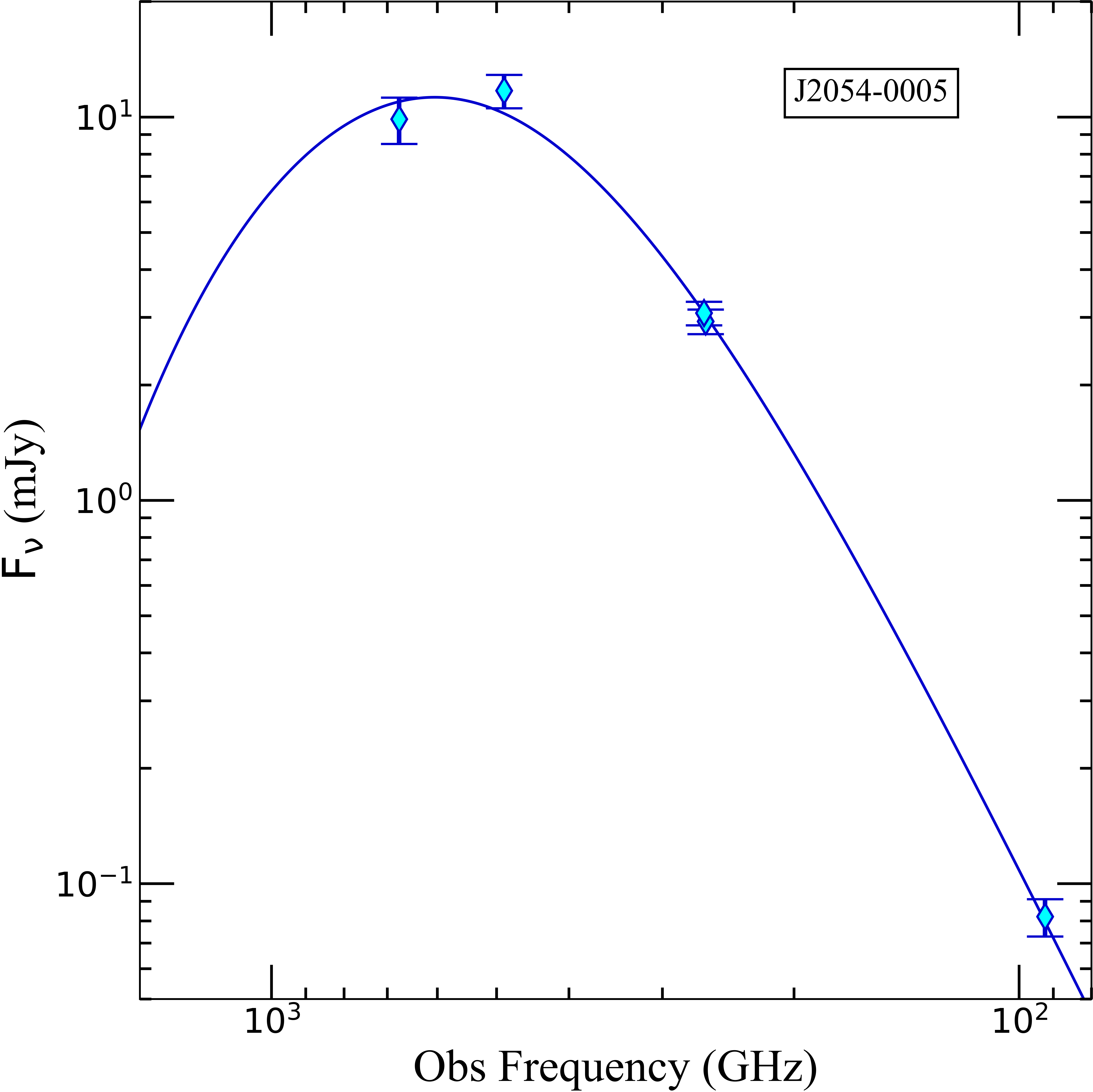}
		\includegraphics[width=0.45\linewidth]{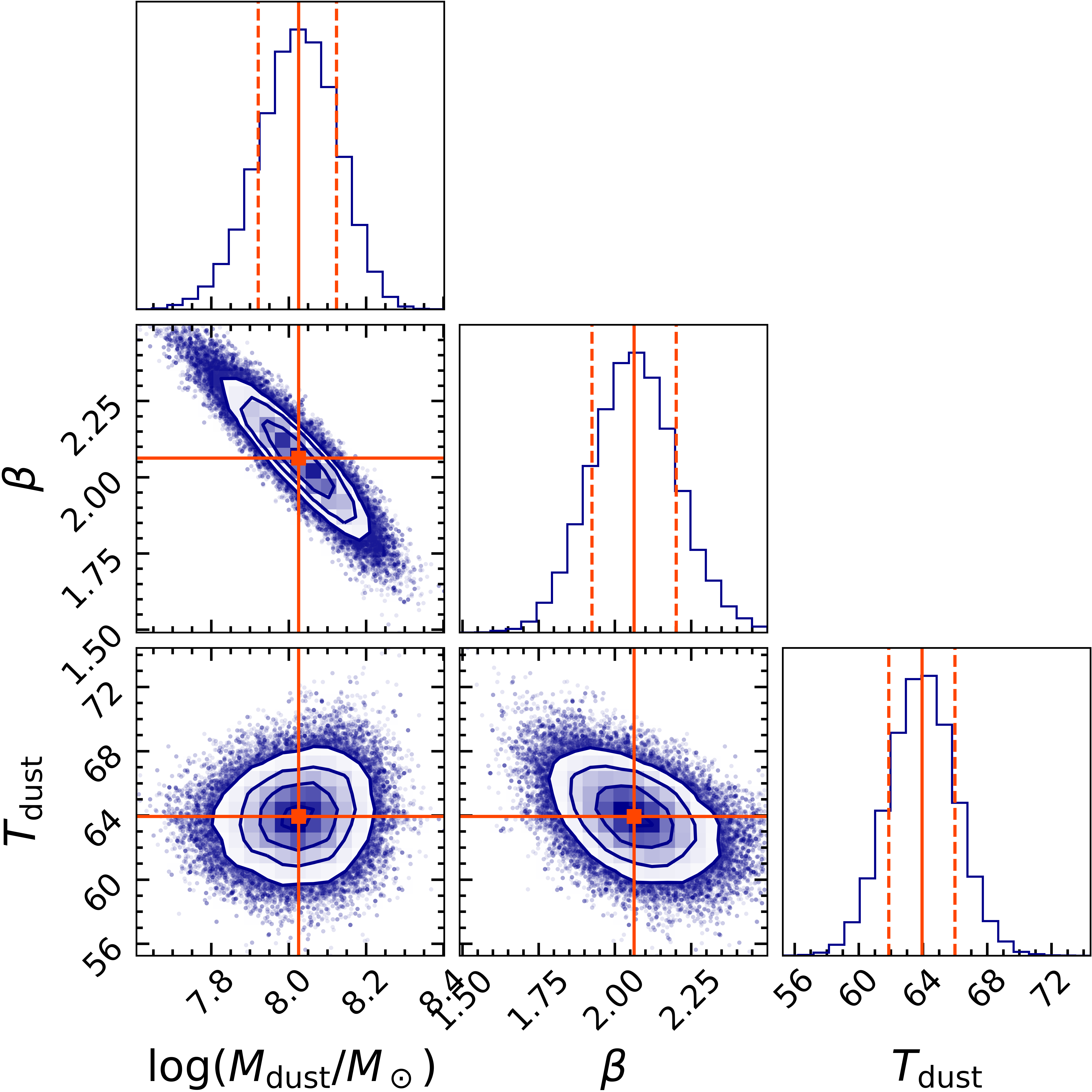}
		\caption{Same as Fig. \ref{fig:c3-sedj036} for QSO J2054-0005.}
		\label{fig:c3-sedj2054}
	\end{figure*}

	Therefore, the MBB model has three fitting parameters: the dust temperature ($T_{\rm dust}$), the dust mass ($M_{\rm dust}$), and $\beta$. We explored the 3D parameter space using a Markov chain Monte Carlo (MCMC) algorithm implemented in the \texttt{EMCEE} package \citep{foreman2013}. We assumed uniform priors for the fitting parameters: $10{\ \rm K} <T_{\rm dust}<300$ K, $10^{5} \ {\rm M_{\odot}}<M_{\rm dust}<10^{9}\ {\rm M_{\odot}}$, and $1.0<\beta<3.0$. The best-fitting $T_{\rm dust}$, $M_{\rm dust}$, and $\beta$, obtained from a MCMC with 40 chains, 6000 trials and a burn-in phase of $\sim 100$ for each QSO, are reported in Tab. \ref{tab:c3-sed-res}. The errors on the best-fitting parameters were computed considering the 16th and 84th percentiles of the posterior distribution of each parameter. Overall, we found $T_{\rm dust} = 50 - 80$ K, $M_{\rm dust}\sim 10^8\ \rm M_\odot$, and $\beta = 1.6-2.5$. Fig. \ref{fig:c3-sedj036} shows the observed SEDs with the best-fitting function and the posterior distributions of $T_{\rm dust}$, $M_{\rm dust}$, and $\beta$ for J036+03, J0224-4711, and J231-20, and Fig. \ref{fig:c3-sedj2054} shows this for J2054-0005. 
	
	We estimated the total infrared (TIR) luminosity for the best-fit model for each source by integrating from $8$ to $1000\ \mu$m rest-frame, and we derived the SFR considering that ${\rm SFR}\propto L_{\rm TIR}$. The proportionality factor between SFR and $L_{\rm TIR}$ depends on the initial mass function (IMF) chosen. We adopted a Kroupa IMF \citep[][]{kroupa2003}. The values of $L_{\rm TIR}$ and SFR derived for our QSO sample are reported in Tab. \ref{tab:c3-sed-res}. An Salpeter or Chabrier IMF \citep{salpeter1955,chabrier2003} would imply an SFR that is higher by a factor of 1.16 or lower by factor of 0.67, respectively. When comparing with results of the SFR from the literature, we rescaled the SFR according to the Kroupa IMF (see also \citealt{kennicutt2012}):
	
	\begin{equation}
		\label{eq:sfr}
		{\rm SFR}\ ({\rm M_\odot\ yr^{-1}})= 1.496 \times 10^{-10}\ L_{\rm IR\ 8-1000\mu m} (L_\odot).
	\end{equation}
	
	As a word of caution, we recall that we adopted a lower limit for the flux in B8 of J231-20, and thus,  $T_{\rm dust}$ and SFR can be underestimated in principle. By performing another fit considering the B8 flux not corrected for the contribution of the companion (which is an upper limit to the flux of the QSO), we indeed derived $T_{\rm dust}=61$ K and SFR$=1913 \ \rm M_\odot ~yr^{-1}$, which is a factor of 2 higher than that obtained considering the lower limit in B8, while the dust mass and emissivity index agree well with the previous values. The best-fitting curve is displayed as a dashed line in the bottom left panel of Fig. \ref{fig:c3-sedj036}, along with the uncorrected B8 flux as an empty diamond. Hereafter, we conservatively use the results of the SED fitting with the B8 flux corrected for the companion emission, considering that the SFR can vary of a factor of 2 at most.

	We adopted a contribution of the AGN emission to the dust heating of 50\% considering that all our sources are hyperluminous QSOs at $z\gtrsim 6$, likely sharing similar properties with the hyperluminous QSOs at $z=2-4$ in \citet{duras2017} and with J1148+5251 in \citet{schneider2015}, which also belongs to the HYPERION sample. This assumption yields SFR$=466\pm 75\ \rm M_{\odot}\ yr^{-1}$ for J036+03, an SFR$=2485^{+1682}_{-768} \ \rm M_{\odot}\ yr^{-1}$ for J0224-4711, an SFR$= 496\pm 118$ for J231-20, and an SFR$=730 \pm 75$ for J2054-0005. The role of AGN in heating the dust is further discussed in Sect. \ref{subsec:AGNeffect}.
	
	\begin{figure*}[t]
		\centering
		\includegraphics[width=0.45\linewidth]{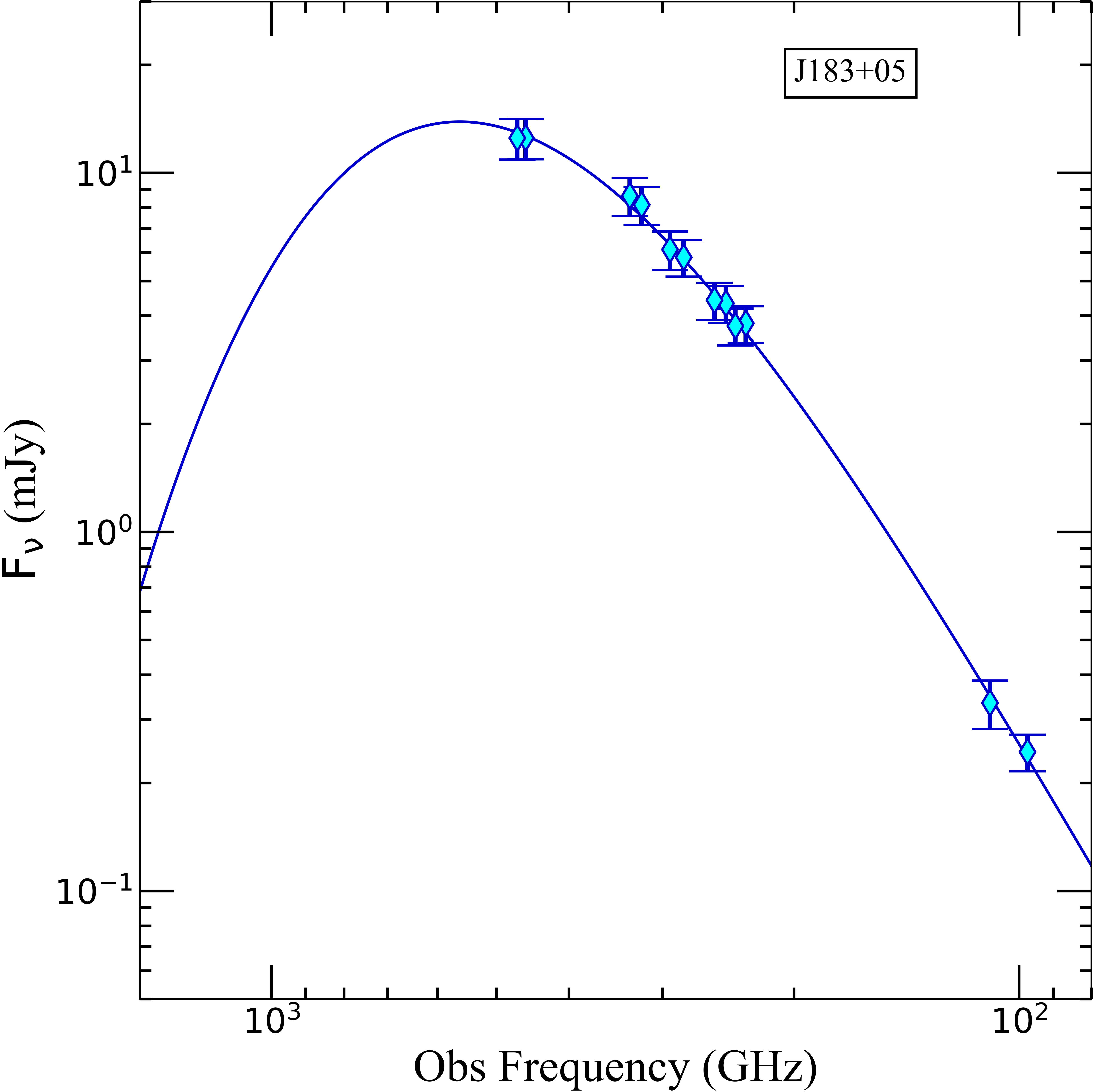}
		\includegraphics[width=0.45\linewidth]{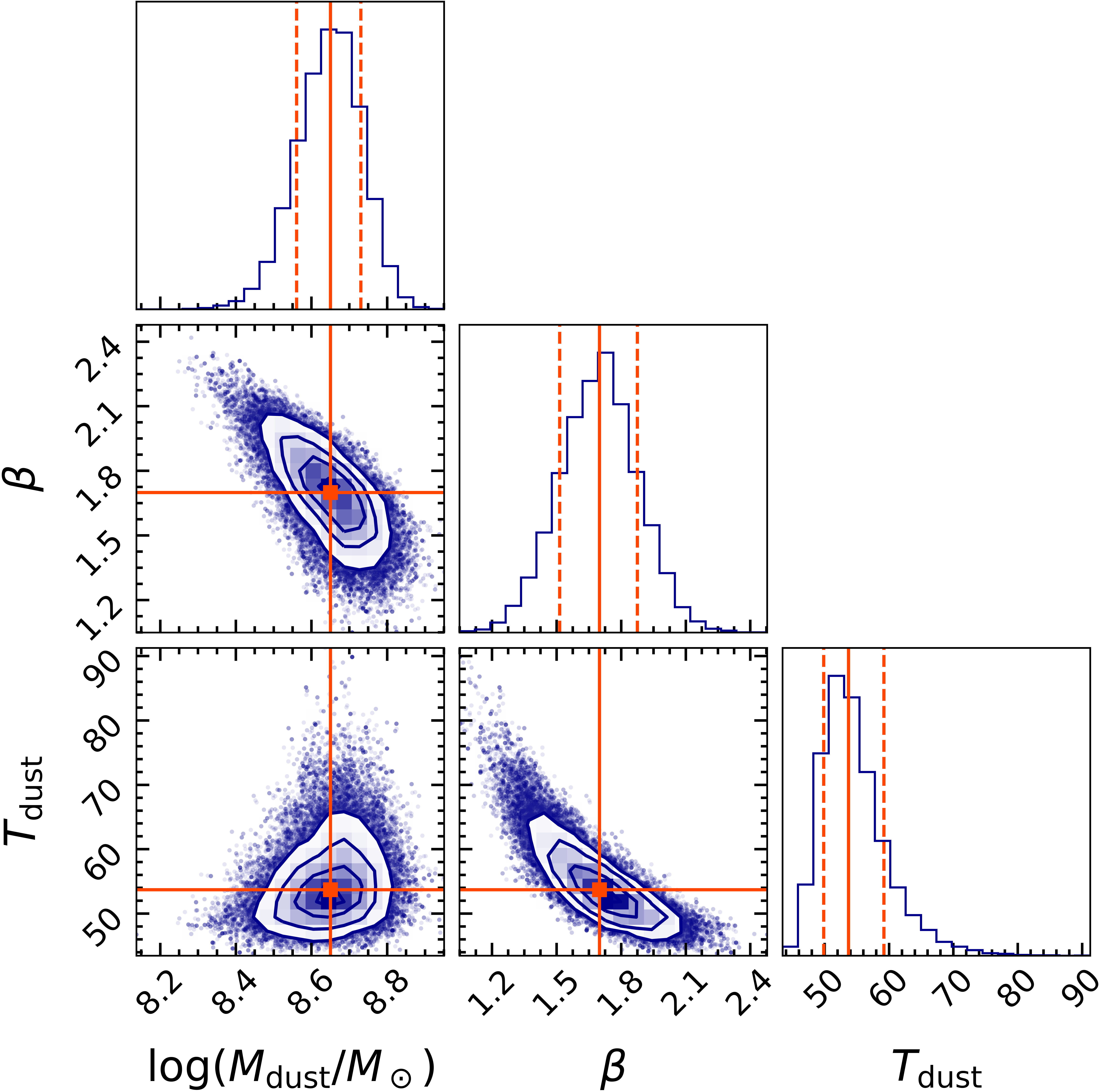} 
		\includegraphics[width=0.45\linewidth]{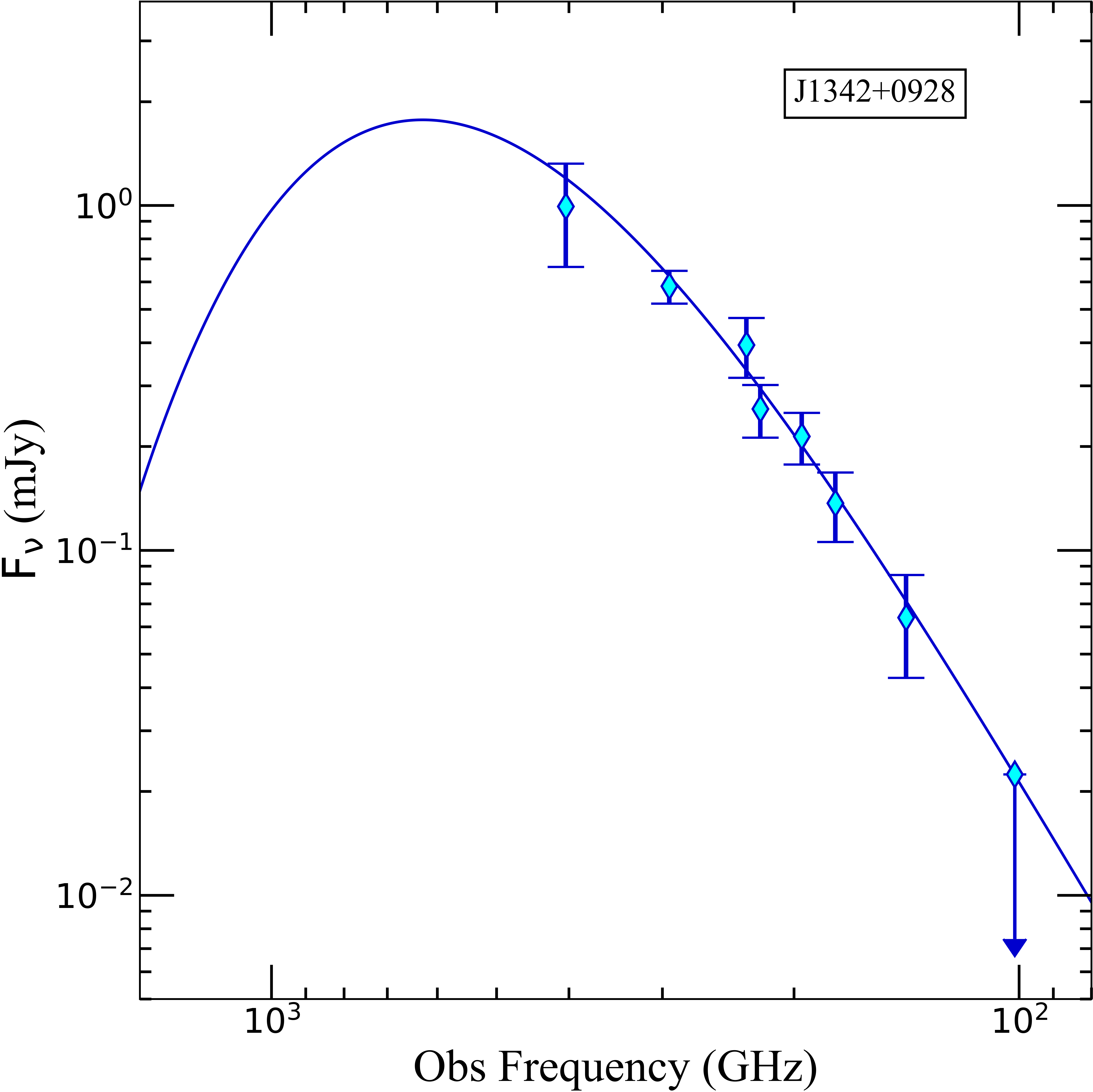}
		\includegraphics[width=0.45\linewidth]{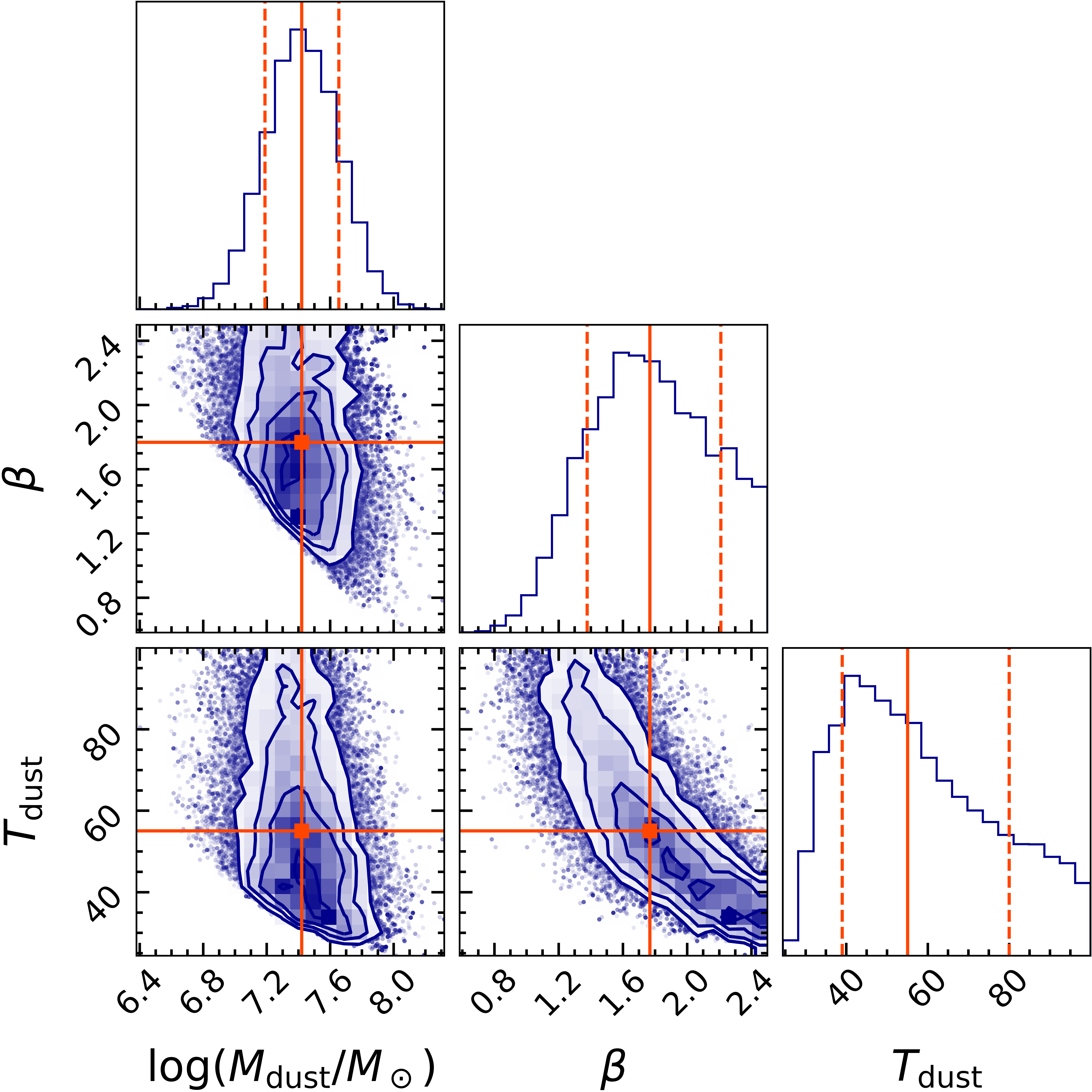}
		
		\caption{Observed SEDs and best-fitting results. Left panels: Observed SEDs of QSOs J183+05, J1342+0928 (top and bottom row) as cyan diamonds. The best-fitting curve is shown as a solid blue line. Right panels: Corner plot showing the posterior probability distributions of $T_{\rm dust}, M_{\rm dust}, {\rm ~and} ~\beta$. The solid orange lines indicate the best-fitting value for each parameter, and the dashed lines mark the 16th and 84th percentiles for each parameter.}
		\label{fig:c3-sedothers}
	\end{figure*}
	
	We wished to perform a statistically sound investigation of the dust properties in a sample of $z>6$ QSOs and therefore explored the archives and the literature in our search for other suitable candidates for our analysis of cold-dust SED, that is, with ALMA and/or NOEMA observations of the continuum emission probing both the Rayleigh-Jeans regime and the peak region of the SED. To our knowledge, only four other QSOs at $z>6$ currently have multiple-frequency observations sampling the Rayleigh-Jeans part and, barely, the peak of the cold-dust SED: QSOs J1319+0950, J1148+5251, J1342+0928, and J183+05. Their SEDs were analyzed in \citet{carniani2019} (J1319+0950 and J1148+5251), in \citet{novak2019, witstok2023} (J1342+0928), and in \citet{decarli2023} (J183+05). Since we adopted the same method as \citet{carniani2019}, we used the results for the dust properties and SFRs of J1319+0950 and J1148+5251 found in Carniani's work, which are reported here in Table \ref{tab:c3-sed-res}. For consistency, we modeled the observed SEDs of the other two QSOs using the procedure described in the previous paragraph because the method and/or the opacity models adopted in \citet{novak2019, witstok2023, decarli2023} were different from ours. The best-fitting SEDs are shown in Fig. \ref{fig:c3-sedothers}, and the results can be found in Tab. \ref{tab:c3-sed-res}. Within the uncertainties, our results agree well overall with those in the previous papers. The discrepancies in the best-fitting values arise from the different regime assumed (optically thin in \citealt{novak2019}, and thick in \citealt{witstok2023, decarli2023}) and in the different opacity models. 
	
	It is worth noting that the uncertainties on $T_{\rm dust}$ (SFR) increase to $\sim 40\%$ ($>80\%$ for SFR) when the peak of the cold-dust SED is not probed: This is the case of J1319+0950 and J1342+0928, which lack ALMA observations in bands 8 and 9, and of J0224-4711, for which even band 9 is unable to reach the peak of the SED given that this QSO has an extremely bright dust emission. In the latter case, an ALMA B10 observation is required. The strongest effect is seen in J1342+0928, for which $T_{\rm dust}$ is determined with an uncertainty of $\sim 40\%$, leaving the SFR basically unconstrained. To investigate the impact of B8-B9 observations on the uncertainties of $T_{\rm dust}$, we again fit the SEDs for all the QSOs in our sample, excluding the data available at $\lambda_{\rm rest}>100\ \mu$m for each source, that is, corresponding to ALMA B8 and B9. We found that in this case, the uncertainties on $T_{\rm dust}$ become $\sim 70\%$ on average (in some cases, up to 90\%). This highlights the importance of high-frequency observations (ALMA bands 8, 9, and 10) in providing precise and reliable estimates of the dust properties and SFR.
	
	\begin{figure}[t]
		\centering
		\includegraphics[width=1\linewidth]{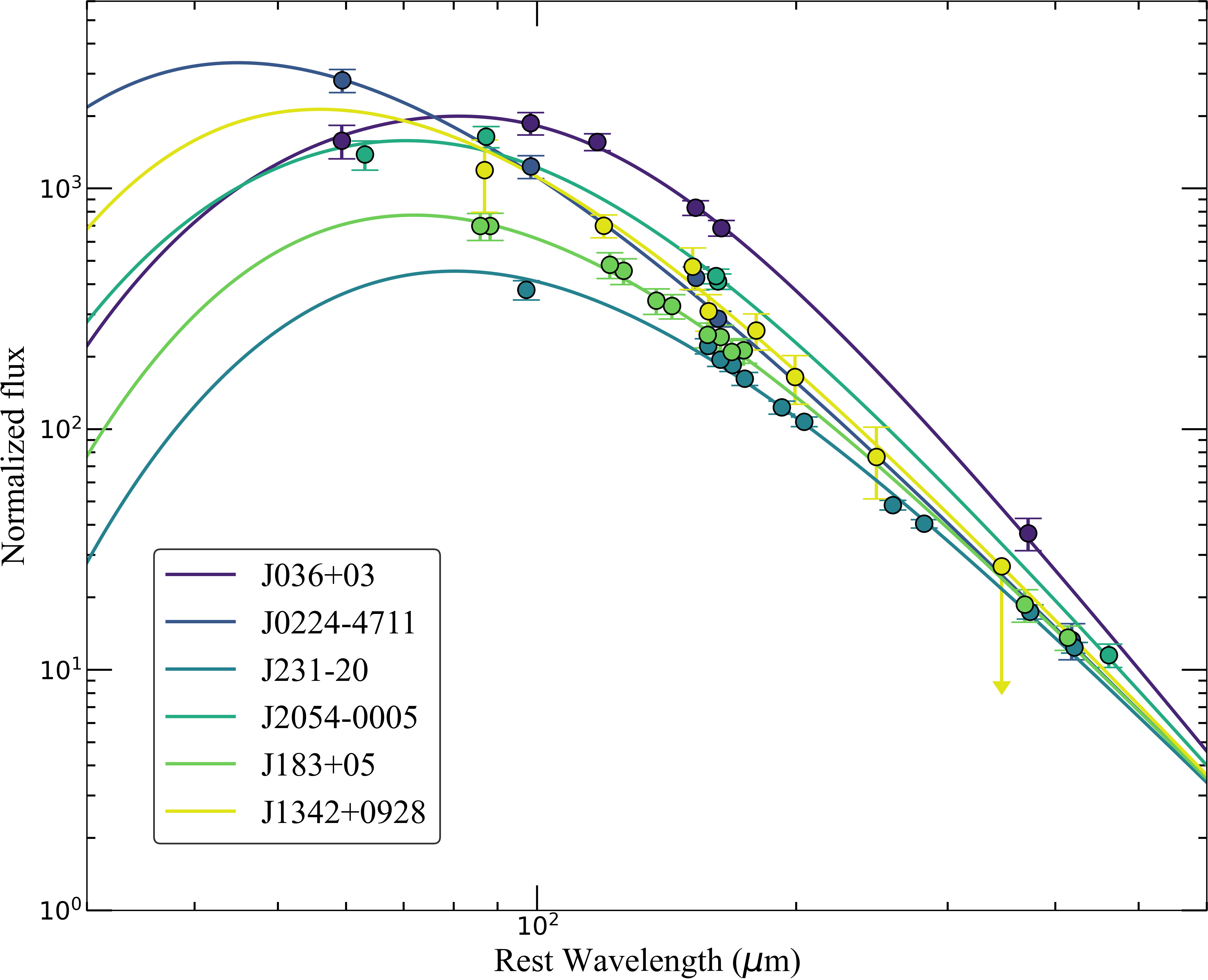}
		\caption{Compilation of cold-dust SEDs of the six QSOs analyzed in this work, normalized at 850 $\mu$m.}
		\label{fig:sed-all}
	\end{figure}
	
	In Fig. \ref{fig:sed-all} we show the cold-dust SEDs analyzed in this work. All SEDs were normalized at $\lambda = 850\ \mu$m rest-frame in order to facilitate the comparison. It is worth noting the difference in slope and position of the peak, which are clearly related to the variation in the dust properties, namely $T_{\rm dust}$, $M_{\rm dust}$, and $\beta$. 
	
\begin{figure}[t]
		\centering
		\includegraphics[width=1\linewidth]{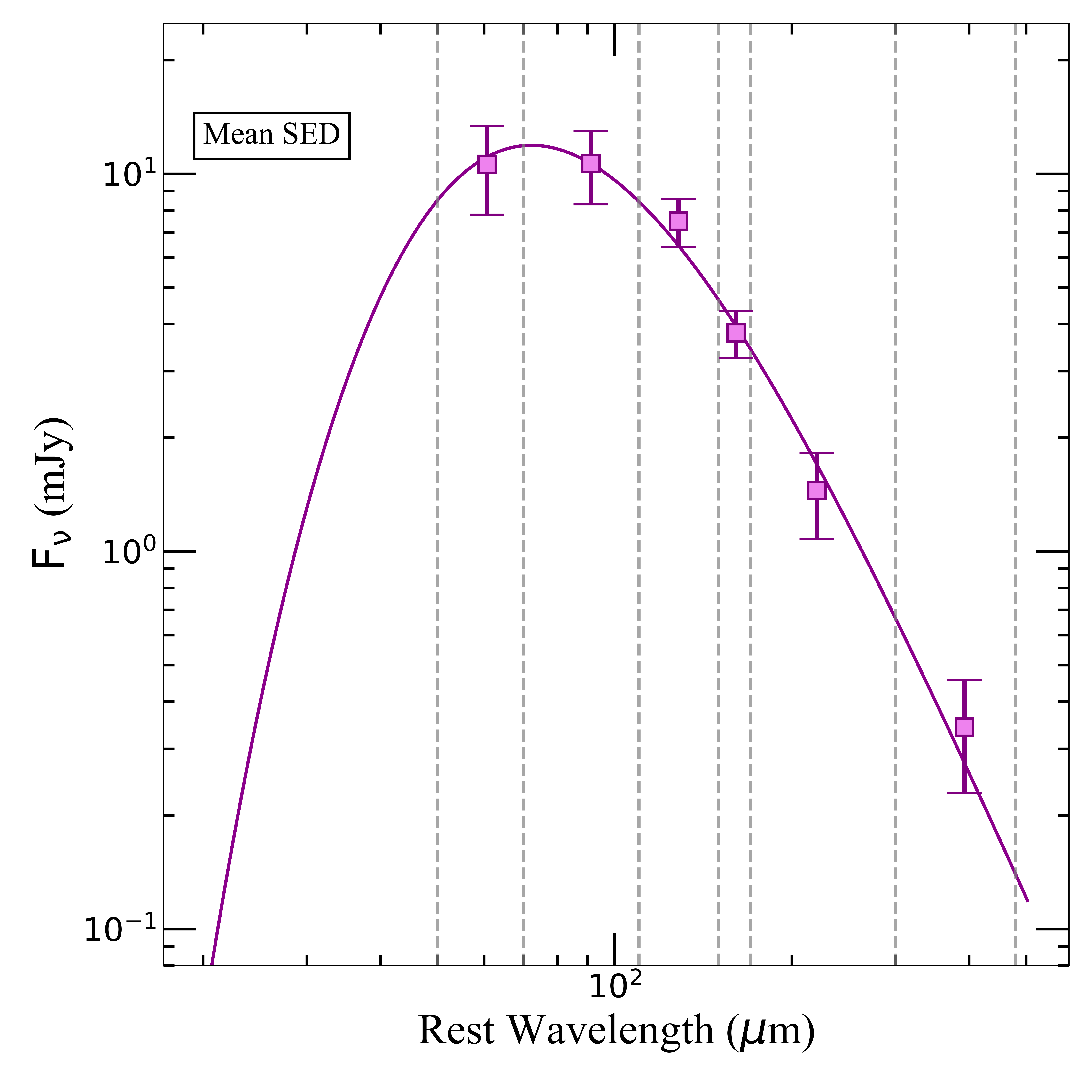}
		
		\includegraphics[width=1\linewidth]{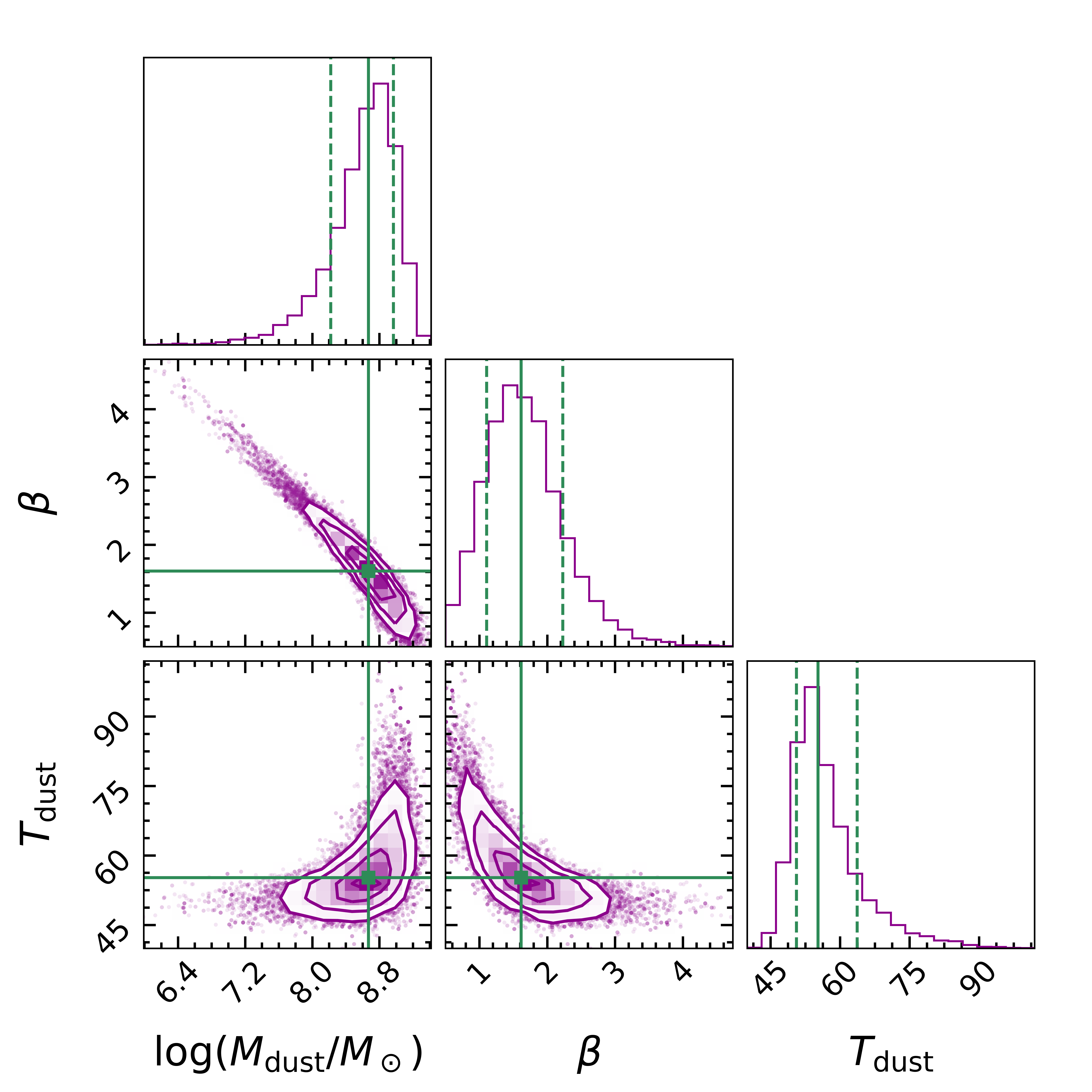}
		\caption{Results for the observed mean cold-dust SED. Left panel: Observed mean fluxes and best-fitting function. The dashed gray lines mark the six bins we used to develop the mean SED. Right panel: Corner plot showing the posterior probability distributions of $T_{\rm dust}, M_{\rm dust}, {\rm ~and} ~\beta$. The solid green lines indicate the best-fitting value for each parameter, while the dashed lines mark the 16th and 84th percentiles for each parameter.}
		\label{fig:sed-mean}
	\end{figure}

	Additionally, we developed an observed mean cold-dust SED by averaging (mean) the observational data available for all the QSOs in our sample in six wavelength bins\footnote{The bins were defined considering the different ALMA bands and trying to have a similar number of data in each bin.}. They are shown as dashed gray lines in Fig. \ref{fig:sed-mean}, along with the resulting averaged fluxes as violet squares. The error bars are the standard errors on the mean associated with each flux. We fit this mean SED following the procedure described above, and we obtained $T_{\rm dust}=54^{+8}_{-4}$ K, $M_{\rm dust}=(4.6\pm 3.0)\times 10^8\ \rm M_\odot$, and $\beta=1.6 \pm 0.55$. Fig. \ref{fig:sed-mean} shows the best-fitting function and the posterior distribution of each fitting parameter.
	
	\subsection{Molecular gas mass}
	\label{sec:gasmass}
	
	In this section, we infer the molecular gas mass of QSOs J0224-4711, J1319+0950, and J2054-0005  from molecular tracers such as the CO(7-6) and CO(6-5) emission lines. The molecular gas mass estimates in high-$z$ QSO host galaxies rely on CO observations, and especially on intermediate (J$_{\rm up}$=5-7) transitions \citep{venemans2017a, Yang2019, decarli2022}, which are found to be at the peak of the CO spectral line energy distribution \citep[CO SLED,][]{li2020, feruglio2023}. In principle, low-J transitions should be preferred, as they are less sensitive to uncertainties on the CO excitation. However, these are quite challenging to detect because of their intrinsic faintness.
	
\begin{figure}[t]
		\centering
		\includegraphics[width=1\linewidth]{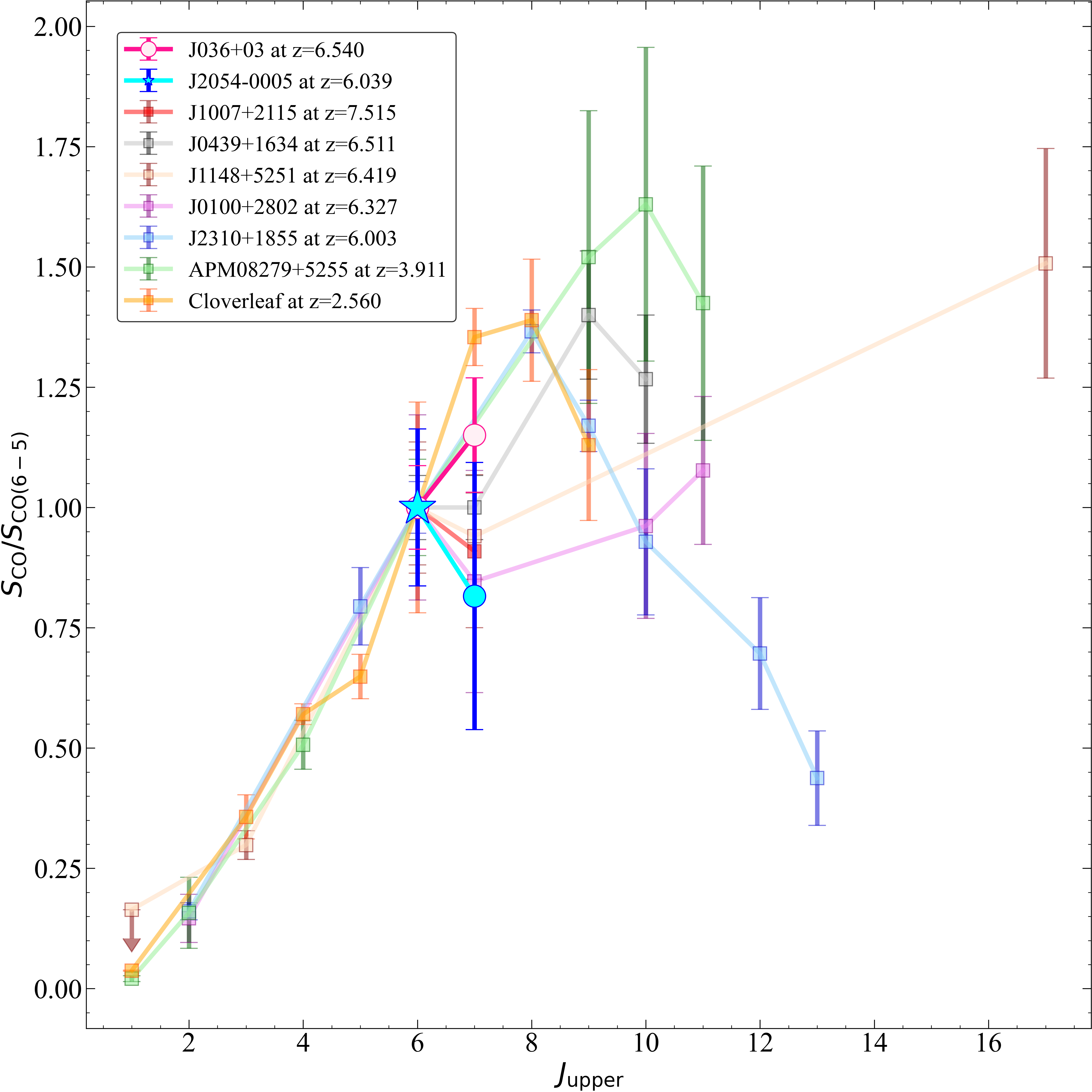}
		\caption{CO SLED of J036+03 and J2054-0005 compared with those of other QSOs at $z>6$ and at lower redshift. The CO SLEDs for J036+03 and J2054-0005 are shown as pink circles and cyan markers, respectively. For J2054-0005, the star refers to the CO(6-5) line studied in this work and the circle to the CO(7-6) in \citet{decarli2022}. The results for J036+03 are taken from \citet{decarli2022}. All other sources are displayed as shadowed squares: J1007+2115 at z=7.5419 in red \citep{feruglio2023}; J0439+1634 at $z=6.511$ in gray \citep{Yang2019}; J1148+5251 at $z=6.419$ in brown  \citep{riechers2009, gallerani2014}; J0100+2802 at $z=6.327$ in purple \citep{wang2019}; J2310+1855 at $z=6.003$ in blue \citep{li2020}; APM08279+5255 at $z=3.911$ in green \citep{papado2001, weiss2007}; and Cloverleaf at $z=2.560$ in orange \citep{bradford2009, uzgil2016}.}
		\label{fig:c2-cosled}
	\end{figure}

	\noindent The molecular gas mass was therefore derived as
	
	\begin{equation}
		M_{\rm H2, CO} = \alpha_{\rm CO} r^{-1}_{\rm J-(J-1)} L_{\rm CO(J-(J-1))},
	\end{equation}
	
	\noindent where $\alpha_{\rm CO}=0.8~ \rm M_{\odot}~ (K\,km\,s^{-1} \,pc^{2})^{-1}$ is the CO-to-H$_2$ conversion factor typical for ULIRG and QSOs \citep{carilli2013}, and $r_{\rm J-(J-1)}$ is the CO(J-(J-1))/CO(1-0) luminosity ratio. Fig. \ref{fig:c2-cosled} shows the CO SLEDs normalized to the J$_{\rm up}$=6 transition of J036+03 \citep{decarli2022} and J2054-4711, limited to the (6-5) and (7-6) transitions \citep[for CO(7-6) of J2054-0005 see ][]{decarli2022}, compared with other $z>6$ QSOs, and QSO APM0879+5255 at $z=3.9$ and QSO Cloverleaf at z=2.56 \citep{li2020, feruglio2023}, for which the CO SLED is well constrained down to J$_{\rm up}$=1. We note that the CO SLED for QSO at $z>6$ presents a flattening at the CO(6-5) and (7-6) transitions on average, and that all CO SLEDs have similar slopes from CO(6-5) to CO(2-1) transitions, while at higher J, the CO SLED are very different. 
	
	So far, APM0879+5255 and Cloverleaf are the only objects for which the CO(1-0) transition is detected and the ratio of the transition CO(2-1) and (1-0) is consistent between the two sources. The upper limit on the CO(1-0) of J1148+5251 also indicates a similar CO(2-1)-to-CO(1-0) ratio. Therefore, in order to estimate the molecular gas mass for J1319+0950, and J2054-0005 from CO(6-5), we assumed $r_{65}=$CO(6-5)/CO(1-0)=$1.23\pm 0.66$, considering the average between the maximum and minimum values of $r_{65}$ in APM0879+5255 and Cloverleaf. We then obtained $M_{\rm H2}=(1.5 \pm 0.2)\times 10^{10}$ for J1319+0950, and $M_{\rm H2}=(6.4 \pm 1.0) \times 10^9\ \rm M_\odot$ for J2054-0005. Regarding J0224-4711, we assumed that CO(7-6)$\sim$CO(6-5), given the averaged CO SLED of high-$z$ QSO (see Fig. \ref{fig:c2-cosled}), and $r_{76}=0.76\pm 0.41$, considering the average between APM0879+5255 and Cloverleaf. This gave $M_{\rm H2}=(1.0 \pm 0.1) \times 10^{10}\ \rm M_\odot$ for J0224-4711. 
	
	For the purposes of this work, we also investigated the properties of QSO J036+03, and thus, we computed its molecular gas mass from CO(6-5) consistently with the rest of the sample. \citet{decarli2023} reported the detection of the CO(6-5) emission line of J036+03, finding $L'_{\rm CO(6-5)}=(1.27\pm 0.11)\times 10^{10}~ \rm K\,km\,s^{-1} \,pc^{2}$. This yielded $M_{\rm H2}=(7.5 \pm 0.64) \times 10^9\ \rm M_\odot$ following our assumptions.
	
	Our analysis revealed that the gas masses in our sample of QSOs, estimated with the smallest statistical uncertainties, are $\sim 10^{10}\ \rm M_\odot$ on average. However, it is worth noting that our assumptions for the CO(6-5)-to-CO(1-0) (or CO(7-6)-to-CO(1-0)) line ratio introduce significant systematic uncertainties to our estimates of  $\gtrsim 50\%$. Our results agree well on average with the gas masses found in \citet{neeleman2021} derived from converting the [CII] underlying continuum flux into a dust mass and then assuming a constant GDR (=100). In contrast, the gas mass estimated from converting the [CII] luminosity directly into a molecular mass using the conversion of \citet{zanella2018} mostly is higher by one order of magnitude than ours. Specifically, molecular gas masses for QSOs J1319+0950 and J2054-0005 were derived in \citet{neeleman2021}, who reported higher values than ours using the GDR and the Zanella conversion factor. In the former case, the GDR assumed in \citet{neeleman2021} (GDR=100) may not be valid for every QSO (see the results and discussion of the GDR in our sample in Sect. \ref{subsec:GDR}). In the latter, the Zanella conversion factor, which is calibrated for $z\sim 2$ main-sequence galaxies, may not be applicable for high-$z$ QSO hosts. This was also reported in \citet{Tripodi2022} for the case of QSO J2310+1855 at $z\sim 6$.
	
	The molecular gas masses and line properties are summarized in Tab. \ref{tab:c2-colines}. The uncertainties reported on the gas mass are only statistical. Systematics uncertainties are induced by the choice of $\alpha_{\rm CO}$ (a factor of 0.2-0.3 dex) and by the scaling from high-J CO transition to J=1 using the CO excitation ladder (a factor of 20-30\%).
	
	\section{Discussion}
	\label{sec:disc}
	
	In this section, we first discuss our findings concerning the dust and gas within our high-$z$ QSO sample by conducting a comparative analysis with the properties observed in lower-z sources. Finally, we investigate the evolutionary paths of the ten QSOs in our sample, exploiting the results for the properties of their host galaxies derived in Sects. \ref{sec:sed} and \ref{sec:gasmass}.
	
	\subsection{Properties of the QSO host galaxies}
	
	Our final sample comprises ten QSOs at $6\lesssim z<7.5$, out of which six belong to the HYPERION sample. In the previous sections, we analyzed the cold-dust SED in a homogeneous way for all quasars, and the results are summarized in Tab. \ref{tab:c3-sed-res}. In the following, we briefly describe the compilation of galaxy and AGN host samples we used for the comparison.
	
	\begin{figure*}[t]
		\sidecaption
		\includegraphics[width=12cm]{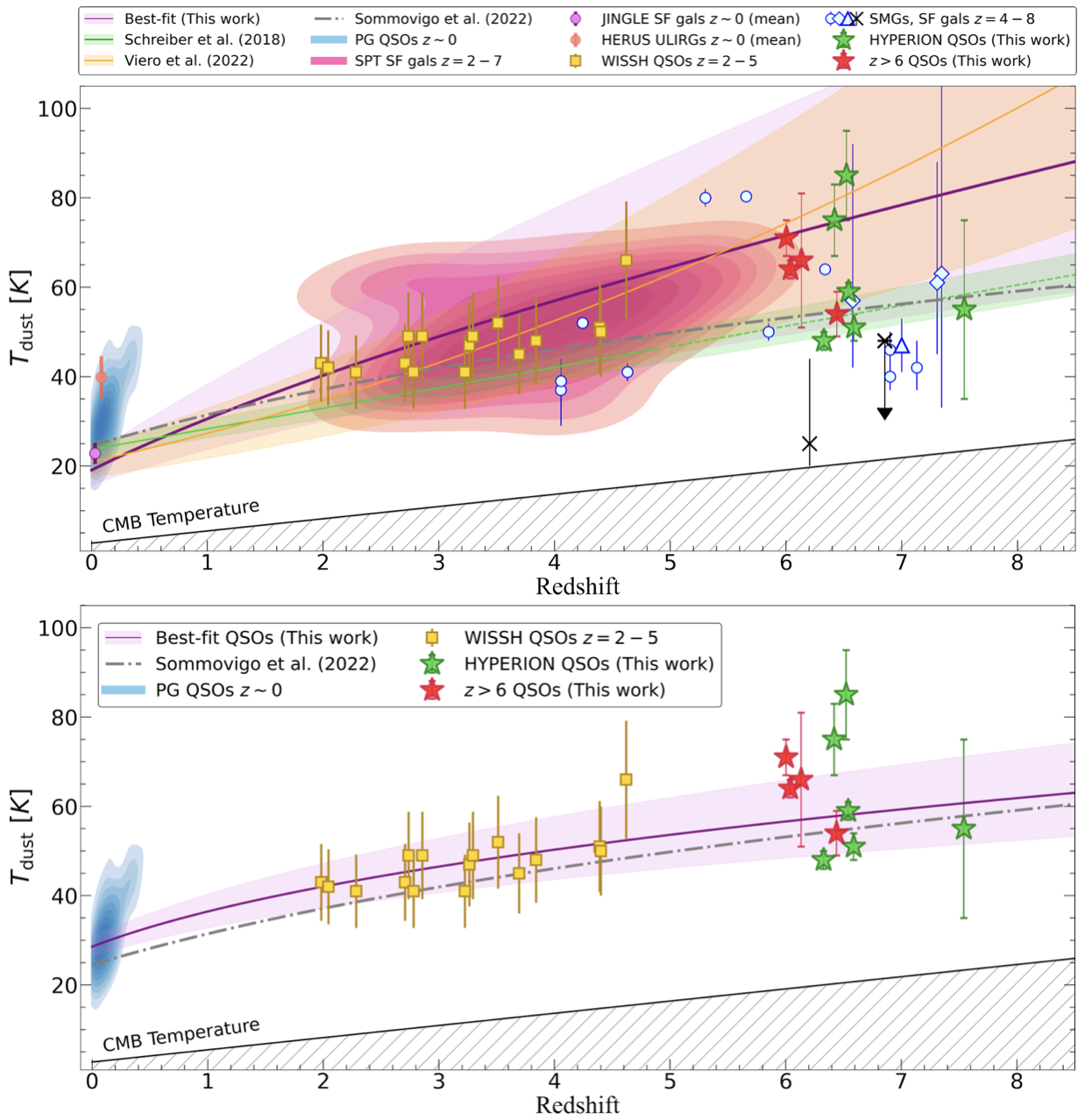}
		\caption{Dust temperature as a function of redshift. Our sample is shown as stars (red for Hyperion QSOs, green otherwise). Top panel: Comparison of our results with samples of local QSOs \citep[PG, blue colormap with contours, ][]{petric2015}, local star-forming galaxies \citep[JINGLE, mean value as pink dot, ][]{lamperti2019}, local ULIRGs \citep[HERUS, mean value as orange dot, ][]{clements2018}, $2<z<7$ star-forming galaxies \citep[SPT, pink colormap with contours ][]{reuter2020}, $2<z<5$ QSOs \citep[WISSH, yellow squares, ][]{duras2017}, $4<z<7.5$ SMGs and star-forming galaxies \citep[light blue dots, ][]{witstok2023}, three $z>6.7$ REBELS galaxies \citep[light blue diamonds, ][]{algera2023}, an average-REBELS galaxy \citep[light blue triangle, ][]{sommovigo2022}, and two galaxies with very low dust temperatures at $z>6$ \citep[black crosses, ][]{witstok2022, harikane2020}. The observed trends inferred by \citet{viero2022} and \citet{schreiber2018} are shown as an orange and green lines and shaded regions, respectively. Our best-fitting curve is shown as a purple line with a shaded region. The theoretical relation found in \citet{sommovigo2022} is shown as a dashed gray line. The solid black line marks the CMB temperature level. Bottom panel: Focus on the comparison to the samples of local PG QSOs (blue color map with contours) and $2<z<5$ WISSH QSOs. Our best-fitting curve considering QSOs alone is shown as a purple line with a shaded region.}
		\label{fig:c3-tdust-z}
	\end{figure*}
	
	\noindent For the local Universe, we considered the JCMT dust and gas the In Nearby Galaxies Legacy Exploration (JINGLE) survey, the \textit{Herschel} (U)LIRG Survey (HERUS), and a sample of QSOs selected from the Palomar-Green (PG) survey \citep[as also done in][]{witstok2023}. 
	The JINGLE sample is composed of 192 nearby ($0.01<z<0.05$) galaxies, for which \citet{lamperti2019} studied their cold-dust SED using photometric data in the 22-850 $\mu$m range from \textit{Herschel}, applying a hierarchical Bayesian fitting approach. We used the cold-dust properties derived from their two MBB models, which also take into account the warm dust component in the optically thin regime. These results agreed well with those from a single MBB model. The HERUS sample comprises 43 $z<0.3$ (ultra)-luminous infrared galaxies or (U)LIRGs observed with the \textit{Infrared Astronomical Satellite} (IRAS) and \textit{Herschel} \citep{sanders2003,clements2018}. \citet{clements2018} derived the dust properties for this sample assuming an MBB function in the optically thin regime. \citet{witstok2023} recalculated the dust properties of the JINGLE and HERUS datasets, employing a method analogous to ours (see Sect. \ref{sec:sed} and \citealt{witstok2022}), except for the variation in the chosen opacity model. Their results agreed with those presented in \citet{lamperti2019} and \citet{clements2018} within the uncertainties. Therefore, for the sake of simplicity, we used the results from the original papers. The dust properties of the PG sample, consisting of 85 nearby ($z<0.5$) QSOs, were obtained by \citet{petric2015} through modeling the photometry taken by \textit{Herschel} by means of an MBB function with $\beta$ fixed to 2.0 and the dust model of \citet{draine2007}\footnote{They also derived the dust properties using the dust model of \citet{compiegne2011} with $\beta=1.91$ finding that the dust masses are systematically larger by about 20\%-40\% when adopting the model of \citet{draine2007}.}. 
	
	At higher redshift, we selected a sample of 81 $2<z<7$ strongly gravitationally lensed, dusty star-forming galaxies identified by the South Pole Telescope (SPT). \citet{reuter2020} analyzed the cold-dust SEDs of the objects in this sample employing an MBB function in the optically thick regime with $\beta$ fixed to 2. We also included the dust properties inferred for 16 QSOs belonging to the WISE-SDSS Selected Hyper-luminous (WISSH) sample at $2<z<5$ \citep{duras2017}. Their SEDs were analyzed accounting for the cold dust, dusty torus, and warm dust emission components. In particular, the cold-dust emission was modeled by an MBB function in the optically thin regime with $\beta=1.6$. Finally, at $4<z<7.5$, we considered a sample composed of three submm galaxies (SMGs) and eight star-forming (SF) galaxies from \citet{witstok2023}. As stated before, they adopted a model for the SED fitting analogous to ours.
	For the QSO host galaxy samples, we note that the PG QSOs have a lower bolometric luminosity $L_{\rm bol}$ compared to our $z\gtrsim 6$ sample ($L_{\rm bol, PG} \sim 10^{44-47}\rm ~erg ~s^{-1}$), whereas the WISSH quasars are the most luminous, with $L_{\rm bol, WISSH} \gtrsim 10^{47.5}\rm ~erg ~s^{-1}$ \citep[see, e.g., ][]{vietri2018}.
	
	\subsubsection{Dust temperature}
	\label{subsec:tdust}
	
	In the past few years, the trend of $T_{\rm dust}$ as a function of time has been the object of multiple studies \citep[e.g.,][]{faisst2017,schreiber2018,sommovigo2020,bakx2021,viero2022}. However, the behavior of the $T_{\rm dust}-z$ relation, which increases or plateaus at $z>4$, has remained unclear. The top panel of Fig. \ref{fig:c3-tdust-z} presents the redshift distribution of the dust temperatures in our sample compared to the low-z samples described above. Additionally, we display the average $T_{\rm dust}$ derived for REBELS\footnote{`Reionization Era Bright Emission Line Survey' (REBELS; PI: Bouwens) is an ALMA large program targeting 40 of the brightest known galaxies at $z>6.5$ \citep{bouwens2022}.} galaxies by \citet{sommovigo2022} as a light blue triangle, and $T_{\rm dust}$ found for three REBELS galaxies at $z>6.7$ by \citet{algera2023} as light blue diamonds. As a word of caution, the dust temperatures for REBELS galaxies derived in \citet{sommovigo2022} carry large uncertainties since they are based on a combination of a single photometric point in band 6 and the [CII] line emission. Hence, in this context, we exclusively present the mean value. \citet{algera2023} analyzed band 8 and 6 observations of REBELS-12\footnote{The Band 8 continuum of REBELS-12 was undetected, therefore they derived an upper-limit.}, REBELS-25, and REBELS-38 considering both optically thin and thick cases. We present the results for the optically thick case, fixing $\beta=1.5$\footnote{Fixing $\beta$ introduces an uncertainty of $\sim 25\%$ in the estimate of $T_{\rm dust}$.}. Because we relied on only two photometric points per galaxy, the inferred $T_{\rm dust}$ are still uncertain ($\Delta T_{\rm dust}=30-60\%$); the results obtained in the thin and thick regime agree within the large uncertainties. For clarity, we show the mean values of the temperature distribution for the JINGLE and HERUS samples with their corresponding standard deviation\footnote{For both samples, the mean value of the distribution corresponds to the median.}. Interestingly, there is no significant difference in $T_{\rm dust}$ between QSOs and normal SF galaxies at fixed redshift, also considering that our sample is biased toward high luminosity and therefore possibly higher dust temperatures. 
	
	We observe an increasing trend of $T_{\rm dust}$ with redshift that is naturally expected from a theoretical perspective as a result of decreasing gas depletion times, as seen in \citet{sommovigo2022}. Sommovigo et al. theoretically derived that $T_{\rm dust}\propto t_{\rm dep}^{-1/6}\propto (1+z)^{5/2(4+\beta)}$, where $\beta$ is the dust emissivity index. This $T_{\rm dust}-z$ relation is shown in Fig. \ref{fig:c3-tdust-z} (dash-dotted gray line) for $\beta=2.03$ and assuming optical depth and metallicity (in solar unit) both equal unity, as presented in \citet{sommovigo2022}, implying $T_{\rm dust}\propto (1+z)^{0.42}$. This theoretical relation is able to reproduce the trend observed in many SF galaxies (e.g., considering REBELS and ALPINE galaxies), and in \citet{schreiber2018} from the stacking of star-forming galaxies at $0.5<z<4$ in the deep CANDELS fields. The stacked SEDs were fit with a library of template SEDs generated with $\beta=1.5$. \citet{schreiber2018} found a linear trend with redshift, and we also show an extrapolated curve at $z>4$ as dashed green line. 
	
	\begin{figure}
		\centering
		\includegraphics[width=1\linewidth]{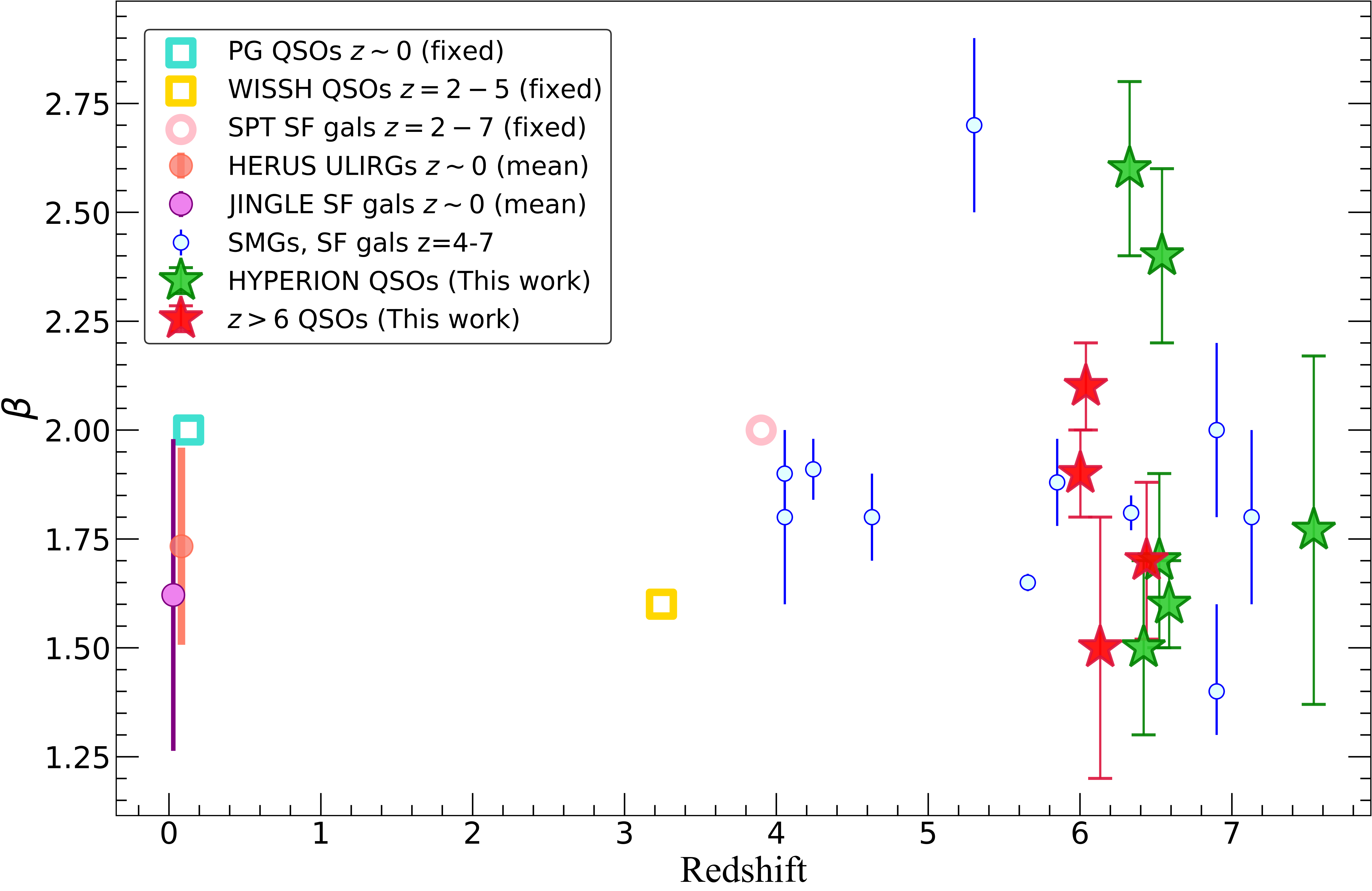}
		\caption{Dust emissivity index, $\beta$, as a function of redshift. The symbols and colors are the same as in Fig. \ref{fig:c3-tdust-z}. For the PG, WISSH, and SPT samples, the symbols are empty to indicate that the value of $\beta$ in these cases was fixed and not derived from SED analyses.}
		\label{fig:c3-beta-z}
	\end{figure}
	
	\begin{figure}
		\centering
		\includegraphics[width=1\linewidth]{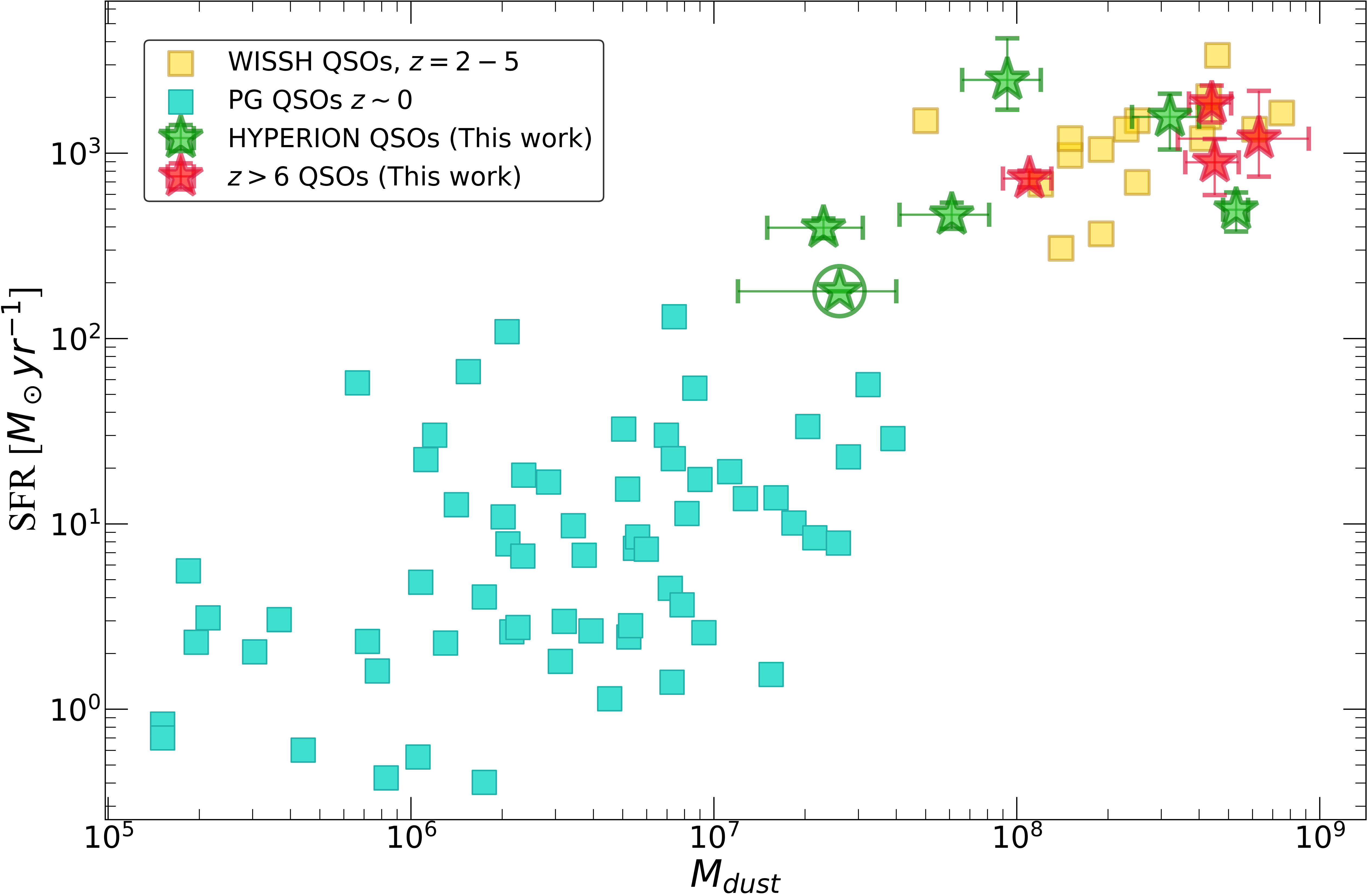}
		\caption{Star formation rate as a function of dust mass. The symbols and colors are the same as in Fig. \ref{fig:c3-tdust-z}. The SFRs for the WISSH and our sample are corrected for a factor of 50\% to take the AGN contribution to the dust heating into account.}
		\label{fig:c3-SFR}
	\end{figure}
	
	On the other hand, the population of SPT SF galaxies, QSOs ,and SF with $T_{\rm dust}>60$ K shows a steeper increase in $T_{\rm dust}$ with z, which is well captured by the observed relation inferred by \citet{viero2022}. They employed stacked maps in the FIR/submm of 111227 galaxies at $0<z<10$ from the COSMOS-MOS2020/FARMER catalog to derive dust temperatures at different redshifts. Their observed relation for $T_{\rm dust}(z)$, which is a second-order polynomial in $z$, agrees well with the observed trend of the WISSH QSOs, with the sources at $z>5$ that shows the most extreme temperatures ($\sim 70-80$ K), and with the lower limit of \citet{bakx2020} and estimate of \citet{behrens2018} based on data from \citet{laporte2017} \citep[see][]{viero2022}. However, it fails to describe the bulk of the PG QSOs and HERUS ULIRGs and the sources at $z>5$ with temperatures below $60$ K (\citealt{faisst2020,bethermin2020,hashimoto2019,sommovigo2022}, see \citealt{viero2022}). As a word of caution, these latter low-temperature estimates are very uncertain because they were derived from ALMA observations that did not bracket the peak of the SED, and they are therefore not adequate for modeling hotter dust components. 
	
	In order to find a general relation that can be applied to SF galaxies, SMGs, and QSO host galaxies at high $z$, we fit the observed data (excluding the stacked results) adopting the parameterization of $T_{\rm dust}(z)$ that was theoretically found in \citet{sommovigo2022}, therefore using a power law of the form 
	
	\begin{equation}
		T_{\rm dust} (z) = A \times (1+z)^B,
	\end{equation}
	
	\noindent where $A$ and $B$ are free fitting parameters. The latter is linked with $\beta$, since $\beta=-4+5/2B$ (considering the derivation of \citealt{sommovigo2022}). We found $A=19\pm 2$ K and $B=0.7\pm 0.1$, implying a value of $\beta$ that is unphysical. This underlines that $T_{\rm dust}$ does not depend uniquely on redshift, especially when considering different galaxy populations. \citet{sommovigo2022} pointed out that at fixed redshift, the scatter in $T_{\rm dust}$ derives from variations either in optical depth or in metallicity. This is a likely scenario for different galaxy populations. Our best-fitting function, shown as a purple line and shaded region in the top panel of Fig. \ref{fig:c3-tdust-z}, is slightly flatter than Viero's at $z>1$, but still agrees with it pretty well in both the low- and high-$z$ regimes. It captures both the flattening and the increasing trends at $z>4$, that is, it describes the populations with both higher and lower $T_{\rm dust}$. 
	
	Taking advantage of our robust estimates for $T_{\rm dust}$ in our QSO sample, we also performed a fit only to the observed QSO dust temperatures at different redshift in order to investigate the $T_{\rm dust}-z$ relation in QSOs for the first time. We found $A=29\pm 2$ K and $B=0.35\pm 0.04$. This trend is flatter than Sommovigo's, but still agrees well with observed SF galaxies. The majority of the sources are within the uncertainties, including those from \citet{schreiber2018}. Three sources with low $T_{\rm dust}$ belonging to the sample of \citet{witstok2023} (SPT0311-58W, SPT0311-58E, and A1689-zD1) agree within $2\sigma$ with our relation, along with another three sources that are well known for their low dust temperatures \citep[J0217-0208, COS-3018555981, and REBELS-25, see][]{witstok2022, algera2023, harikane2020}. As a word of caution, the dust temperature estimated for J0217-0208 and COS-3018555981 has large uncertainties because the data available for the two sources are poor (one or two detections and an upper limit) and because of the assumption of $\beta=1.5$. In particular, for COS-3018555981, we show the upper limit on $T_{\rm dust}$ derived in \citet{witstok2022}. Even if these two sources were excluded from our fit of the $T_{\rm dust}-z$ relation, our results would not change. As discussed in \citet{sommovigo2022}, the scatter seen in $T_{\rm dust}$ at fixed redshift can be explained by variations in column density and metallicity within sources for optically thin and thick galaxies, respectively. This best-fitting relation is shown as a purple line with shaded region in the bottom panel of Fig. \ref{fig:c3-tdust-z}.

	\subsubsection{Dust emissivity}
	\label{subsec:beta}

	We now focus on the investigation of the dust emissivity index, $\beta$, which is physically related to the microscopic properties of dust. Fig. \ref{fig:c3-beta-z} shows the redshift distribution of $\beta$ for our sample and the comparison samples. As was also found in \citet{witstok2023}, $\beta$ does not evolve with redshift, with a mean value of $\sim 1.6\pm 0.2$ \citep[see also][]{beelen+2006}, indicating that the average dust properties do not change drastically. Interestingly, other studies of SMGs found a value of $\beta$ that was closer to 2 on average \citep[e.g.,][]{dacunha2021,cooper2022,bendo2023,mckay2023,liao2023}. There is no difference (on average) in this case either between the QSOs belonging to the HYPERION sample and the others. As a word of caution, we note that high-$z$ samples are biased toward high luminosity, and this can introduce a bias toward lower values of $\beta$ \citep[see also][]{witstok2023}.  $\beta$ is significantly higher than 2 in only two exceptions, all of which lie in the high-$z$ regime: J0100+2802, J036+03 from our sample, and GN10 from \citet{witstok2023}. 
	
	It is not straightforward to assess the physical reasons of these high $\beta$ value. In principle, $\beta$ depends on the physical properties and chemical composition of the grains, and possibly on environment and temperature. There are cases in which $\beta$ can be higher than 2 \citep[see, e.g.,][]{valiante2011, galliano2018}. Spatially resolved studies in low-z galaxies showed a spread of $\beta$ within a single galaxy, probably due to the temperature mixing, the different properties of the grain populations, or a combination thereof \citep{galliano2018}. \citet{liao2023} discussed how to interpret $\beta$ in terms of dust properties and suggested that the grain composition plays a significant role in determining $\beta$. Simulations conducted by
	\citet{hirashita2014} indicated that a $\beta$ value of 2 can result from emission by either graphite or silicate grains. However, the correlation between the exact grain composition and $\beta$ is unclear. In our case, a physical explanation for the high value of $\beta$ would require studies of the dust grain properties and/or a detailed analysis of the temperature mixing, which is beyond the scope of this work. 
	
	\subsubsection{Dust mass and star formation ratio}
	\label{subsec:dustmass}

	In Fig. \ref{fig:c3-SFR}, we compare the SFR versus dust mass of our sample with results from the PG and WISSH sample of QSOs. In our sample, five QSOs are the first at $z>6$ for which the SFR was derived with the smallest statistical uncertainty based on high-frequency observations. As already mentioned in Sect. \ref{sec:sed}, the uncertainty of $\sim 40\%$ on $T_{\rm dust}$ of J1342+0928 strongly affects the determination of the SFR, which has a large uncertainty. Therefore, this QSO is marked with a green circle in the plot. A correlation between the dust content and the star formation activity is evident in all the three samples, which are at different redshifts, with some scatter. This correlation is thought to be a consequence of the Schmidt-Kennicutt relation, linking the SFR to the dust content through the GDR \citep[see, e.g.,][]{santini2014}. Overall, the SFR is higher in the $z>2$ samples (by $\sim 2$ orders of magnitudes on average), supporting the well-known concept that high-$z$ QSOs are hosted in highly star-forming galaxies. The dust masses are higher at high $z$ by $\lesssim 2$ orders of magnitudes on average as well. 
	
	Observations of galaxies at low and high $z$ \citep{dunne2011,driver2018,pozzi2020,beeston2018} found that the amount of dust in galaxies has decreased by a factor of $\sim 2-3$ during the last $\sim 8$ Gyr, and this was recently reproduced by \citet{parente2023} using the semi-analytic model (SAM) L-GALAXIES2020. Considering an average value, we also observe a mild increase in the dust mass from our HYPERION sample ($M_{\rm dust, HYP}^{\rm mean}=1.8 \times 10^8\ \rm M_\odot$) to the WISSH sample at lower-z ($M_{\rm dust, WISSH}^{\rm mean}=3 \times 10^8\ \rm M_\odot$), and followed by a drop of 2 orders of magnitude at $z\sim 0$. Interestingly, here, we observe a difference within our sample: Four HYPERION QSOs have lower dust mass than the other $z>6$ QSOs, up to more than about one order of magnitude and a factor of 2 on average. However, this is a preliminary result because two-thirds of the sample still has to be analyzed, and we will design future ALMA and NOEMA observations to complete the overview of the sample. 
	
	Even though we focused on QSOs, we recall that the relation between SFR and dust mass has also been studied in normal galaxies \citep{santini2014,dacunha2010,hjorth2014} and in SMGs at $z=1-4$ \citep[e.g.,][]{dudzevi2020}. \citet{santini2014} used \textit{Herschel} observations to estimate the dust mass of a large sample of galaxies extracted from the GOOD-S, GOOD-N, and COSMOS fields, and they performed a stacked analysis on a grid of redshifts, stellar masses, and SFR. Similarly to us, they found correlations between SFR and dust mass at different redshifts, from the local Universe out to $z=2.5$. Their analysis revealed no clear evolution of the dust mass with redshift at a given SFR and stellar mass, indicating that galaxies with similar properties (in terms of SFR and stellar mass) do not show a significant difference in terms of dust content across the cosmic epochs, out to $z=2.5$. 
	
	It is challenging to assess whether or not there is an evolution of $M_{\rm dust}$ with z at fixed SFR for the samples of QSOs we considered because the statistics is still poor at high $z$, especially for the low $M_{\rm dust}$ regime. Moreover, we are not able to explore the relation involving the stellar mass in our sample because $M_{\rm *}$ estimates are not yet available for our high-$z$ sample. Future JWST campaigns devoted to the investigation of the stellar content in QSOs and galaxies at $z\gtrsim 6$ will certainly allow us to perform these studies in detail \citep[as done in, e.g.,][]{harikane2023,santini2023}. 
	
	\subsubsection{Star formation efficiency}
	\label{subsec:SFE}
	
	\begin{figure}
		\centering
		\includegraphics[width=1\linewidth]{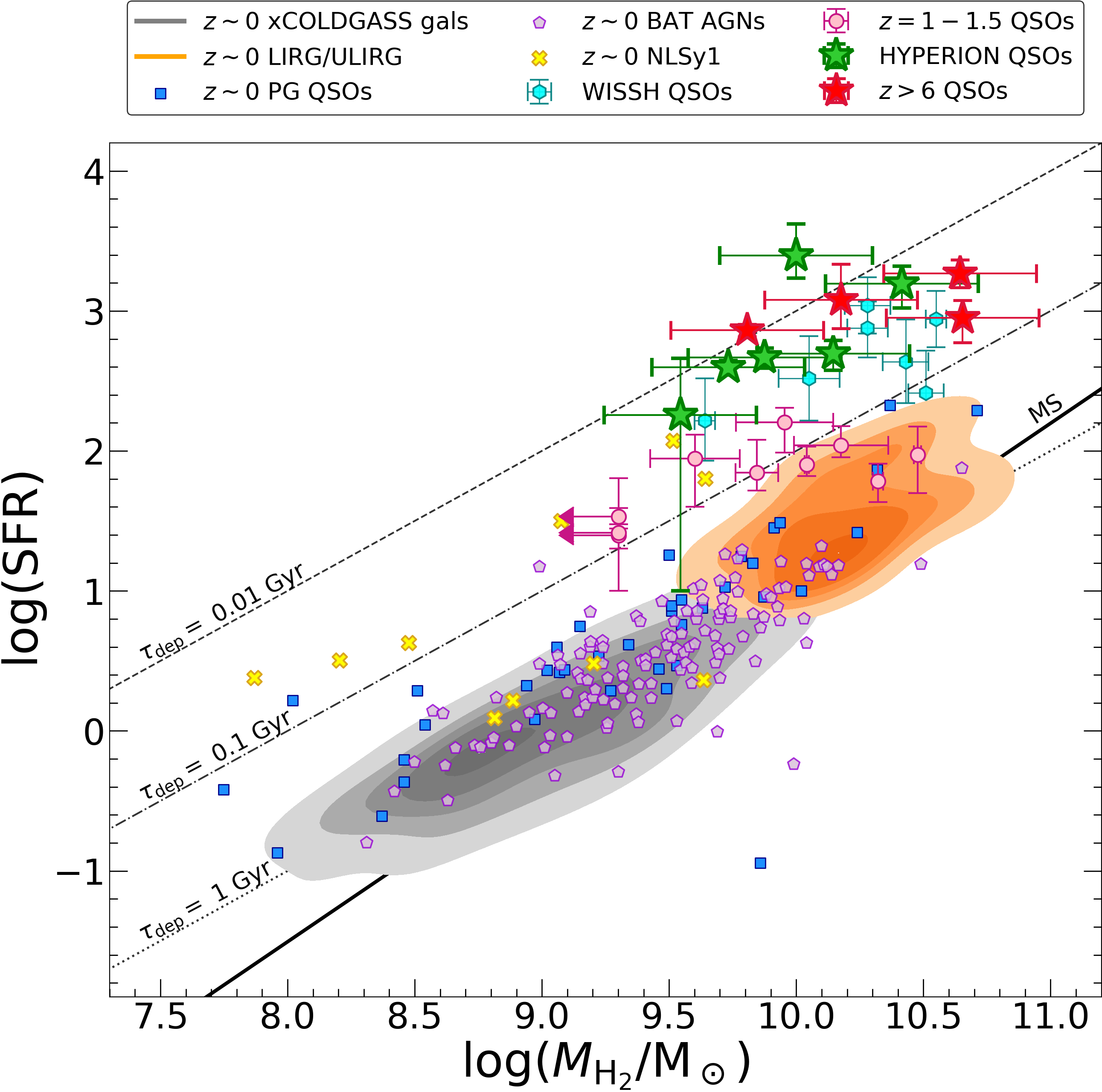}
		\caption{SFR as a function of the molecular gas mass, $M_{\rm H_2}$. The symbols for our sample are the same as in Fig. \ref{fig:c3-tdust-z}. We compare our results with the WISSH sample \citep{bischetti2021}, with $1<z<1.5$ QSOs (pink dots), and with $z\sim 0$ sources such as galaxies, ULIRGs, QSOs, AGNs, and narrow-line Seyfert 1 galaxies (symbols and colors in the legend; \citealt{saintonge2011,saintonge2017,salome2023,shangguan20a,shangguan20b,rosario18,koss21}). The gray lines represent fixed values of the gas depletion time (i.e., the inverse of SFE), which is reported at the top of the line. The solid black line is the galaxy main sequence derived by \citet{sargent2014} considering massive galaxies ($M_*>10^{10}\rm ~M_\odot$) up to $z\sim 2$.}
		\label{fig:SFE}
	\end{figure}
	
	The efficiency with which gas is converted into stars may vary by some orders of magnitude depending on the QSO and galaxy properties, such as luminosity, mass, the presence of a disk, bulges or mergers, and on cosmic time. At fixed redshift, the observed strong variation in the gas star formation efficiency (${\rm SFE}={\rm SFR}/M_{\rm H_2}$) has usually been attributed to the existence of two distinct star formation laws: a secular mode in main-sequence (MS) galaxies, and a star-bursting (SB) mode characterized by short gas depletion timescales ($\tau_{\rm dep}= 1/{\rm SFE}\lesssim 100$ Myr; e.g., \citealt{solomon2005, sargent2014, daddi2010, scoville2023, genzel2010, vallini2024}). The Kennicutt-Schmidt (KS) relation between the $\Sigma_{M_{\rm H2}}$ and $\Sigma_{\rm SFR}$ \citep{kennicutt1998} is the most common method for studying the SFE in galaxies. However, we lack spatially resolved observations for all the QSOs in our sample. As an alternative, we discuss the integrated KS relation, considering $M_{\rm H_2}$ and $\rm SFR$.
	
	The wide range of SFEs, and thus $\tau_{\rm dep}$, that is spanned by different galaxies at different cosmic times is shown in Fig. \ref{fig:SFE}. We compared our results (red and green stars) with the $2<z<5$ WISSH QSO sample, $1<z<1.5$ QSOs \citep{castillo2024}, a sample of narrow-line Seyfert 1 galaxies at $z<0.1$ \citep{salome2023}, 23 PG QSOs at $z<0.1$ \citep{shangguan20a,shangguan20b}, $0.025<z<0.05$ xCOLDGASS galaxies \citep{saintonge2011,saintonge2017}, $0.002<z<0.09$ LIRGs and ULIRGs with CO(1-0) detections from the literature \citep{tacconi2018}, and AGNs at $z<0.05$ selected from the \textit{Swift}-BAT all-sky catalog \citep{rosario18,koss21}. As a reference, we also plot the gas MS computed by \citet{sargent2014} considering a sample of galaxies at $0<z<2$ (solid black line). While discussing this comparison, we recall the possible sources of systematic uncertainties (see also Sects. \ref{sec:gasmass}, \ref{subsec:AGNeffect}): the assumption of the $\alpha_{\rm CO}$, the scaling from the CO ladder for the CO($J\rightarrow J-1$) transition with $J>1$, and the assumption of $T_{\rm dust}$ for the cold-dust SED fitting and of the AGN contribution to the dust heating (for AGN-QSOs only).
	
	Our sample exhibits $\tau_{\rm dep}\sim 10^{-2}$ Gyr, which is smaller by $1-2$ orders of magnitude than those of the comparison samples up to $z\sim 2$. As pointed out in Sect. \ref{subsec:tdust}, this may be connect to the high dust temperatures found in our QSOs. \citet{vallini2024} modeled the variation in the dust temperature in terms of gas depletion timescales, optical depth, and metallicity. Based on the values of $T_{\rm dust}$ in Tab. \ref{tab:c3-sed-res}, the implied $\tau_{\rm dep}$ of the \citet{vallini2024} models agrees with those found from the observed SFR and $M_{\rm H_2}$ (see Tab. \ref{tab:c3-sed-res}). 
	
	Moreover, considering the high-z MS derived by \citet{scoville2023}, our QSO host galaxies are found to be star-bursting sources. The star-bursting nature of our sample is also supported by the fact that $\tau_{\rm dep}$ is lower than that found considering the relation of \citet{tacconi2020} for $\tau_{\rm dep}(z)$ in SB galaxies extrapolated up to $z\sim 6$ \citep[see also][]{vallini2024}. According to the results of \citet{scoville2023}, for SB galaxies such as ours, 70\% of the increased SFR relative to the MS is due to the elevated SFEs and not to the increased gas masses at early epochs.

	\subsubsection{Gas-to-dust ratio}
	\label{subsec:GDR}
	
	\begin{figure}
		\centering
		\includegraphics[width=1\linewidth]{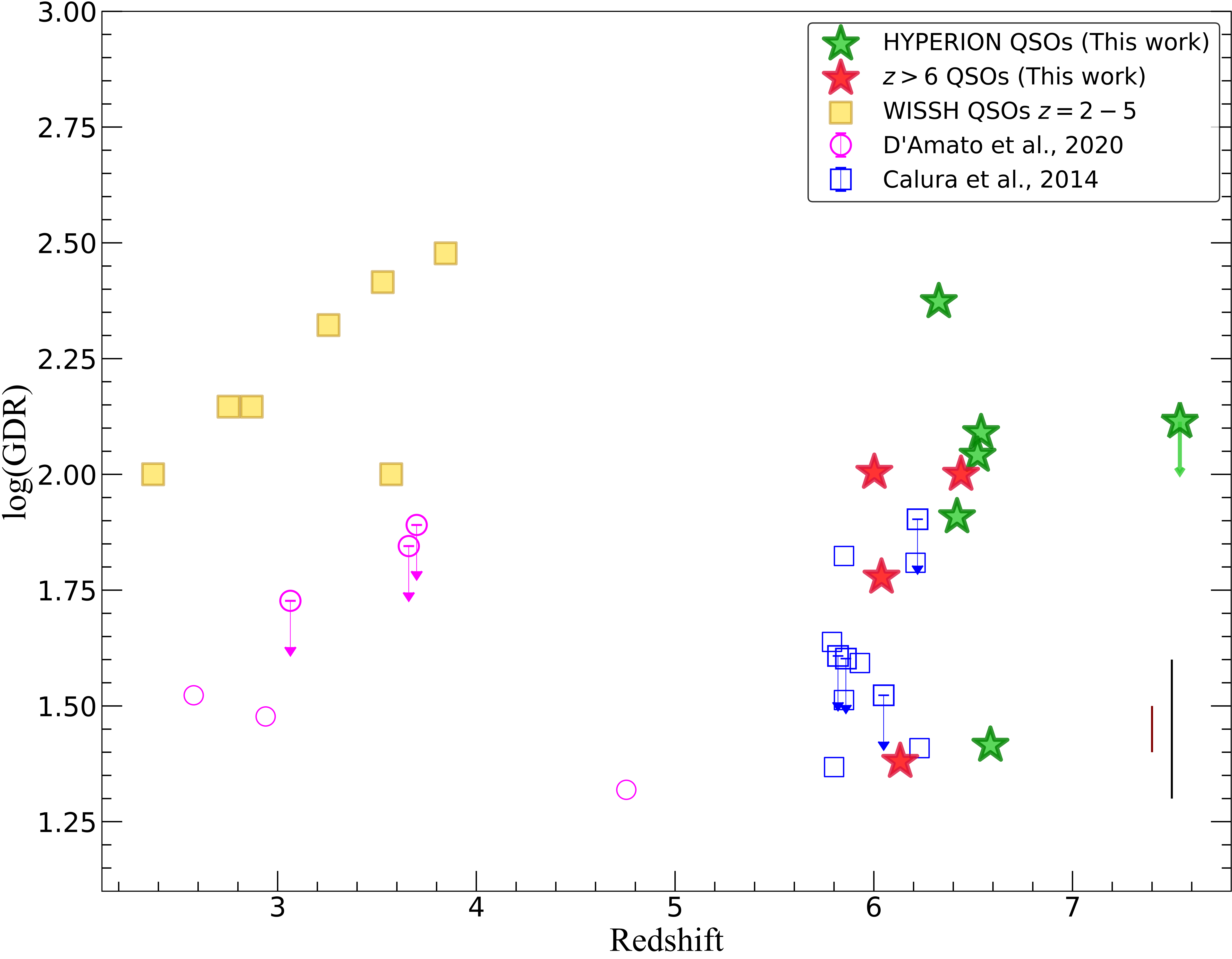}
		\caption{Redshift distribution of the GDR. The symbols and colors for the WISSH and our sample are the same as in Fig. \ref{fig:c3-tdust-z}. We compare our results with the WISSH sample \citep{bischetti2021}, a sample of $2<z<5$ star-forming galaxies hosting a heavily obscured AGN in the Chandra Deep Field-South \citep[magenta dots, ][]{damato2020}, and a sample of $z>5.5$ QSOs \citep[blue squares][]{calura2014}. The two vertical lines at the bottom right side are the systematic uncertainties induced by the choice of $\alpha_{\rm CO}$ and dust mass estimation (black line, $\sim$ 0.3 dex) and $r_{65}$ or $r_{76}$ in computing the gas mass (brown line, $\sim$ 0.1 dex).}
		\label{fig:c3-GDR}
	\end{figure}
	
	\begin{table*}[t]
		\caption{Properties of QSOs in our sample}
		\centering
		\begin{tabular}{lccccccccc}
			\hline
			QSO & z & $\log(M_{\rm BH}/M_\odot)$ & $\log(M_{\rm dyn}/M_\odot)$ & SFR & $G_{\rm BH}$  & $G_{\rm GAL}$ & angles & Class & Refs\\
			& & & & [$\rm M_\odot \ yr^{-1}$] & [$10^{-8}\rm \ yr^{-1}$] & [$10^{-8}\rm \ yr^{-1}$] & [deg] & \\
			\hline
			J0100+2802 & 6.327 & 10.04 $\pm$ 0.27 & 10.51 $\pm$ 0.1 & 396 & 2.5 & 1.8 & 54 & BG (S) & [1],[2] \\
			J036+03 & 6.540 & 9.49 $\pm$ 0.12 & 10.46 $\pm$ 0.1 & 466 & 1.1 & 1.8 & 31 & GG & [1],[2] \\
			J0224-4711 & 6.522 & 9.36 $\pm$ 0.08 & 10.58 $\pm$ 0.3 & 2485 & 2.3 & 6.9 & 19 & GG & [1],[3] \\
			J231-20 & 6.587 & 9.50 $\pm$ 0.09 & 10.15 $\pm$ 0.2 & 496 & 1.0 & 4.5 & 13 & GG & [1],[2] \\
			J1342+0928 & 7.540 & 8.90 $\pm$ 0.14 & 10.56 $\pm$ 0.3 & 180 & 3.1 & 0.5 & 81 & BG & [1],[2] \\
			J1148+5251 & 6.419 & 9.74 $\pm$ 0.03 & 10.18 $\pm$ 0.3 & 1570 & 1.1 & 16 & 4 & GG &  [1],[4] \\
			\hline
			\hline
			J2310+1855 & 6.003 & 9.67$^{+0.06}_{-0.08}$ & 10.72 $\pm$ 0.1 & 1855 & 1 & 4 & 15 & GG & [5],[6] \\[0.1cm]
			J2054-0005 & 6.390 & 9.17 $\pm$ 0.3 & $>10.46$ & 730 & 1.1 & 2.7 & 23 & GG & [2] \\[0.1cm]
			J1319+0950 & 6.133 & 9.53$^{+0.05}_{-0.11}$ & 11.09 $\pm$ 0.1 & 1197 & 0.9 & 1.0 & 43 & S & [5],[2] \\[0.1cm]
			J183+05 & 6.439 & 9.41$^{+0.21}_{-0.41}$ & $>11.11$ & 894 & 1.0 & 0.7 & 54 & BG (S) & [5],[2] \\[0.1cm]
			
			\hline
		\end{tabular}
		\label{tab:c4-prop}
		\flushleft 
		\footnotesize {{\bf Notes.}  Columns: QSO name; redshift; BH mass; dynamical mass; SFR computed in Sect. \ref{sec:sed}; slope of the relation $G_{\rm BH}-G_{\rm gal}$, exemplified by the slope of the arrows in Fig. \ref{fig:c4-mbh-mdyn}; classification of the evolutionary state of each QSO; references for BH and dynamical masses. QSOs above the double black line belong to the HYPERION sample. SFRs are corrected by a factor of 50\% to account for the possible contribution of the AGN to the dust heating (see Sect. \ref{subsec:AGNeffect}). The dynamical mass for J0224-4711 was preliminarily estimated using the [CII] FWHM from an archival ALMA observations in band 6. This work is still in progress (Tripodi in prep.). Classification: BG = BH growth regime, S = symbiotic growth regime, and GG = galaxy growth regime. J0100+2802 and J183+05 are classified twice; see text (Sect. \ref{subsec:evol}). References: [1] \citet{zappacosta2023}; [2] \citet{neeleman2021}; [3] Tripodi in prep.; [4] \citet{riechers2009}; [5] \citet{mazzucchelli2023}; [6] \citet{Tripodi2022}. For each line with multiple references, the first reference is for $M_{\rm BH}$ and $L_{\rm bol}$, which enters in $\dot{M}_{\rm BH}$, and the second reference is for $M_{\rm dyn}$.}
	\end{table*}
	
	We computed the GDR for the QSOs in our sample using the dust and gas masses reported in Tab. \ref{tab:c3-sed-res}, which were either estimated in this work or were taken from previous works in the literature (all references are reported in the table). In Fig. \ref{fig:c3-GDR}, we present the redshift distribution of the GDR in our sample compared with the WISSH sample \citep{bischetti2021}, a sample of $2<z<5$ star-forming galaxies hosting a heavily obscured AGN in the Chandra Deep Field-South \citep[magenta dots, ][]{damato2020}, and a sample of $z>5.5$ QSOs \citep[blue squares][]{calura2014}. \citet{damato2020} derived the dust mass by modeling the SEDs with an MBB in the optically thin regime assuming $\beta=2.0$, and \citet{calura2014} also adopted a MBB in the optically thin regime, but assuming both $T_{\rm dust}=47$ K and $\beta=1.6$ because these two works both mostly relied on a single data point at 250 GHz. Therefore, these latter estimates for $M_{\rm dust}$ are highly uncertain and, for instance, $M_{\rm dust}$ (GDR) would be lower (higher) by $\sim 2.4$ times if a lower temperature, $T_{\rm dust}=33$ K, were assumed. For the comparison samples in Fig. \ref{fig:c3-GDR}, the vertical black line represents an average uncertainty on the GDR of about 0.3 dex, which accounts for the uncertainties on the cold-dust SED modeling. For our sample, the dust masses are in constrast constrained with uncertainties smaller than 30\% on average. Systematic uncertainties are also significant when the gas mass is derived from CO transitions higher than J=1-0 because this requires assuming a scaling between the luminosity of the considered CO transition and of CO(1-0) (see Sect. \ref{sec:gasmass}). This correction depends on the CO excitation ladder, which may vary from one source to the next depending on the ISM conditions. Given the large systematics in the gas mass determination (0.2-0.3 dex for $\alpha_{\rm CO}$ and 20-30\% for $r_{65}$ or $r_{76}$), we did not include error bars in the GDR plot. We stress, however, that the dust masses in our sample are derived with the smallest statistical uncertainties. This has never been achieved before in a sample of QSOs at $z\gtrsim 6$. In low-z galaxies, a GDR$\sim 100$ is typically observed \citep[e.g., ][]{draine2007,leroy2011}, while studies of massive star-forming galaxies and SMGs out to $z\sim 3-5$ found a GDR that might increase with z, with a typical GDR$\sim 120-250$ at $z \sim 2-4$ \citep[e.g., ][]{saintonge2013,miettinen2017}. In some cases, the GDR was also found to be <100, as in recent studies of SMGs \citep{birkin2021,liao2023}. Overall, the WISSH sample and our sample of QSOs show GDRs above 100 on average (this value is also commonly assumed when deriving the dust mass). In particular, our HYPERION QSOs show the highest GDRs on average (GDRs$>100$) of the sources at $z>6$. Two QSOs, J1319+0950 and J231-20, exhibit particularly low ($<50$) GDRs that are comparable with other QSOs from \citet{calura2014}. The tail of low GDRs in our sample, that is, GDR$<50$, can mostly be attributed to the different dust mass values because the gas masses are $\sim 10^{10}\ M_{\odot}$ for all the sources (see Sect. \ref{sec:gasmass}). J1319+0950 and J231-20 both have the highest dust masses in our sample, $M_{\rm dust}\sim (5-6) \times 10^8\ \rm M_\odot$.
	
	\subsubsection{Caveat: Effect of active galactic nuclei on the dust heating}
	\label{subsec:AGNeffect}
	
	As mentioned in Sect. \ref{sec:sed}, when the SFR is derived, the presence of an AGN at the center of the host galaxy likely plays a role in heating the surrounding dust. In theory, the galaxy is characterized by a distribution of dust temperatures that rise toward the center. Both observations and simulations have shown that in the innermost region of the host galaxy ($\sim$ a few hundred parsec), $T_{\rm dust}$ can be as high as some hundred K, and this is mainly due to the presence of the AGN as a heating source \citep{walter2022, dimascia2021, dimascia2022}. The resolution of our observations prevents us from performing a spatially resolved study of the dust emission, and therefore, from isolating the warmer central dust component. We therefore modeled the dust emission with a single temperature BB, that is, we mixed the dust emission that is mainly heated by the stellar distribution with the emission that is mainly heated by the central AGN. This implies that the SFR derived above from the best-fit MBB might be contaminated by the contribution of the AGN to the heating of the dust in the center. 
	
	To overcome this problem, there are two different possibilities that can also be used simultaneously. First, observations with a resolution of $\sim 100-300$ pc can be used to disentangle the warmer dust component, which mainly resides in the center of the colder and more extended component. Then, it is possible to fit the warm and cold dust component separately, either using two MBBs or with an MBB for the cold dust emission and a radiative transfer modeling for the warm dust heated by the AGN. This approach has been adopted at high-$z$ only by \citet{tsukui2023} for a QSO host galaxy at $z=4.4$. They found a warm-dust component with $T_{\rm dust, warm}=87$ K and an AGN contribution to the dust heating of $\sim 60\%$. 
	
	\begin{figure*}[t]
		\sidecaption
		\includegraphics[width=12cm]{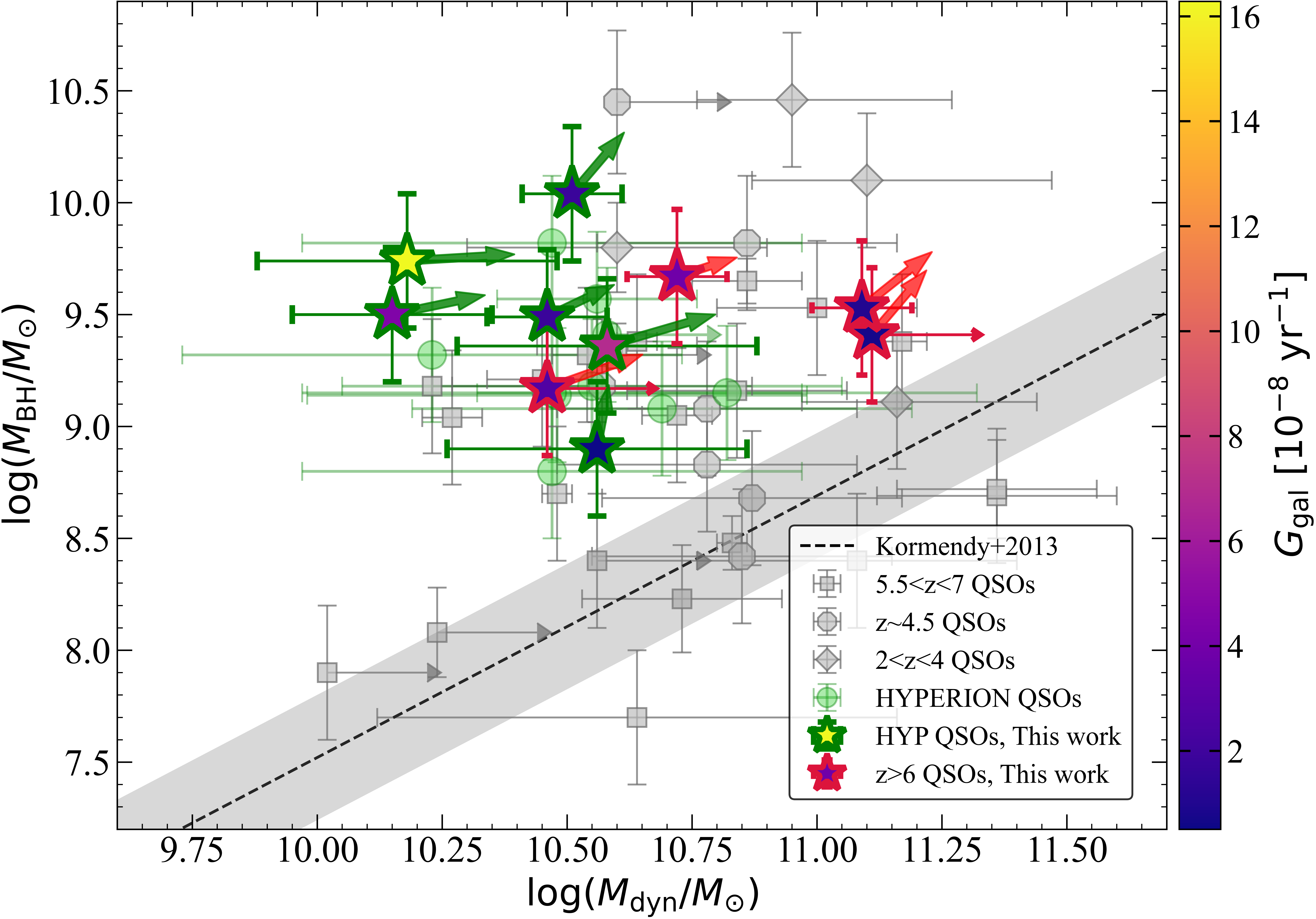}
		\caption{BH mass vs. dynamical mass for our sample (stars with green contours for HYPERION QSOs, and red contours otherwise), WISSH QSOs at $z\sim 2-4$ (gray diamonds; \citealt{bischetti2021}), and luminous $z\sim 4-7$ QSOs (gray dots and gray squares; \citealt{venemans2016,venemans2017a,willott2013,willott2015,willott2017,kimball2015,trakhtenbrot2017, feruglio2018, mortlock2011, derosa2014,kashikawa2015, neeleman2021}). The dashed black line (and shaded area) is the local relation from \citet{kormendy2013}. The light green dots are the remaining HYPERION QSOs for which we were not yet able to perform a detailed study of the dust properties due to a lack of observations in the submm regime. The stars are color-coded based on the value of $G_{\rm gal}$. The thin red arrows indicate upper limits on the dynamical mass.}
		\label{fig:c4-mbh-mdyn}
	\end{figure*}
	
	The availability of very high-resolution data is still quite rare at high $z$, and this approach has indeed not yet been applied to any QSO at $z>4.4$. As an alternative to the spatially resolved study of the SED, radiative transfer models can be used to determine the AGN contribution to the dust heating. \citet{schneider2015} used a radiative transfer code to follow the transfer of radiation from the central source and from stellar sources through the dusty environment of the host galaxy of QSO SDSS J1148+5251 at $z=6.4$. For the stellar sources in the host galaxy, they adopted the SED computed with the PÉGASE population synthesis model using as input the star formation histories, age, and metallicities of the stellar populations predicted by
	GAMETE/QSOdust. To account for the emission of the dust in the host at longer wavelength, they considered two heating sources, (1) the stellar component and (2) the central AGN. This latter can account for up to $\sim 70\%$ of the dust emission, supporting the idea that AGN have a strong impact on heating the surrounding dust. The exact amount of the contribution of AGN to the dust heating depends on the prescription adopted to model the central source and the torus. \citet{schneider2015} tested alternative but still reasonable models and found that the AGN contribution can vary from $30\%$ to $70\%$. \citet{duras2017} also investigated the effect of the AGN to the FIR emission in the WISSH quasars at $1.8<z<4.6$, using the same approach as in \citet{schneider2015}. In particular, considering the least and most luminous QSOs in their sample, they found that the AGN contribution to the FIR fluxes is $43\%$ and $60\%$, respectively, pointing toward a mild trend with luminosity. Therefore, they assumed an average contribution of $\sim 50\%$ to the total FIR luminosity, which also accounts for the uncertainties in the radiative transfer model and is in line with the average result found in \citet{schneider2015}. 
	
	Considering that the dynamic range of bolometric luminosity, redshift, and properties of the host galaxy of the QSOs in our sample is quite small and that these properties are remarkably like those of J1148+5251, the impact of the AGN on the dust heating may be similar. Therefore, we considered an AGN contribution of $\sim 50\%$ to the dust heating. A precise estimate of the AGN impact on the FIR emission would require either a spatially resolved study of the SED or radiative transfer modeling. The former is not yet feasible because of the lack of high-resolution data, and the latter is beyond the scope of this project, but will be the goal of future investigations.
	
	\subsection{Evolutionary paths}
	\label{subsec:evol}

	As mentioned in Sect. \ref{sec:intro}, the BH masses of luminous QSOs at $z\gtrsim 6$ can be extremely high for their cosmic epoch, of about $\gtrsim 10^9\ \rm M_\odot$. The temporal window within which they could have accreted their mass spans approximately 650-750 Myr (considering a possible seeding era at $z=15-22$). These masses are not lower than those of hyperluminous QSOs at lower redshift, meaning that the BH growth had to be a fast process and that the process had to stop with a similarly high efficiency after the rapid build-up. QSOs at $z\gtrsim 6$ indeed appear to lie above the local $M_{\rm BH}-M_{\rm dyn}$ correlation, and thus the BH growth seems to precede that of its host galaxy \citep{volonteri2012,pensabene2020,fan2022}.
	
	Focusing on the BH dominance evolutionary scenario, which seems to be the most likely formation path of local massive galaxies, we can distinguish two large regimes: the first regime is characterized by an intense and predominant growth of the BH, and the second regime is marked by an intense and predominant growth of the host galaxy\footnote{As a note of caution, we must emphasize that during the BH- (galaxy-) dominated regime, the galaxy (BH) is still growing, but more slowly and/or less efficiently than the BH (galaxy).}. Additionally, a transition phase can be imagined in which the growth of the two components is in almost perfect balance (a symbiotic growth). Already at $6<z<8$, SMBHs have reached masses similar to those observed in the most massive local galaxies. This implies that the SMBH growth has to slow down. At these redshifts, we may therefore be able to witness QSOs in their transition phase toward the galaxy regime or in the galaxy regime, for instance, moving toward the local relation in the $M_{\rm BH}-M_{\rm dyn}$ plane.

	In the following, we quantitatively characterize the three different regimes outlined above (BH dominated, galaxy dominated, and symbiotic).
	For this purpose, we defined the growth efficiency of the galaxy as $G_{\rm gal}={\rm SFR}/M_{\rm gal}$, where $M_{\rm gal}= M_{\rm dyn}-M_{\rm BH}$ is the mass of the galaxy, and the SFR was corrected for the QSO contribution. This term is a lower limit to the specific SFR of the galaxy defined as SFR$/M_*$, which might be a better probe of the galaxy growth in principle. We did not use the specific SFR because the stellar mass is not yet available for most high-redshift QSOs. We derived the BH growth efficiency as $G_{\rm BH}=(1-\epsilon)\dot{M}_{\rm BH}/M_{\rm BH}$, where $\epsilon$ is the radiative efficiency, and $\dot{M}_{\rm BH}$ is the BH accretion rate that depends on the bolometric luminosity of the source\footnote{$\dot{M}_{\rm BH}=L_{\rm bol}/(\epsilon ~c^2)$, where $c$ is the speed of light.}. We assumed $\epsilon=0.1$ \citep[e.g., ][]{marconi2004}, and we used the BH mass derived from the MgII emission line. Comparing these two terms, we distinguished among (1) $G_{\rm BH}>G_{\rm gal}$ or black-hole-dominance regime, (2) $G_{\rm BH}=G_{\rm gal}$ or symbiotic growth, and (3) $G_{\rm BH}<G_{\rm gal}$ or galaxy-dominance regime. The proportionality factor between $G_{\rm BH}$ and $G_{\rm gal}$ that allowed us to distinguish among these different regimes can be straightforwardly translated into an angle, that is, an angle $>45^{\circ}$ corresponds to the BH-dominance regime, an angle $\sim 45^{\circ}$ to the symbiotic growth, and an angle $< 45^{\circ}$ to the galaxy regime. The properties derived for our sample of QSOs are summarized in Tab. \ref{tab:c4-prop}. In particular, the SFRs were derived in this work, the BH masses and bolometric luminosities for $\dot{M}_{\rm BH}$ \citep{zappacosta2023, mazzucchelli2023}, and the dynamical masses \citep{neeleman2021, riechers2009, Tripodi2022} were taken from the literature as reported in the last column of Tab. \ref{tab:c4-prop}.
	
	Before we discuss our results, it is important to briefly illustrate the possible caveats. As reported in Tab. \ref{tab:c4-prop}, both $M_{\rm BH}$ and $M_{\rm dyn}$ carry large systematic uncertainties\footnote{The errors reported on $M_{\rm dyn}$ take into account the systematics due to the circular velocity and the radius of the galaxy used to derive $M_{\rm dyn}$ (see also the discussion in \citealt{neeleman2021}). For the majority of the QSOs in our sample, $M_{\rm dyn}$ was derived in \citet{neeleman2021} in a consistent way.} (0.1-0.3 dex). Additionally, even though the SFR was derived with high statistical accuracy and corrected for a reasonable AGN contribution (50\%; see Sect. \ref{subsec:AGNeffect}), this contribution may in principle vary from $\sim$ 30\% to 70\%. These uncertainties (a factor of $\sim$ 2 for $M_{\rm BH}$ and $M_{\rm dyn}$, and a factor of $\sim$ 1.5 for the SFR) affect the inferred evolutionary scenario because $G_{\rm BH}$ may vary by a factor of 2 and $G_{\rm gal}$ by a factor of 3 at most. Keeping in mind these systematics, we discuss our results below, based on the best-fitting values for $M_{\rm BH}$, $M_{\rm dyn}$, and SFR (reported in Tab. \ref{tab:c4-prop}).
	
	We evaluated the evolutionary state of the QSOs in our sample. Fig. \ref{fig:c4-mbh-mdyn} presents our results, where the HYPERION QSOs in our sample are shown as a star with green contours, and the other QSOs in our sample as a star with red contours. The stars are color-coded as a function of $G_{\rm gal}$. The slope of the arrows associated with each star corresponds to the angle reported in Tab. \ref{tab:c4-prop}, that is, to the proportionality factor of the $G_{\rm BH}-G_{\rm gal}$ relation. Given the large systematics quoted above, we did not draw the uncertainty of the arrows for clarity. Overall, the growth of the QSOs in our sample is mainly dominated by the galaxy, or it is symbiotic (see also Tab. \ref{tab:c4-prop}). In particular, the color-code of the stars shows that the closer a QSO to the local relation, the slower the growth of the galaxy on average. 
	
	In \citet{Tripodi2022,tripodi2023b}, we investigated the evolutionary path of J2310+1855 and J0100+2802 in detail. For J0100+2802, we found $G_{\rm BH}>G_{\rm gal}$, suggesting that the BH still dominates the process of BH-galaxy growth in this QSO at $z=6.327$\footnote{The slope of the arrow in \citet{tripodi2023b} is higher than reported in this work. This is because the SFR was computed using the Chabrier IMF rather then the Kroupa IMF (as in this work), which yielded an SFR that was lower by a factor of $\sim 0.7$ than in this work, and therefore, to a lower $G_{\rm gal}$ and a higher slope of the arrow.}. On the other hand, in QSO J2310+1855 at $z\sim6$, AGN feedback might be slowing down the accretion onto the SMBH, while the host galaxy grows fast \citep{Tripodi2022,bischetti2022}. One of the likely causes of the slow-down of the SMBH accretion is radiatively driven AGN winds that impact on the accreting matter, providing enough momentum to stop further accretion, and which can further propagate outward on the scale of the host galaxy. In J2310+1855, the SMBH accretion may be limited by the ionized wind traced by a CIV broad absorption line (BAL) system \citep{bischetti2022}. J2310+18655 also shows evidence of a [CII] outflow approximately located in the central kiloparsec, with an outflow mass $M_{\rm out} = 3.5\times 10^8\ \rm M_{\odot}$. Additionally, \citet{shao2022, butler2023} detected molecular outflows traced by OH and OH+. QSO J231-20, which shows a galaxy growth similar to J2310+1855, is also a BAL QSO \citep{bischetti2022}. Additionally, evidence of high-velocity [CII] emission wings and/or OH absorption wings indicating powerful and fast outflows was found in J2054-0005 \citep{salak2023}, J1148+5251 \citep{maiolino2012, cicone2015}, and tentatively in J1319+0950 \citep{herrera-camus2019}. For all of these QSOs, the angles of the $G_{\rm BH}-G_{\rm gal}$ relation are $\lesssim 45^\circ$. Therefore, the onset of the symbiotic or galaxy dominated regime may be linked to a phase of strong feedback that hampers the BH growth. Three QSOs still experience a BH-dominated growth: J0100+2802, J1342+0928, and J183+05. J0100+2802 and J1342+0928, both belonging to the HYPERION sample, have the highest and lowest BH mass in our sample, respectively. Given that the locally observed SMBH masses do not exceed $10^{11}\ \rm M_\odot$ and that the BH of J0100+2802 is the most massive observed at $z>6$ with already $10^{10}\ \rm M_\odot$, a substantial BH growth is an unlikely prospect. BH-dominated growth therefore is indeed a peculiar scenario for J0100+2802. Conversely, it is likely that J1342+0928 is in the process of strong BH growth given its low BH mass. Interestingly, J183+05 is closest to the local relation. Both in J0100+2802 and J183+05, high-velocity wings were detected in [CII] and OH, respectively \citep{tripodi2023c, butler2023}. The outflows here might also be related to the fact that both QSOs may be approaching a phase of symbiotic growth. For the uncertainties on $M_{\rm BH}, M_{\rm dyn}$, and SFR quoted above, both J0100+2802 and J183+05 would fall in the symbiotic-growth regime on average. The other QSOs in our sample would keep the same classification as reported in Tab. \ref{tab:c4-prop}.
	
	To summarize, our study suggests that QSOs at $z\gtrsim 6$ experience a phase of intense galaxy growth. This may be connected to the emergence of strong outflows that are able to regulate the BH growth. On the $M_{\rm BH}-M_{\rm dyn}$ plane, high-$z$ QSOs appear to converge toward the massive end of the local relation. This suggests that they are viable and plausible candidate progenitors of the massive galaxies found in the local Universe. 
	
	\section{Summary}
	\label{sec:sum}
	
	We exploited newly acquired and archival ALMA observations to perform an analysis of the cold-dust SED in a sample of ten QSOs at $z\gtrsim 6$, deriving the dust properties and SFR with the smallest statistical uncertainty ($<10-25\%$). We also investigated the molecular gas in our sample in order to estimate the GDR. The final goal was to discuss the evolutionary path of the QSOs in our sample by assessing whether high-$z$ QSOs can be considered the progenitors of local massive galaxies. We divided our sample into two subsamples depending on whether they belonged to the HYPERION sample. The HYPERION QSOs experienced an intense and rapid SMBH growth, and we therefore aimed to examine whether they showed similar or distinct properties in terms of host galaxies compared to other QSOs at the same redshift. We summarize our main findings below.
	
	\begin{itemize}
		\item The analysis of the CO(6-5) and CO(7-6) emission lines in select QSOs provides insights into their molecular gas masses, which average around $10^{10}\ \rm M_\odot$, which is consistent with typical values for high-redshift QSOs. Our findings support the picture in which high-$z$ QSO host galaxies have large gas reservoirs that constitute the fuel for star formation. The combination of this information with the SFR and dust masses estimated from the analysis of the SEDs can provide crucial insights into the galaxy assembly and evolution.
		
		\item Proprietary and archival ALMA observations in bands 8 and 9 enabled precise constraints on the dust properties and SFR of four QSOs at $z>6$ for the first time. Taking advantage of these new results, we developed a mean cold-dust SED by combining all the observations available at $50\mu m<\lambda<500\mu m$ for the ten QSOs in our sample. This offered a comprehensive view of the dust properties in the QSO hosts. 
		
		\item Our investigation of the redshift distribution of dust temperatures revealed a large scatter in dust temperature between QSOs and typical star-forming galaxies at fixed redshift, but indicated a general trend of increasing $T_{\rm dust}$ with redshift, as theoretically expected. Considering the whole population of galaxies at $0<z<7$, our best-fitting $T_{\rm dust}-z$ relation is of the form $T_{\rm dust}\propto (1+z)^{0.7\pm 0.1}$, which is steeper than expected from theory ($T_{\rm dust}\propto (1+z)^{0.4}$ (see \citealt{sommovigo2022}). This implies that the variation in $T_{\rm dust}$ in different sources has non-negligible dependences on other physical properties, such as optical depth and metallicity. When we focused on QSOs alone, we recovered the theoretical results with $T_{\rm dust}\propto (1+z)^{0.35\pm 0.04}$.
		
		\item Investigating the variation in the dust emissivity index $\beta$ with redshift, we found that it is approximately constant with $z$ ($\beta\sim 1.6$), indicating that all sources share similar dust properties. Two QSOs, J0100+2802 and J036+03, show $\beta>2$, suggesting that they may have peculiar dust properties. The analysis of the dust extinction curves in the rest-frame UV-optical could provide complementary information on the properties of dust in these QSOs. 
		
		\item All the QSOs in our sample are highly star forming, with an ${\rm SFR}\sim 200-2000\ \rm M_\odot ~yr^{-1}$. Compared with local QSOs, the SFR is higher at $z>2$ by $\sim 2$ orders of magnitudes on average, ranging from few hundred to thousands of $\rm M_\odot ~yr^{-1}$. This supports the well-known concept that high-$z$ QSOs are hosted in highly star-forming galaxies. The dust masses are higher at high $z$ as well, by $\lesssim 2$ order of magnitudes on average. As a preliminary result, the dust masses of the HYPERION QSOs in our sample are lower on average (about a factor of 2) than the other QSOs in our sample. 
		
		\item The observed high SFR in our sample yields high SFEs, and thus, very low gas depletion timescales ($\tau_{\rm dep}\sim 10^{-2}$ Gyr). The latter is connected to the observed high dust temperatures and indicates that the nature of our sample's host galaxies is indeed star bursting. In order to derive firm conclusions on the SFE in high-z QSOs, we need a systematic study of the SFE in a larger sample that includes QSOs at lower luminosities. 
		
		\item We found a large scatter of the GDR in our sample, from 30 to 250. The lowest measured GDRs are due to massive reservoirs of dust, $M_{\rm dust} \sim 5-6 \times 10^8\ \rm M_\odot$, which pose challenges to theoretical modeling of dust formation. Interestingly, the HYPERION QSOs show the highest GDRs in our sample owing to their lower dust masses, $M_{\rm dust} \sim 2-9 \times 10^7\ \rm M_\odot$, whereas their $H_2$ gas reservoirs are in line with those previously found in QSOs at the same z ($M_{\rm H_2}\sim 10^{10}\ \rm M_\odot$). 
		
		\item We were able to investigate and discuss the evolutionary path of our sample of ten QSOs with an accurate determination of the dust properties and SFR. Our study suggests that QSOs at $z\gtrsim 6$ experience a phase of rapid galaxy growth. This may be connected to the emergence of strong outflows that are able to regulate the BH growth. BALs were indeed detected in J2310+1855 and J231-20 \citep{bischetti2022}, and evidence of powerful and fast outflows was found in J2310+1855 \citep{Tripodi2022, shao2022, butler2023}, J2054-0005 \citep{salak2023}, J1148+5251 \citep{maiolino2012, cicone2015}, and tentatively in J1319+0950 \citep{herrera-camus2019}. On the $M_{\rm BH}-M_{\rm dyn}$ plane, high-$z$ QSOs appear to be converging toward the massive end of the local relation. This makes high-$z$ QSOs viable and plausible candidate progenitors of massive galaxies found in the local Universe. Interestingly, the average pathway pursued by high-$z$ QSOs to end up as local massive galaxies involves an intense BH growth, which is supported by the upward offset from the local $M_{\rm BH}-M_{\rm dyn}$ relation, followed by a substantial growth of the galaxy. This is in contrast with the picture of the formation of massive local galaxies via symbiotic growth. Our scenario is further supported by the evidence of a stellar bulge in J2310+1855 \citep{tripodi2023a}, which indicates that the structure of QSOs at $z\sim 6$ is surprisingly similar to that typical of local massive galaxies.
	\end{itemize}

	\vspace{1cm}
	\noindent \textit{Acknowledgments.} We thank the referee for providing insightful comments that helped the clarity and robustness of this work.
	This paper makes use of the following ALMA data: ADS/JAO.ALMA\#2019.1.01633.S, 2015.1.00399.S, 2018.1.01790.S, 2021.2.00151.S, 2017.1.01472.S, 2018.1.01188.S, 2021.1.00934.S, 2021.2.00064.S, 2018.1.01289.S, 2019.1.00672.S, 2017.1.01195.S, 2016.1.01063.S. ALMA is a partnership of ESO (representing its member states), NFS (USA) and NINS (Japan), together with NRC (Canada), MOST and ASIAA (Taiwan) and KASI (Republic of Korea), in cooperation with the Republic of Chile. The Joint ALMA Observatory is operated by ESO, AUI/NRAO and NAOJ. The project leading to this publication has received support from ORP, that is funded by the European Union’s Horizon 2020 research and innovation programme under grant agreement No 101004719 [ORP]. RT acknowledges financial support from the University of Trieste. RT, CF, FF, MB, EP acknowledge support from PRIN MIUR project “Black Hole winds and the Baryon Life Cycle of Galaxies: the stone-guest at the galaxy evolution supper”, contract \#2017PH3WAT. C.-C.C. acknowledges support from the National Science and Technology Council of Taiwan (NSTC 111-2112M-001-045-MY3), as well as Academia Sinica through the Career Development Award (AS-CDA-112-M02). EP, LZ and CF acknowledge financial support from the Bando Ricerca Fondamentale INAF 2022 Large Grant "Toward a holistic view of the Titans: multi-band observations of $z>6$ QSOs powered by greedy supermassive black-holes". RM acknowledges ERC Advanced Grant 695671 QUENCH, and support from the UK Science and Technology Facilities Council (STFC). RM also acknowledges funding from a research professorship from the Royal Society. SC acknowledges support from the European Union (ERC, WINGS,101040227). 
	\textit{Facilities:} ALMA. \textit{Software:} astropy \citep{astropy2022}, Matplotlib \citep{matplotlib2007}, SciPy \citep{scipy2023}, CASA (v5.1.1-5, \citealt{casa2022}).
	
	\vspace{5mm}
	
	% WARNING
	%-------------------------------------------------------------------
	% Please note that we have included the references to the file aa.dem in
	% order to compile it, but we ask you to:
	%
	% - use BibTeX with the regular commands:
	%   \bibliographystyle{aa} % style aa.bst
	%   \bibliography{Yourfile} % your references Yourfile.bib
	%
	% - join the .bib files when you upload your source files
	%-------------------------------------------------------------------
	
	\bibliography{biblio}
	\bibliographystyle{aa} 
	
	\appendix
	
	\section{Individual objects}
	\label{app:a}
	
	\subsection{SDSS J010013.02+280225.8 (HYPERION)}
	
	\noindent For QSO SDSS J010013.02+280225.8 (hereafter J0100+2802) at $z_{\rm [CII]}=6.327$ \citep{wang2019}, \citet{wu2015} estimated a bolometric luminosity of $L_{\rm bol}=4.29\times 10^{14}\rm \ L_{\odot}$ and a BH mass of $M_{\rm BH}=1.24\times10^{10}\ \rm M_{\odot}$, making it the most optically luminous QSO with the most massive SMBH known at $z>6$. Both measurements have been recently confirmed by JWST \citep{eilers2022}. \citet{wang2019} performed a multi-frequency analysis of the dust SED, but they could not obtain a precise determination of the dust properties, concluding that J0100+28 has either a high dust emissivity ($\beta \gtrsim 2$) or a high dust temperature ($T_{\rm dust} \gtrsim 60$ K), or a combination of thereof. 
	
	\subsection{PSO J036.5078+03.0498 (HYPERION)}
	
	\noindent PSO J036.5078+03.0498 \citep[hereafter J036+03, ][]{venemans2015} at $z=6.5405$ was observed for the first time in the Pan-STARRS1 survey \citep{venemans2015}, and it has a BH mass of $M_{\rm BH, MgII}=(2.69-3.09) \times 10^{9}\ \rm M_\odot$ from the analysis of MgII emission line and a bolometric luminosity of $L_{\rm bol}=(2.13-3.16)\times 10^{47}$ erg s$^{-1}$ \citep{zappacosta2023,mazzucchelli2023}. \citet{decarli2022} studied the CO(6-5), CO(7-6) and [CI]$_{2-1}$ emission lines of this QSO host with NOEMA observations, and they found $L_{\rm CO(6-5)}=12.7\times 10^9\ \rm K ~km ~s^{-1} ~pc^2$, $L_{\rm CO(7-6)}=10.7\times 10^9\ \rm K ~km ~s^{-1} ~pc^2$ and $L_{\rm [CI]}=5.7\times 10^9\ \rm K ~km ~s^{-1} ~pc^2$, implying a molecular gas mass of $M_{\rm H2, CO}=5.0^{+0.5}_{-0.6} \times 10^{10}\ \rm M_\odot$ and $M_{\rm H2, [CI]}=7.1^{+1.6}_{-1.4} \times 10^{10}\ \rm M_\odot$. The [CII] emission shows ordered motion, with a clear and regular velocity gradient in the moment-1 map and $v_{\rm rot}/\sigma>3$, and it has a $L_{\rm [CII]}=3.38\times 10^9\ \rm L_\odot$ and a size of $\sim 2.4\times 1.6$ kpc$^2$ \citep{venemans2020, neeleman2021}. From the modeling of the velocity rotation they estimated a dynamical mass of $M_{\rm dyn}=2.9^{+1.1}_{-0.7}\times 10^{10}\ \rm M_\odot$, and from the [CII] underlying continuum emission they derived a gas mass of $M_{\rm H2, cont}=2.8^{+15}_{-1.1} \times 10^{10}\ \rm M_\odot$, assuming a gas-to-dust ratio of 100 and a molecular-to-total gas mass fraction of 0.75. \citet{greiner2021} did not find any companion brighter than $M_{1450}(AB)<-26$ mag within 0.1-3.3$h^{-1}$ comoving Mpc search radius, using the simultaneous seven-channel Gamma-ray Burst Optical/Near-infrared Detector, confirming the results of \citet{venemans2020}, who did not detect any companion for this source using ALMA observation of [CII] emission. 
	
	\subsection{VDES J022426.54-471129.4 (HYPERION)}

	\noindent VDESJ022426.54-471129.4 \citep[hereafter J0224-4711, ][]{reed2017} at $z=6.5222$, firstly discovered by \citet{reed2017}, is one of the most X-ray luminous QSOs at $z>5.5$ and the most X-ray luminous QSO at $z>6.5$ \citep{pons2020,zappacosta2023}. It belongs to the HYPERION sample, the XQR-30 sample and the ASPIRE survey \citep{yang2023}. It has a bolometric luminosity of $L_{\rm bol}=3.47\times 10^{47}$ erg s$^{-1}$, and BH mass of $M_{\rm BH, MgII}=(1.30-2.29) \times 10^{9}\ \rm M_\odot$ from the analysis of MgII emission line \citep{reed2019,wang2021,zappacosta2023,mazzucchelli2023} and $M_{\rm BH, H\beta}=2.15 \times 10^{9}\ \rm M_\odot$, from the analysis of H$\beta$ emission line \citep{yang2023}. It has the most extreme broad and blueshifted [OIII] lines observed to date, even compared to observations of lower-redshift QSOs, with a velocity shift of $-1690$ km s$^{-1}$ relative to the narrow [OIII], suggesting powerful ionized outflows \citep{yang2023}.
	
	\subsection{PSO J231.6576-20.8335 (HYPERION)}
	
	\noindent PSO J231.6576-20.8335 (hereafter J231-20) at $z=6.587$ has been discovered using the Pan-STARRS1 survey \citep{mazzucchelli2017}, and it is one of the brightest objects at $z>6.5$. \citet{decarli2017} detected a [CII]-bright nearby companion at $<$10 kpc separation. \citet{Pensabene2021} performed an extensive study of both the QSO and its companion, detecting [NII], CO(7-6), CO(10-9) emission lines, two OH transitions and their underlying continuum in both of them. Additionally, the CO(15-14), CO(16-15) and three transitions of H$_2$O emission line have been detected for the central QSO. Analyzing the cold-dust SED of both the QSO and its companion, they derived $T_{\rm dust}=54$ K, $M_{\rm dust}=5.1\times 10^8\ \rm M_\odot$ and $T_{\rm dust}=35-46$ K, $M_{\rm dust}=(2.3-3.4)\times 10^8\ \rm M_\odot$, respectively. The estimates on the dust temperature suffer of high uncertainties given that they lack of high-frequency observations to probe the peak of the SED. \citep{neeleman2021} found a dynamical mass $M_{\rm dyn}=1.4\times 10^{10}\ \rm M_\odot$ from the modeling of the velocity rotation curve, and reported a BH mass of $M_{\rm BH} = 4.1\times 10^9\ \rm M_\odot$. Finally, \citet{bischetti2022} classified this as a broad absorption line (BAL) QSO, indicating that J231-20 may be caught in a phase of strong BH feedback.
	
	\subsection{ULAS J134208.10+092838.35 (HYPERION)}

	\noindent ULAS J134208.10+092838.35 (hereafter J1342+0928) at $z=7.54$, the most distant QSO known to date, was discovered by \citet{banados2018} who reported an absolute AB magnitude $M_{1450}=-26.8$, bolometric luminosity of $L_{\rm bol}=10^{13}\ \rm L_\odot$, and an SMBH mass of $8\times 10^8\ \rm M_\odot$. It was followed-up with NOEMA resulting in the detection of bright [CII] emission and upper limits on several CO lines \citep{venemans2017a}. \citet{novak2019} presented ALMA observations of the dust continuum and the ISM of the host galaxy J1342+0928. They well constrained the Rayleigh-Jeans tail of the dust SED, deriving a $M_{\rm dust}=3.5\times 10^7\ \rm M_\odot$ and a SFR$\sim 150 \rm \ M_\odot ~yr^{-1}$, fixing the temperature at 47 K. They also detected many atomic fine structure line, such as [CII], [NII], [OIII], and limits on [CI], [OI] and multiple CO transitions (with a tentative stack detection).
	
	\subsection{SDSS J114816.64+525150.3 (HYPERION)}

	\noindent SDSS J114816.64+525150.3 (hereafter J1148+5251) at $z=6.42$ was discovered by \citet{fan2003} who reported an absolute magnitude of $M_{1450}=-27.82$. It is one of the most studied QSOs at high-$z$. It was observed by Subary Telescope \citep{iwamuro2004}, \textit{Spitzer} \citep{jiang2006, hines2006} and \textit{Herschel} \citep{leipski2013}, and therefore this allowed a full modeling of the SED of this QSO \citep{li2008, valiante2011, schneider2015,carniani2019}. In particular, \citet{schneider2015} derived that the AGN contribution to the dust heating in this QSO can be between 30\% and 70\%. \citet{gallerani2014} detected an exceptionally strong CO(17–16) line in this QSO with the Plateau de Bure interferometer (PdBI) and performed a detailed analysis of the CO SLED using previously detected lower CO transitions \citep{bertoldi2003a,walter2003,riechers2009}.
	
	\subsection{SDSS J231038.88+185519.7}
	
	\noindent QSO SDSS J231038.88+185519.7 (hereafter J2310+1855), first discovered in SDSS \citep{jiang2006,wang2013}, is one of the most FIR-luminous QSOs and one of the brightest optical QSOs known at $z\sim 6$, with $L_{\rm bol}=9.3 \times 10^{13}\ \rm L_\odot$. The redshift measured with the QSO rest-frame UV line emission is $z = 6.00\pm 0.03$ \citep{wang2013}. \citet{feruglio2018} detected and analyzed the CO(6-5) and [CII] emission lines and the submm continuum of J2310+1855, deriving a size of the dense molecular gas of $2.9\pm 0.5$ kpc and of $1.4 \pm 0.2$ kpc for the 91.5 GHz dust continuum and a molecular gas mass of $M({\rm H}_2)=(3.2\pm 0.2)\times 10^{10}\rm M_{\odot}$. They estimated a dynamical mass of $M_{\rm dyn} =  (4.1^{+9.5}_{-0.5})\times 10^{10}\rm M_\odot$, measuring a disk inclination of $i\sim50$ deg. They also inferred the BH mass from the CIV emission line, measured in the X-shooter/VLT spectrum of the QSO, obtaining $M_{\rm BH}=(1.8\pm 0.5)\times 10^9\rm M_\odot$. \citet{shao2019} presented a detailed analysis of the FIR and submm SED and derived a dust temperature of $T\sim 40$ K, a dust mass of $M_{\rm dust}= 1.6\times 10^9\rm M_\odot$, a FIR luminosity $L_{\rm FIR}^{8-1000 \mu m}=1.6\times  10^{13}\ \rm L_{\odot}$, and an SFR$= 2400-2700\ \rm M_\odot yr^{-1}$. \citet{dodorico2018} detected a very metal-poor, proximate damped Lyman $\alpha$ system (DLA) located at z=$5.938646 \pm 0.000007$ in the X-shooter/VLT spectrum of J2310, which was associated with a CO emitting source at $z = 5.939$. This source, called Serenity-18, was detected through its CO(6-5) emission line at [RA, DEC] =  23:10:38.44, 18:55:21.95.

	\subsection{SDSS J205406.49-000514.8}

	\noindent SDSS J205406.49-000514.8 (hereafter J2054-0005) at $z=6.39$ was selected from SDSS stripe 82 with $m_{1450}$ = 20.60, that is, about one magnitude fainter than the objects discovered from the SDSS main survey \citep{jiang2008}. \citet{Leipski2014} reported observations in band z, y, J, H, K and with \textit{Herschel}, however they were not able to fully study its SED due to lack of observations in the mm/submm regime. \citet{wang2013} reported a BH mass of $M_{\rm BH} = 8.6 \times 10^8\rm \ M_\odot$, later updated to $M_{\rm BH} = 1.48 \times 10^9\ \rm M_\odot$ using the MgII emission line \citep{neeleman2021}. \citet{neeleman2021} studied the rotation curve of J2054-0005 using a high-resolution ALMA observation of the [CII] emission of this object, and they determined a lower limit for the dynamical mass of $M_{\rm dyn} > 2.9 \times 10^{10}\ \rm M_\odot$.
	
	\subsection{ULAS J131911.29+095051.4}
	
	\noindent ULAS J131911.29+095051.4 (hereafter J1319+0950) at $z=6.133$ was discovered in the UKIRT Infrared Deep Sky Survey (UKIDSS) with $m_{1450} = 19.65$, which lies in the typical magnitude range of the optically bright $z\sim 6$ quasars selected from the SDSS main survey \citep{mortlock2008}. \citet{wang2011} detected the CO(6-5) emission line and its underlying continuum, deriving a gas mass of $M_{\rm gas}=1.5 \times 10^{10}\ \rm M_\odot$. Later, \citet{wang2013} analyzed the [CII] line emission and underlying continuum, deriving a dynamical mass of $M_{\rm dyn}=12.5 \times 10^{10}\rm \ M_\odot$ (also confirmed by \citealt{shao2017}). \citet{shao2017} estimated a BH mass $M_{\rm BH} = (2.7 \pm 0.6) \times 10^9 \rm \ M_\odot$ from the MgII line, which contributes 2\% of the dynamical mass of the system. \citet{carniani2019} performed a detailed study of the cold-dust SED of J1319+1950 deriving a dust mass of $\log(M_{\rm dust}/M_\odot)=8.8\pm 0.2$ and a dust temperature of $T_{\rm dust}=66^{+15}_{-10}$ K. \citet{herrera-camus2020} tentatively detected the OH 119 $\mu$m doublet in absorption, which is blueshifted with a median velocity that suggests the presence of a molecular outflow, although characterized by a modest molecular mass loss rate of $\sim 200\ \rm M_\odot ~yr^{-1}$.

	\subsection{PSO J183.1124+05.0926}
	
	\noindent PSO J183.1124+05.0926 (hereafter J183+05) at $z = 6.439$ was discovered by \citet{banados2018} using color-color selections from the Pan-STARRS1 database \citep{chambers2016} and follow-up photometric and spectroscopic observations. The [CII] luminosity in this source is the highest among 27 quasars at $z>6$ surveyed in \citep{decarli2018}. J183+05 showed a clear velocity gradient however, given the resolution of the observation, only an upper limit to dynamical mass has been derived $M_{\rm dyn} >1.3 \times 10^{11}\ \rm M_\odot$ from the study of its rotation curve \citep{neeleman2021}. It has a BH mass of $M_{\rm BH} = 3.0\times 10^9\ \rm M_\odot$ derived from MgII emission line. Recently, \citet{decarli2023} analyzed the cold-dust SED of this QSO in detail, deriving $T_{\rm dust}=47 \pm 2$ K, $M_{\rm dust}= (8.7 \pm 1.1) \times 10^{8} \ \rm M_\odot$ and ${\rm SFR}=1330\ \rm M_\odot ~yr^{-1}$. They also presented a multi-line study of this object, reporting detections of [CII], [OIII], [NII], CO(7-6), OH and two H$_2$O transitions.
	
	\vspace{0.2cm}
	
	\section{Analysis of continuum emission from archival observations}
	\label{app:b}
	
	\subsection{QSO J036+03} 
	
	\noindent We analyzed the three ALMA observations available in B6 and B7 for J036+03, and we derived continuum flux densities and sizes at 243.11 GHz, 260.53 GHz and 338.71 GHz, performing a 2D Gaussian fit with CASA for each observation. All continuum emissions at these frequencies are spatially resolved and the values for flux densities, peak fluxes and sizes are reported in Tab. \ref{tab:c3-table-obs}. 
	
	In order not to miss the fainter and more extended flux, we tapered the higher resolution observations at 243.1 GHz and 260.5 GHz, performing the imaging with \texttt{uvtaper}=[$0.9$ arcsec], reaching a resolution of $\sim 0.7\times 0.7$ arcsec$^2$ and of $\sim 0.88\times 0.83$ arcsec$^2$ respectively. The sources in the tapered maps were fitted with the CASA 2D Gaussian fit, and the new continuum fluxes are reported in bold within brackets in Tab. \ref{tab:c3-table-obs}. At 243.1 GHz, the higher resolution observation missed $\sim 20\%$ of the flux. Similar flux losses were also seen by \citet{wang2019} when analyzing high resolution observations of QSO J0100+2802. We checked that there was no further flux gain if tapering at even lower resolution.

	\subsection{QSO J0224-4711}
	
	\noindent We analyzed the three ALMA observations available in B3 and B6 for J0224-4711, and we derived continuum flux densities and sizes at 95.33 GHz, 245.67 GHz, and 260.51 GHz, performing a 2D Gaussian fit with CASA for each observation. The two B6 continuum emissions are spatially resolved and the values for flux densities, peak fluxes and sizes are reported in Tab. \ref{tab:c3-table-obs}. The emission in B3 is not resolved (see left panel of Fig. \ref{fig:c2-co76}), therefore we considered the peak flux as total flux of the source, which is $0.103 \pm 0.009$ mJy/beam, analogously to B8 and B9.
	
	Also for this object, we tapered the higher resolution observations at 245.67 GHz and 260.51 GHz, performing the imaging with \texttt{uvtaper}=[$0.7$ arcsec], reaching a resolution of $\sim 0.7\times 0.7$ arcsec$^2$. The sources in the tapered maps were fitted with the CASA 2D Gaussian fit, and we found no gain in flux even when tapering both observations at even lower resolution.
	
	\subsection{QSO J2054-0005}
	
	\noindent We analyzed the continuum emission at different frequencies of J2054-0005 from the archival observations reported in Tab. \ref{tab:c3-table-obs}. The analyses were carried analogously to the ones of J036+03 and J0224-4711, and the info about the fluxes and sizes obtained are reported in Tab. \ref{tab:c3-table-obs}.
	
	We tapered the higher resolution observations at 92.26 GHz, 262.6 GHz, and 488.31 GHz using \texttt{uvtaper}=[0.7 arcsec], in order to account for the more extended and fainter emission. We achieved a resolution of 1.0 $\times$ 0.89 arcsec$^2$, 0.88 $\times$ 0.82 arcsec$^2$, and 0.88 $\times$ 0.81 arcsec$^2$ for the 92.26 GHz, 262.6 GHz and 488.31 GHz observation, respectively. We did not find any further emission for the lowest frequency observation, while we gained $\sim 5\%$ of the flux in the other two observations.
	
	\subsection{QSO J231-20}
	
	\begin{figure}
		\centering
		\includegraphics[width=0.95\linewidth]{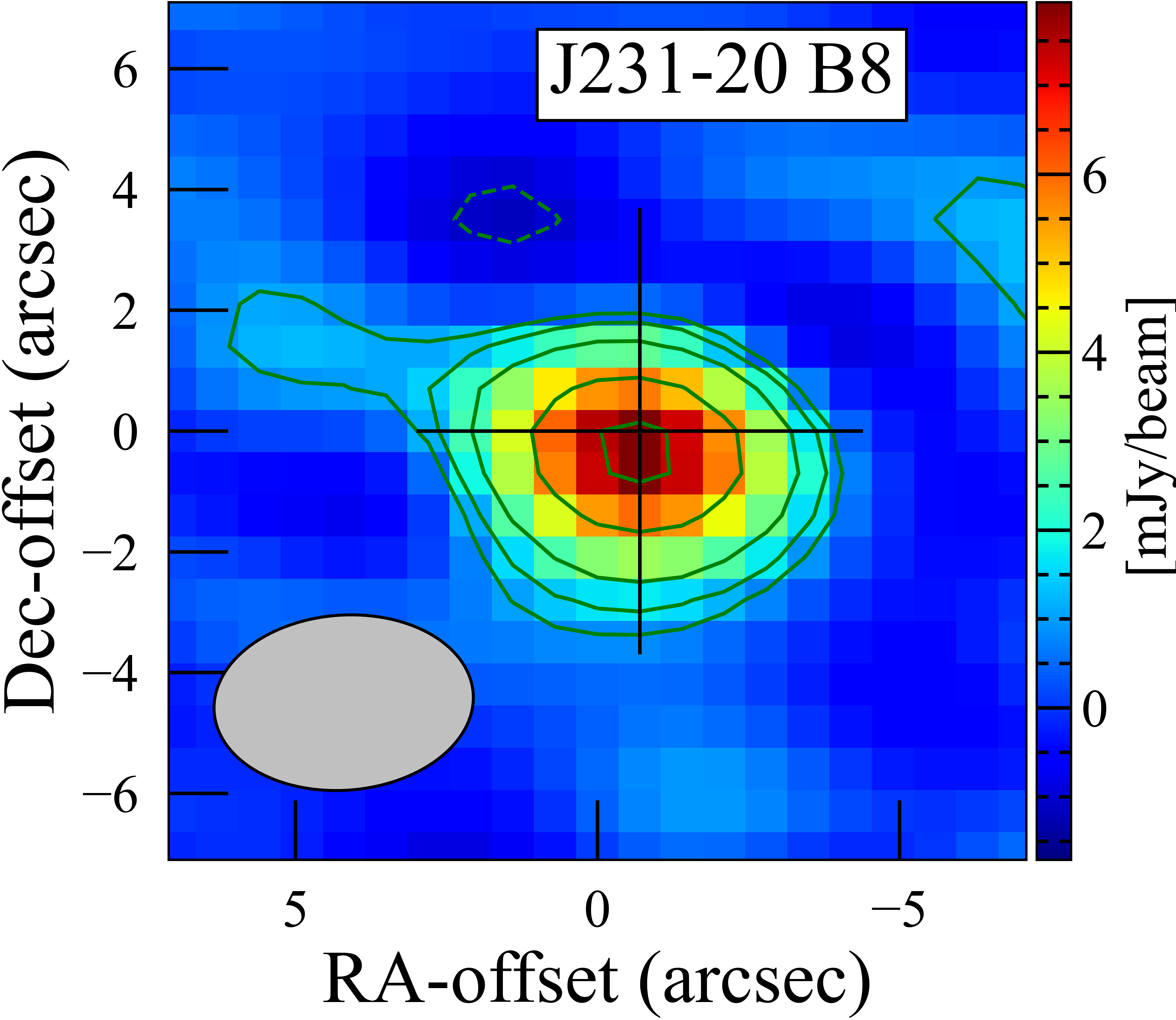}
		\caption{406 GHz dust continuum map of QSO J231-20 (levels $-3,-2,2,3,5,10,\text{and }15\sigma$, $\sigma = 0.5$ mJy/beam). The clean beam ($4.3\times 2.9\rm \ arcsec^2$, PA=-85.8$^\circ$) is indicated in the lower left corner of the diagram. The cross indicates the position of the continuum peak. }
		\label{fig:app-j231}
	\end{figure}
	
	\noindent Fig. \ref{fig:app-j231} shows the B8 continuum emission of QSO J231-20. Details on the analysis can be found in Sect. \ref{subsec:j231}.

 \section{Tables}
 \label{app:tab}

 \begin{sidewaystable*}
		\caption{Continuum emission at different frequencies for J036+03, J0224-4711, J231-20, and J2054-0005}
		\centering       
		\label{tab:c3-table-obs}                        
		\begin{tabular}{c c c c c c c c c c}        
			\hline\hline  
			QSO & Obs freq  & Beam cont & r.m.s. cont & Flux density & Peak flux & Size & Size  & Refs & Project ID \\ 
			& (GHz) & [arcsec$^2$] & [mJy/beam] & [mJy] & [mJy] & [arcsec$^2$] & [kpc$^2$] & & \\
			\hline  
			J036+03 & 106.97 & 6.1 $\times$ 3.5 & 0.038 & 0.13 $\pm$ 0.02 & -- & -- & -- & [1]  & S17CD (NOEMA)\\
			&  243.11 & 0.12 $\times$ 0.08 & 0.007 & 2.05 $\pm$ 0.1  & 0.54 $\pm$ 0.02 & 0.17 $\times$ 0.14 & 0.94 $\times$ 0.78 & TP & 2019.1.01633.S\\
			& & & & ({\bf 2.43 $\pm$ 0.06}) & ({\bf 2.22 $\pm$ 0.03}) & & & & \\
			& 260.53 & 0.18 $\times$ 0.17 & 0.01 & 2.84 $\pm$ 0.13 & 1.28 $\pm$ 0.04  & 0.22 $\times$ 0.18 & 1.22 $\times$ 1.00 & TP, [2] & 2015.1.00399.S \\
			& & & & ({\bf 2.95 $\pm$ 0.03}) & ({\bf 2.71 $\pm$ 0.01}) & & & & \\
			& 338.71 & 0.77 $\times$ 0.54 & 0.03 & 5.54 $\pm$ 0.05 & 4.69 $\pm$ 0.03 & 0.31 $\times$ 0.22 & 1.72 $\times$ 1.22 & TP & 2018.1.01790.S\\
			& 404.99 & 2.10 $\times$ 1.43 & 0.60 & 6.63 $\pm$ 0.39 & 6.63 $\pm$ 0.39 & -- & -- & TP & 2021.2.00151.S\\
			& 670.92 & 1.12 $\times$ 0.97 & 0.50 & 5.60 $\pm$ 0.69 & 5.60 $\pm$ 0.69 & -- & -- & TP & 2021.2.00151.S\\
			\hline  
			J0224-4711 & 95.33 & 3.95 $\times$ 2.32 & 0.016 & 0.094 $\pm$ 0.010 & 0.094 $\pm$ 0.010 & -- & -- & TP & 2017.1.01472.S \\
			& 245.67 & 0.56 $\times$ 0.56 & 0.024 & 2.03 $\pm$ 0.05 & 1.49 $\pm$ 0.02 & 0.35 $\times$ 0.33 & 1.95 $\times$ 1.84 & TP & 2018.1.01188.S \\
			& 260.51 & 0.13 $\times$ 0.11 & 0.013 & 3.01 $\pm$ 0.25 & 0.80 $\pm$ 0.05  & 0.20 $\times$ 0.18 & 1.11 $\times$ 1.00 & TP & 2021.1.00934.S \\
			& 405.19 & 3.87 $\times$ 2.79 & 0.88 &  8.73 $\pm$ 0.38 & 8.73 $\pm$ 0.38 & -- & -- & TP & 2021.2.00151.S\\
			& 670.96 & 1.08 $\times$ 0.95 & 1.3 & 19.9 $\pm$ 0.96 & 19.9 $\pm$ 0.96 & -- & -- & TP & 2021.2.00151.S\\
			\hline
			J231-20 & 406.848 & 4.35 $\times$ 2.90 & 0.50 & 8.43 $\pm$ 0.39 & 8.43 $\pm$ 0.39 & -- & -- & TP & 2021.2.00064.S \\
			& & & & 6.74 $\pm$ 0.31$^{(a)}$ & & & & & \\
			
			\hline
			J2054-0005 & 92.26 & 0.42 $\times$ 0.32 & 0.006 & 0.082 $\pm$ 0.009 & 0.066 $\pm$ 0.004 & 0.27 $\times$ 0.07 & 1.57 $\times$ 0.41 & TP & 2018.1.01289.S \\
			& 262.6 & 0.33 $\times$ 0.29 & 0.019 & 2.82 $\pm$ 0.06 & 2.17 $\pm$ 0.03 & 0.19 $\times$ 0.15 & 1.10 $\times$ 0.87 & TP & 2019.1.00672.S \\
			& & & & ({\bf 2.93 $\pm$ 0.07}) & ({\bf 2.74 $\pm$ 0.04}) &  &  & & \\
			& 263.93 & 1.23 $\times$ 1.12 & 0.010 & 3.08 $\pm$ 0.03 & 2.90 $\pm$ 0.02 & 0.33 $\times$ 0.23 & 1.92 $\times$ 1.34 & TP & 2017.1.01088.S \\
			& 488.31 & 0.43 $\times$ 0.37 & 0.064 & 11.32 $\pm$ 0.25 & 9.05 $\pm$ 0.12 & 0.24 $\times$ 0.15 & 1.39 $\times$ 0.87 & TP & 2017.1.01195.S \\
			& & & & ({\bf 11.71 $\pm$ 0.11}) & ({\bf 10.77 $\pm$ 0.06}) &  &  & & \\ 
			& 674.97 & 0.65 $\times$ 0.57 & 0.50 & 9.87 $\pm$ 0.94 & 8.25 $\pm$ 0.48 & 0.29 $\times$ 0.24 & 1.68 $\times$ 1.39 & TP & 2016.1.01063.S \\
			\hline  
		\end{tabular}
		\flushleft 
		\footnotesize{{\bf Notes.} Columns: (1) Target QSO; (2) observed frequency of the continuum emission; (3) clean beam of the continuum map; (4) r.m.s of the continuum map; (5) flux density of the continuum emission at the nominal resolution of the observation and at a tapered resolution when necessary (in bold and brackets); (6) peak flux of the continuum emission at the nominal resolution of the observation and at a tapered resolution when necessary (in bold and brackets); (7) size in arcsec$^2$ of the resolved continuum emission; (8) size in kpc$^2$ of the resolved continuum emission. (9) References: This paper (TP); [1] \citet{decarli2022}; [2] \citet{venemans2020}. (10) project ID of the observation. $^{(a)}$: flux corrected for the contribution of the companion to the QSO emission.}
		\vspace{0.5cm}
	\end{sidewaystable*} 
	
	\begin{sidewaystable*}
		\caption{Details of the observations targeting the CO line emissions}
		\centering       
		\label{tab:c2-table-obs}                        
		\begin{tabular}{c c c c c c c c c}        
			\hline\hline  
			QSO & Obs freq  & Beam cont & r.m.s. cont & line (spw:chans) & Beam line & r.m.s line & chan width & Project ID \\ 
			& (GHz) & [arcsec$^2$] & [mJy/beam] & & [arcsec$^2$] & [mJy/beam] & [km s$^{-1}$] & \\
			\hline
			J0224-4711 &  95.33 & 3.95 $\times$ 2.32 & 0.016 & CO(7-6) (0/1:56-76)$^{(\rm a)}$ & 3.53 $\times$ 2.07 & 0.14 & 43.7 & 2017.1.01472.S \\
			& & & & [CI] (0/1:88-100)$^{(\rm a)}$ & & & & \\
			\hline 
			J1319+0950 & 103.51 & 0.3 $\times$ 0.3 & 0.054 & CO(6-5) (0/1:100-141)$^{(\rm a)}$ & 0.32 $\times$ 0.31 & 0.1 & 24 & 2018.1.01289.S\\
			\hline
			J2054-0005 & 92.26 & 0.42 $\times$ 0.32 & 0.006 & CO(6-5) (0/1:106-130)$^{(\rm a)}$ & 0.39 $\times$ 0.30 & 0.1 & 24 & 2018.1.01289.S\\
			\hline
		\end{tabular}
		\flushleft 
		\footnotesize{{\bf Notes.}  Columns: (1) Target QSO; (2) observed frequency of the observation; (3) clean beam of the continuum map; (4) r.m.s of the continuum map; (5) spw and channels of the detected emission lines; (6) clean beam of the continuum-subtracted cube; (7) r.m.s of the continuum-subtracted cube; (8) channel width of the continuum-subtracted cube; (9) project ID of the observation. $^{(\rm a)}$: spw 0 and spw 1 were combined since they covered almost the same spectral range and in order to maximize the S/N of the emission lines. $^{(\rm b)}$: natural weighting. $^{(\rm c)}$: Briggs weighting with robust=0.5. $^{(\rm d)}$: since the [CII] of J0110+2802 showed emission up to high-velocities ($\sim 1000$ km s$^{-1}$), we combined the two spectral windows in the upper side band to ensure a reliable continuum subtraction.}
	\end{sidewaystable*} 

 \begin{sidewaystable*}
		%\begin{table}
		\caption[Results of the SED fitting for QSOs in our sample]{Results of the SED fitting}
		\centering       
		\label{tab:c3-sed-res}                        
		\begin{tabular}{l l c c c c | c c c c | c c }        
			\hline\hline
			& &  J036+03 & J0224-4711 & J231-20 & J2054-0005 & J183+05 & J1342+0928 & J1319+0950 & J1148+5251 & J0100+2802 & J2310+1855\\
			\hline
			HYP &  & Y & Y & Y & N & N & Y & N & Y & Y & N\\
			z & & 6.540 & 6.522 & 6.587 & 6.390 & 6.439 & 7.540 & 6.133 & 6.419 & 6.327 & 6.003 \\
			$M_{\rm dust} $ & $[10^7\ M_{\odot}]$  & 6.1 ± 2.5 & 9.3 ± 2.7 & 53 ± 5 & 10.6 ± 2.4 & 45 ± 9 & 2.6 ± 1.4 & 63 ± 29 & 32 ± 8 & $2.3\pm 0.8$ & 44 $\pm$ 7\\ [0.1cm] 
			$T_{\rm dust}$ & [K]  & $59\pm 2$ & $85^{+12}_{-8}$ & $51\pm 3$& 64 ± 2 & 54 ± 5 & 55 ± 20 & 66$^{+15}_{-12}$ & 75 ± 8 & $48 \pm 2$ & 71 $\pm$ 4\\[0.1cm]
			$\beta$ & &  2.4 ± 0.2 & 1.7 ± 0.2 & 1.6 ± 0.1 & 2.1 ± 0.1 & 1.7 ± 0.2 & 1.8 ± 0.4 & 1.5 ± 0.3 & 1.5 ± 0.2 & $2.6 \pm 0.2$ & 1.86 $\pm$ 0.12 \\[0.1cm]
			$L_{\rm TIR}$ &  [$10^{12}\ \rm L_{\odot}$]  & 6.2 ± 1.0 & 33.2$^{+22}_{-10}$ & 6.6 ± 1.6 & 9.8 ± 1.0 & 11.9 ± 4.0 & 2.4$^{+3.7}_{-2.3}$ & 16$^{+13}_{-6}$ & 21 ± 7  & 5.3 ± 0.6 & 2.48$^{+0.62}_{-0.52}$\\ [0.1cm]  
			SFR & [$\rm M_{\odot}\ yr^{-1}$] & 466 ± 75 & 2485$^{+1682}_{-768}$ & 496 ± 118 & 730 ± 75 & 894 ± 299 & 180$^{+280}_{-170}$ & 1197$^{+972}_{-449}$ & 1570 ± 524 & 396 ± 48 &  $1855^{+464}_{-389}$\\[0.1cm] 
			Refs-D & & This work & This work & This work & This work & TW,[1] & TW,[2] & [3] & [3] & [4] & [5] \\ 
			\hline
			
			$M_{\rm gas}$ & [$10^{10}\ \rm M_{\odot}$] & 0.7 $\pm$ 0.1 & 1.0 $\pm$ 0.1 & 1.4 ± 0.1 & 0.6 $\pm$ 0.1 & 4.5 & $<0.35$ & $1.5\pm 0.2$ & 2.6 & $0.5\pm 0.2$ & 4.4$\pm$ 0.2\\
			GDR &  & 123 & 110 & 26 & 60 & 100 & $<130$ & 24 & 81 & 236 & 101 \\[0.1cm]
			$\tau_{\rm dep}$ & [Gyr] & 0.01 %0.015
			& 0.004 & 0.03 & 0.008 & 0.05 & $<0.02$ & 0.01 & 0.02 & 0.01 & 0.02 \\[0.1cm]
			Refs-G & & This work & This work & TW,[6] & This work & [1] & [2] & This work & [7] & [1] & [5]\\
			
			\hline                                   
		\end{tabular}
		\flushleft 
		\footnotesize{{\bf Notes}. The results for the QSOs on the left side of the first vertical line are entirely obtained from this work, that is, all observations were analyzed by the author. Between the two vertical lines, the results for J183+05 and J1342+0928 were obtained fitting their observed SEDs presented in \citet{decarli2023} and \citet{novak2019}, respectively; the results for J1319+0950 and J1148+5251 are taken from \citet{carniani2019}. On the right side of the second vertical line, the results for J0100+2802 and J2310+1855 are taken from \citep{tripodi2023b, Tripodi2022}, respectively. The SFRs are computed from $L_{\rm TIR}$ assuming a Kroupa IMF, and they are corrected by a factor of 50\% in order to take the contribution of the AGN to the dust heating into account. The first raw shows whether the QSO belongs to the HYPERION sample. The gas mass for J231-20 is computed from the CO(7-6) luminosity in [6] using the $r_{76}$ derived in Sect. \ref{sec:gasmass} for J1007+2115, and it agrees with that derived from [CI] in [2]. Refs-D: References for the dust properties and SFR. Refs-G: References for the gas mass. Refs: This work (TW); [1] \citet{decarli2023}; [2] \citet{novak2019}; [3] \citep{carniani2019}; [4] \citep{tripodi2023b}; [5] \citet{Tripodi2022}; [6] \citet{Pensabene2021}; [7] \citet{stefan2015}.} 
		
		%\end{table}
	\end{sidewaystable*}

\end{document}